\documentclass[10pt,english]{elsart3p}
\usepackage{natbib}
\usepackage[pdftex]{graphicx}
\usepackage{amssymb}

\newcommand{\ber}{\ensuremath{^7}Be}

\newcommand{\che}{Cherenkov}

\newcommand{\Bipo}{\ensuremath{^{214}}Bi - \ensuremath{^{214}}Po}
\newcommand{\bipo}{\ensuremath{^{212}}Bi - \ensuremath{^{212}}Po}

\begin{document}

\begin{frontmatter}

\title{ The Borexino detector at the Laboratori Nazionali del Gran Sasso }

%
%
\author[Milano]{G. Alimonti},
\author[Lngs,Milano,Deceased]{C. Arpesella},
\author[VT]{H. Back},
\author[Lngs]{M. Balata},
\author[Princeton]{D. Bartolomei},
\author[APC]{A. de Bellefon},
\author[Milano]{G. Bellini},
\author[Princeton2]{J. Benziger},
\author[Genova]{A. Bevilacqua},
\author[Genova]{D. Bondi},
% ...

\vspace{3mm}
\centerline{[Borexino Collaboration]}
\vspace{3mm}

%
%
\address[Milano]{Dipartimento di Fisica, Universit\'a di Milano and INFN Milano,
via Celoria 16, I-20133 Milano, Italy}

\address[Lngs]{Laboratori Nazionali del Gran Sasso, SS 17bis Km 18+910, I-67010
Assergi (AQ), Italy}

\address[VT]{Physics Department, Robeson Hall, Virginia Polytechnic Institute and State University, Blacksburg, VA 24061-0435, USA }

\address[Princeton]{Department of Physics, Princeton University, Jadwin Hall,
Washington Road, Princeton, NJ 08544-0708, USA}

\address[APC]{Astroparticule et Cosmologie APC, 10 rue Alice Domon et L\'eonie Duquet,
75205 Paris cedex 13, France}

\pagebreak

\address[Princeton2]{Department of Chemical Engineering, Princeton University,
Engineering Quadrangle, Princeton, NJ 08544-5263, USA}

\address[Genova]{Dipartimento di Fisica, Universit\'a di Genova and INFN Genova,
via Dodecaneso 33, I-16146 Genova, Italy}

\address[Pavia]{Dipartimento di Fisica, Universit\'a di Pavia and INFN Pavia,
via Bassi 6, I-27100, Pavia, Italy} 

\address[Moscow]{Moscow University, Moscow, Russia}

\address[Monaco]{Technische Universit\"at M\"unchen, James Franck Strasse E15,
D-85747 Garching, Germany}

\address[Peter]{St. Petersburg Nuclear Physics Institute, Gatchina, Russia}

\address[Kurchatov]{RRC Kurchatov Institute, Kurchatov Sq. 1, 123182 Moscow,
Russia}

\address[Dubna]{J.I.N.R., Joliot Curie str. 6, 141980 Dubna (Moscow Region), Russia}

\address[Heidelberg]{Max-Planck-Institut f\"ur Kernphysik,Postfach 103 980,
D-69029 Heidelberg,Germany}

\address[Budapest]{KFKI-RMKI, 1121 Budapest, Hungary}

\address[Cracow]{Institute of Physics, Jagellonian University, ul. Reymonta 4,
PL-30059 Krakow, Poland}

\address[Kiev]{Institute for Nuclear Research, MSP 03680 Kiev, Ukraine}

\address[Perugia]{Dipartimento di Chimica, Universit\'a di Perugia and INFN
Perugia, via Elce di Sotto 8, I-06123 Perugia, Italy}

\address[Deceased]{deceased}

\begin{abstract}

Borexino, a large volume detector for low energy neutrino spectroscopy, 
is currently running underground at the Laboratori Nazionali del Gran Sasso, Italy.
The main goal of the experiment is the real-time measurement of sub MeV solar 
neutrinos, and particularly of the mono energetic (862 keV) $^{7}Be$ electron 
capture neutrinos, via neutrino-electron scattering in an ultra-pure liquid
scintillator.
This paper is mostly devoted to the description of the detector structure, the 
photomultipliers,  the electronics, and the trigger and calibration systems. 
The real performance of the detector, which always meets, and sometimes exceeds, 
design expectations, is also shown. Some important aspects of the Borexino
project, i.e. the fluid handling plants, the purification techniques and the filling 
procedures, are not covered in this paper and are, or will be, published elsewhere (see
Introduction and Bibliography). 

\end{abstract}


\end{frontmatter}

%
%
\section{Introduction}
\label{sec:Intro} 

Borexino is a large volume liquid scintillator detector whose primary 
purpose is the real-time measurement of low energy solar neutrinos. 
It is located deep underground ($\simeq$ 3800~ meters of water equivalent, m w.e.) in the Hall C of the 
Laboratori Nazionali del Gran Sasso (Italy), where the muon flux is 
suppressed by a factor of $\approx 10^6$.

The main goal of the experiment is the detection of the monochromatic neutrinos 
that are emitted in the electron capture decay of $^7Be$ in the Sun \cite{bib:Borex1}. 
This measurement is now in progress, and the very first results 
have been already published in \cite{bib:be7paper}. However, as shown there, the observed 
radioactive background is much lower than expected, which results in a
potential broadening of the scientific scope of the experiment. Particularly, Borexino 
now also aims at the spectral study of other solar neutrino components, such 
as the CNO, pep \cite{bib:c11}  and, possibly, pp and $^{8}$B neutrinos. 

Besides solar physics, the unprecedented characteristics of its apparatus 
make Borexino very competitive in the detection of anti-neutrinos 
($\bar{\nu}$),  particularly those of geophysical origin. The physics goals 
of the experiment also include the detection of a nearby supernova, the 
measurement of the neutrino magnetic moment by means of a powerful neutrino 
source, and the search for very rare events like the electron decay 
\cite{bib:edecay} or the nucleon decay into invisible channels 
\cite{bib:nucleon}.

In Borexino low energy neutrinos ($\nu$)  of all flavors
are detected by means of their elastic scattering of electrons or, in the 
case of electron anti-neutrinos, by means of their inverse beta decay on protons
or carbon nuclei. The electron (positron) recoil energy is converted 
into scintillation light which is then collected by a set of photomultipliers.

This technique has several advantages over both the water \che\ detectors and 
the radiochemical detectors used so far in solar neutrino experiments. 
Water \che\ detectors, in fact, can not effectively detect solar neutrinos
whose energy is below 6 MeV, both because the \che\ light 
yield is low and because the intrinsic radioactive background cannot be 
pushed down to sufficiently low levels. On the other hand, radiochemical experiments 
cannot intrinsically perform spectral measurements and do not detect events in
real time.

An organic liquid scintillator solves the aforementioned problems: the low
energy neutrino detection is possible because of the high light yield,
that in principle allows the energy threshold to be set down to a level of a 
few tens of keV\footnote{However,
the unavoidable contamination of $^{14}$C that is present in any organic
liquid practically limits the "neutrino window" above $\approx$ 200 keV}; the
organic nature of the scintillator, and its liquid form at ambient temperature,
provide very low solubility of ions and metal impurities, and yield the technical 
possibility to purify the material as required. However, no measurement of
the direction of the incoming neutrino is possible and, even more importantly, 
the neutrino induced events are intrinsically indistinguishable from 
$\beta$ and $\gamma$ radioactivity, posing formidable requirements in
terms of radiopurity of the scintillator and of the detector materials.

According to the Standard Solar Model\footnote{Regardless of neutrino 
oscillations which are not relevant at this point}, the order of magnitude 
of sub-MeV solar neutrino interactions is a few tens counts/day for about one hundred tons of 
target material, corresponding to an equivalent activity of a few $\cdot ~10^{-9}$ Bq/kg. 
If one compares this low number with the typical radioactivity of materials (drinking water 
$\simeq$ 10 Bq/kg, air $\simeq$ 10 Bq/kg, rock $\simeq$ 100-1000 Bq/kg) it 
is immediately apparent that the core of the Borexino detector must be 9-10 
orders of magnitude less radioactive than anything on Earth. 
Typical radioactive contaminants in solid materials and water are $^{238}$U 
and $^{232}$Th daughters, and $^{40}$K. Air and therefore normally also 
commercially available nitrogen are typically contaminated by noble gases 
like $^{222}$Rn, $^{39}$Ar and $^{85}$Kr. 

The necessity to measure such a low neutrino flux with a massive detector 
poses severe requirements in terms of radiopurity, not only for the 
scintillator itself, but also for the surrounding materials. Additionally,
the neutrino target (100 t of ``fiducial volume`` in Borexino) must
be almost completely shielded from external $\gamma$ radiation and neutrons
originating from the rock and from the detector materials.

For almost 20 years the Borexino collaboration has been addressing this problem by 
developing suitable purification techniques for scintillator, water, and nitrogen, 
by performing careful material selections, by developing innovative cleaning 
techniques for metal surfaces, and by building and operating a prototype of the
Borexino detector, the Counting Test Facility (CTF). In particular, 
CTF has played a crucial role in this long R\&D phase. It is still the only 
instrument available in the world (except Borexino itself) with the sensitivity 
to measure the radioactive contamination of a liquid scintillator down to levels 
as low as $10^{-16}$ g/g in $^{238}$U and $^{232}$Th\footnote{Here and everywhere in this
paper the unit g/g stands for 1 gram of contamination per gram of solution or material}.
For more details about the specific requirements in terms of radiopurity of the
scintillator and of the detector materials for solar neutrino measurement in 
Borexino see \cite{bib:Borex1} and \cite{bib:Borex2}. For the 
reader's convenience, we summarize here the main requirements:

\begin{itemize}

\item The internal radioactivity of the scintillator must be low enough compared 
to the expected neutrino signal. Particularly, the design goal was 
$<10^{-16}$ g/g in $^{238}$U and $^{232}$Th, $<10^{-14}$ g/g in K$_{nat}$\footnote{K$_{nat}$ is potassium in its natural isotopic abundance.}.

\item The scintillator must be thoroughly sparged with nitrogen gas in order
to remove oxygen (which may deteriorate the optical properties of the scintillator) 
and air borne contaminants (radioactive). The nitrogen purity requirement 
is such that the expected background from $^{222}$Rn, $^{39}$Ar and $^{85}$Kr in
100 t of target scintillator must be less than 1 count/day. This corresponds
to 0.36 ppm for Ar and 0.16 ppt for Kr.

\item The total amount of external $\gamma$ radiation penetrating the central
part of the scintillation volume should be below 1 count/day in 100 t. This puts
stringent requirements on all materials surrounding the detector, the requirements being
more and more stringent for materials closer to the center. 

\end{itemize}

This paper is devoted to the description of the Borexino detector. It is not intended 
to be a complete reference of the Borexino scientific goals, nor will it provide a comprehensive
description of the experiment as a whole. The focus here is the detector, defined
as the collection of scintillator volume, containment vessels, light detection devices 
(photomultipliers and electronics), data acquisition, and calibration systems.
We do not cover here the purification plants (a very large fraction of the
Borexino equipment) nor the purification techniques adopted to purify
scintillator, water and nitrogen. Also, the filling procedures are not covered in
this paper. All these very important parts of the experiment are either already 
published or will be published in the near future.

The paper is structured as follows: section \ref{sec:Borex} gives a general description of 
the detector; section \ref{sec:scintillator} summarizes the main scintillator features; 
section \ref{sec:vessels} describes the Inner Nylon Vessels which contain the 
scintillator and act as ultimate barriers against external contaminations; 
section \ref{sec:InnerDet} describes the main detector with its photomultipliers,
front end electronics, and data acquisition electronics; section 
\ref{sec:OuterDetector} describes the muon detector; sections \ref{sec:Trigger} 
and \ref{sec:Daq} describe the trigger and the data acquisition systems; 
sections \ref{sec:CalibHardware} and \ref{sec:transp} describe the laser
based calibration systems for the photomultipliers and for the monitoring of the
scintillator transparency; section \ref{sec:insertion} describes the insertion 
system for source calibrations.  Finally, the last section provides a brief overview 
of the detector performance on real data. For more details about detector performance
see refs. \cite{bib:be7paper} and \cite{bib:neutrinopaper}.      

\section{General description of the Borexino detector}
\label{sec:Borex}
Borexino is a liquid scintillator detector
designed to provide the largest possible fiducial volume of
ultra-clean scintillator \cite{bib:Borex1},\cite{bib:Borex2}.

\begin{center}
\begin{figure}
\includegraphics[width=0.52\textwidth]{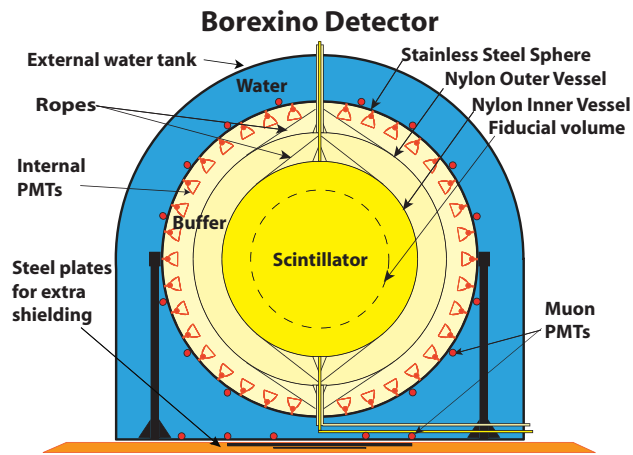}
\caption{Schematic drawing of the Borexino detector.}
\label{fig:detector}
\end{figure}
\end{center}

The detector is schematically depicted in Fig. \ref{fig:detector}. The inner part
is an unsegmented stainless steel sphere (SSS) that is both the
container of the scintillator and the mechanical support of the
photomultipliers. Within this sphere, two nylon vessels separate the
scintillator volume in three shells of radii 4.25~m, 5.50~m and
6.85~m, the latter being the radius of the SSS itself. The inner
nylon vessel (IV) contains the liquid scintillator solution, namely PC
(pseudocumene, 1,2,4-trimethylbenzene $C_6 H_3 (C H_3)_3 $) as a
solvent and the fluor PPO (2,5-diphenyloxazole, $C_{15} H_{11} N O$) 
as a solute at a concentration of 1.5~g/l (0.17~\% by weight). 
The second and the third shell contain PC with a small 
amount (5~g/l) of DMP (dimethylphthalate, $C_6 H_4 (COOC H_3)_2$) that is added
as a light quencher in order to further reduce the scintillation 
yield of pure PC \cite{bib:DMP}.

The PC/PPO solution that we adopted as liquid scintillator satisfies 
specific requirements: high scintillation yield ($\approx 10^4$~photons/MeV), 
high light transparency (the mean free path
is typically 8~m) and fast decay time ($\approx$ 3~ns),
all essential for good energy resolution, precise spatial reconstruction, and
good discrimination between $\beta$-like events and events due $\alpha$ 
particles\footnote{The time profile of the emitted light in an organic scintillator
is usually different for $\beta$-like events and $\alpha$ particles.}. 

Furthermore, several conventional petrochemical
techniques are feasible to purify the hundred of tons of fluids 
needed by Borexino.
The feasibility of reaching the level
of radiopurity required by Borexino was first proven in the tests
performed in the counting test facility (CTF) in 1996
\cite{bib:CTF1-1},\cite{bib:CTF1-2}. Although pure PC is
a scintillator itself, the addition of a small quantity of PPO greatly
improves the time response and shifts the emission wavelength
spectrum to higher values, thus better matching the photomultiplier 
efficiency window.

\begin{center}
\begin{figure}
\includegraphics[width=0.5\textwidth]{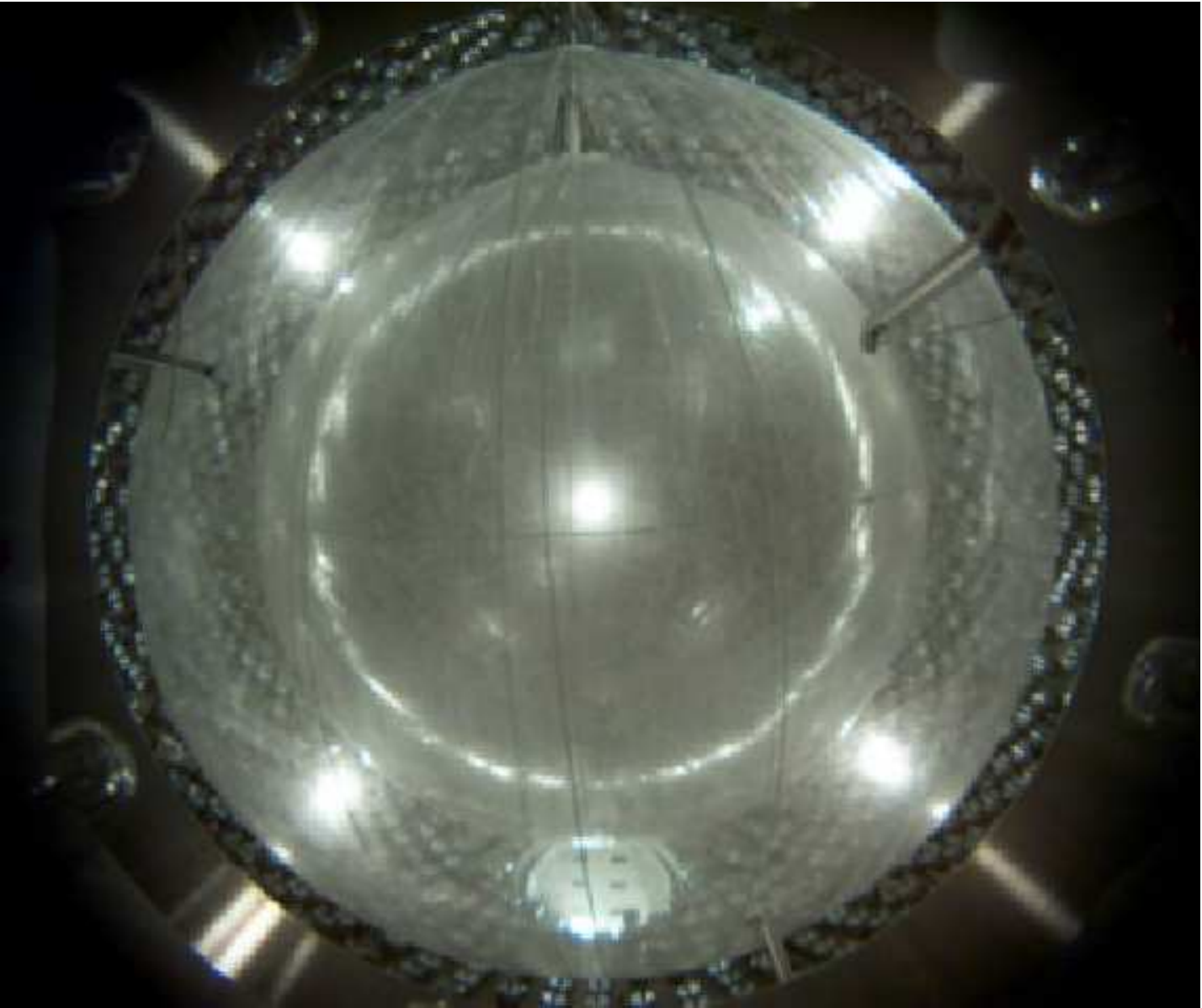}
\caption{The Inner and Outer Nylon Vessels installed and inflated with nitrogen
in the Stainless Steel Sphere. }
\label{fig:nylon-vessels}
\end{figure}
\end{center}

The Inner Vessel is made of 125~$\mu$m thick Nylon-6
carefully selected and handled in order to achieve maximum
radiopurity \cite{bib:nylon}. Since the PC/PPO solution is slightly lighter 
(about 0.4 \%) than the PC/DMP solution, the
Inner Vessel is anchored to the bottom (south pole of the SSS) with
a set of nylon strings. The outer nylon vessel (OV) has a diameter of 11~m 
and is built with the same material as the inner one. The OV
is a barrier that prevents $^{222}Rn$ emanated from
the external materials (steel, glass, photomultiplier materials) to diffuse into
the fiducial volume. Fig. \ref{fig:nylon-vessels} shows the two nylon
vessels inflated in the SSS immediately after their installation.

The buffer fluid between the Inner Nylon Vessel and the SSS (PC/DMP
solution) is the last shielding against external backgrounds.
The use of PC as a buffer is convenient because it matches both the
density and the refractive index of the scintillator, thus
reducing the buoyancy force for the nylon vessel and avoiding optics
aberrations that would spoil the spatial resolution. 

The addition of the DMP quenches the scintillation yield of the buffer 
fluid by a factor of 20.
This is important in order to avoid the unacceptable trigger rate due 
to the radioactivity of the photomultipliers. 

\begin{center}
\begin{figure}
\includegraphics[width=0.49\textwidth]{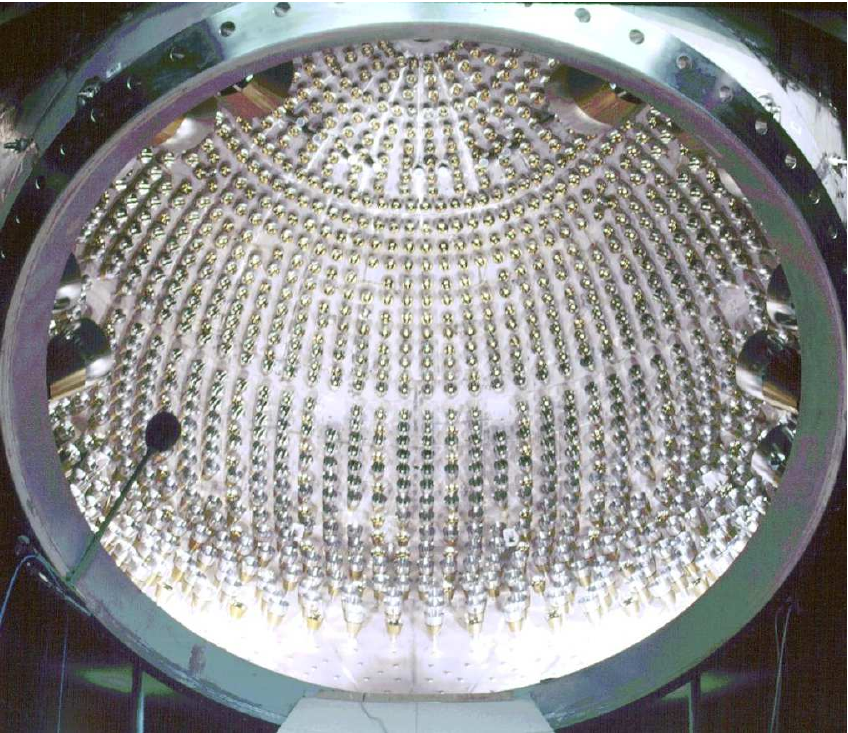}
\caption{Inner surface of the Stainless Steel Sphere. The picture is taken 
from the main SSS door, and shows the internal surface of the sphere with
PMTs evenly mounted inside. The total number of PMTs is 2212. }
\label{fig:SSS-interior}
\end{figure}
\end{center}

The scintillation light is collected by 2212 photomutipliers (PMTs) that
are uniformly attached to the inner surface of the SSS 
(see Fig. \ref{fig:SSS-interior}). All but 384
photomultipliers are equipped with light concentrators that are
designed to reject photons not coming from the active scintillator volume, thus
reducing the background due to radioactive decays originating in 
the buffer liquid or $\gamma$'s  
from the PMTs. The 384 PMTs without 
concentrators can be used to study this background, and to help
identify muons that cross the buffer, but not the Inner Vessel.
The details of the photomultiplier design are described in section 
\ref{sec:pmts_in}.

The SSS is supported by 20 steel legs and  enclosed within a large
tank that is filled with ultra-pure water. 

\begin{center}
\begin{figure}
\includegraphics[width=0.5\textwidth]{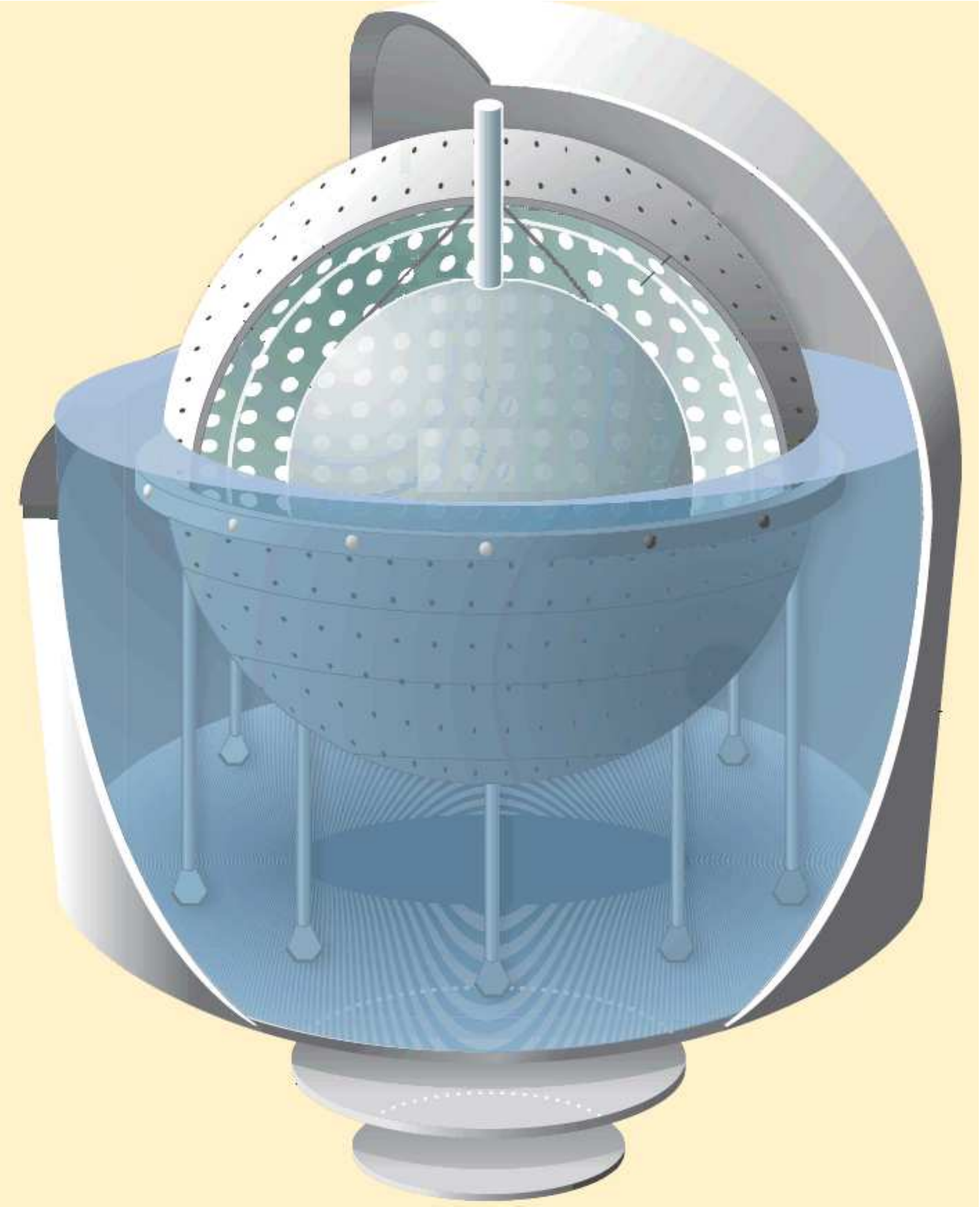}
\caption{A pictorial drawing of the Borexino detector. Inside the Water Tank, the Stainless 
Steel Sphere is supported by 20 steel legs. Within the sphere, the drawings shows
some PMTs (white full circles) and the Inner and Outer Nylon Vessels. The steel plates
beneath the tank improve the shielding against radiation from the rock.}
\label{fig:water-tank}
\end{figure}
\end{center}

The tank (see Fig.
\ref{fig:water-tank}) has a cylindrical base with a diameter of 18~m 
 and a hemispherical top with  a maximum height of 16.9 m.
 
\begin{center}
\begin{figure}
\includegraphics[width=0.46\textwidth]{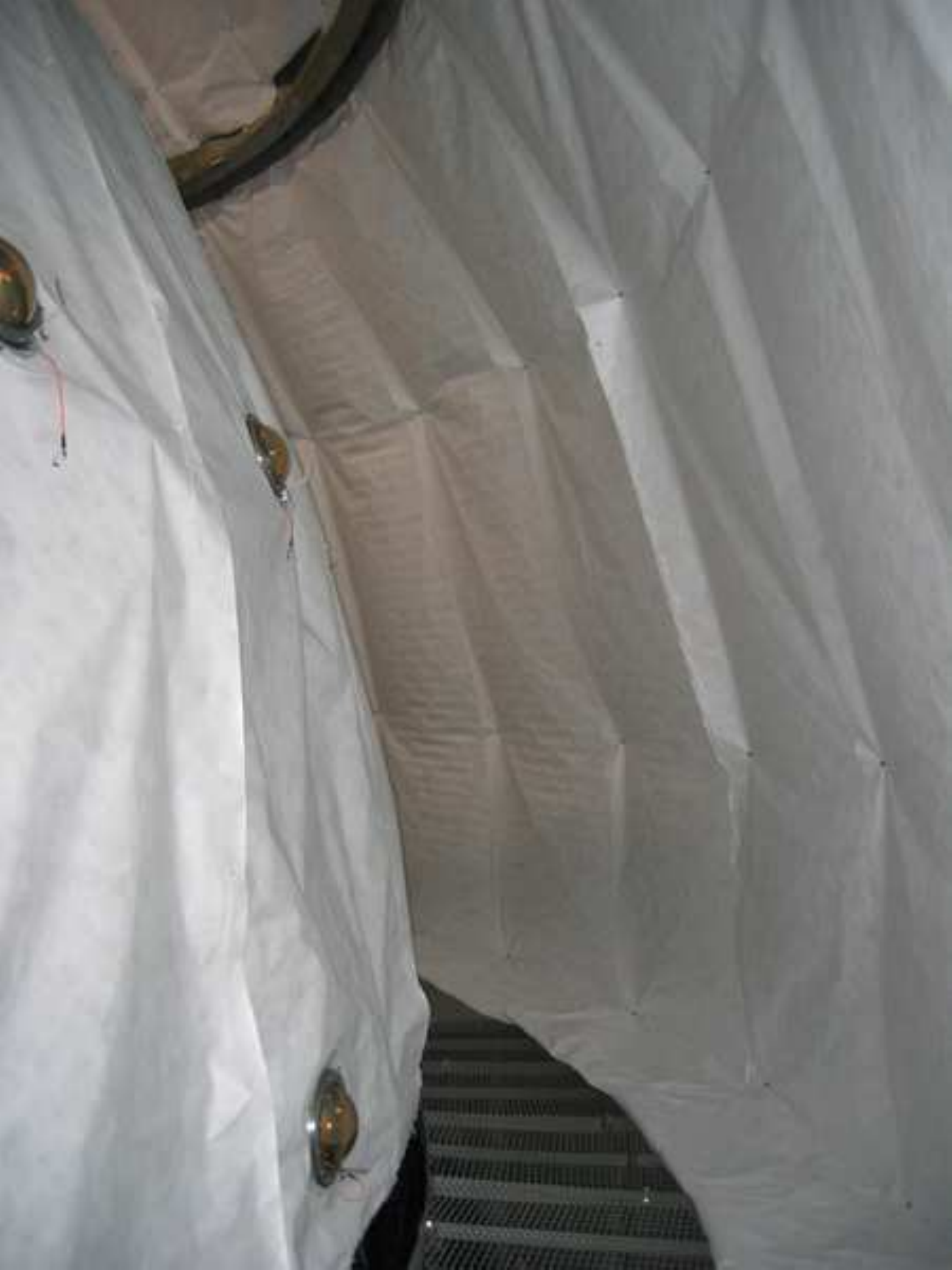}
\caption{The inner surface of the Water Tank covered with a layer of Tyvek. 
The Tyvek sheets improve light collection in the Outer Detector by reflecting 
the photons back into the water. }
\label{fig:tyvek}
\end{figure}
\end{center}

The Water Tank (WT) is a powerful shielding against external background
($\gamma$ rays and neutrons from the rock) and is also
used as a \che\ muon counter and muon tracker. The muon flux, although
reduced by a factor of $10^6$ by the 3800 m w.e. depth of the Gran
Sasso Laboratory, is of the order of 1 m$^{-2}$ h$^{-1}$, corresponding to 
about 4000 muons per day crossing the detector. This flux is well above
Borexino requirements and a strong additional reduction factor 
(about $10^4$) is necessary.
Therefore the WT is equipped with 208 photomultipliers that collect
the \che\ light emitted by muons in water. In order to maximize
the light collection efficiency the SSS and the interior of the 
WT surface are covered with a layer of Tyvek, a white paper-like material 
made of polyethylene fibers (see Fig. \ref{fig:tyvek}).

\section{The Scintillator}
\label{sec:scintillator}

The choice of the scintillator mixture was performed taking into account both
its optical properties and the radiopurity constraints dictated by 
the experiment physics goals. 
The scintillator optical properties have been widely studied on small and 
medium scale samples by using both ultra-violet light as well as $\alpha$, $\beta$
and $\gamma$ radiation \cite{bib:gemma1}, \cite{bib:scint}.
These measurements allowed for the comparison of the main characteristics of different 
scintillators (emission spectrum, time response, light yield, $\alpha/\beta$
discrimination capability) in order to select the most suitable one for Borexino \cite{bib:scint}.

\begin{center}
\begin{figure}
\includegraphics[angle=-90,width=0.55\textwidth]{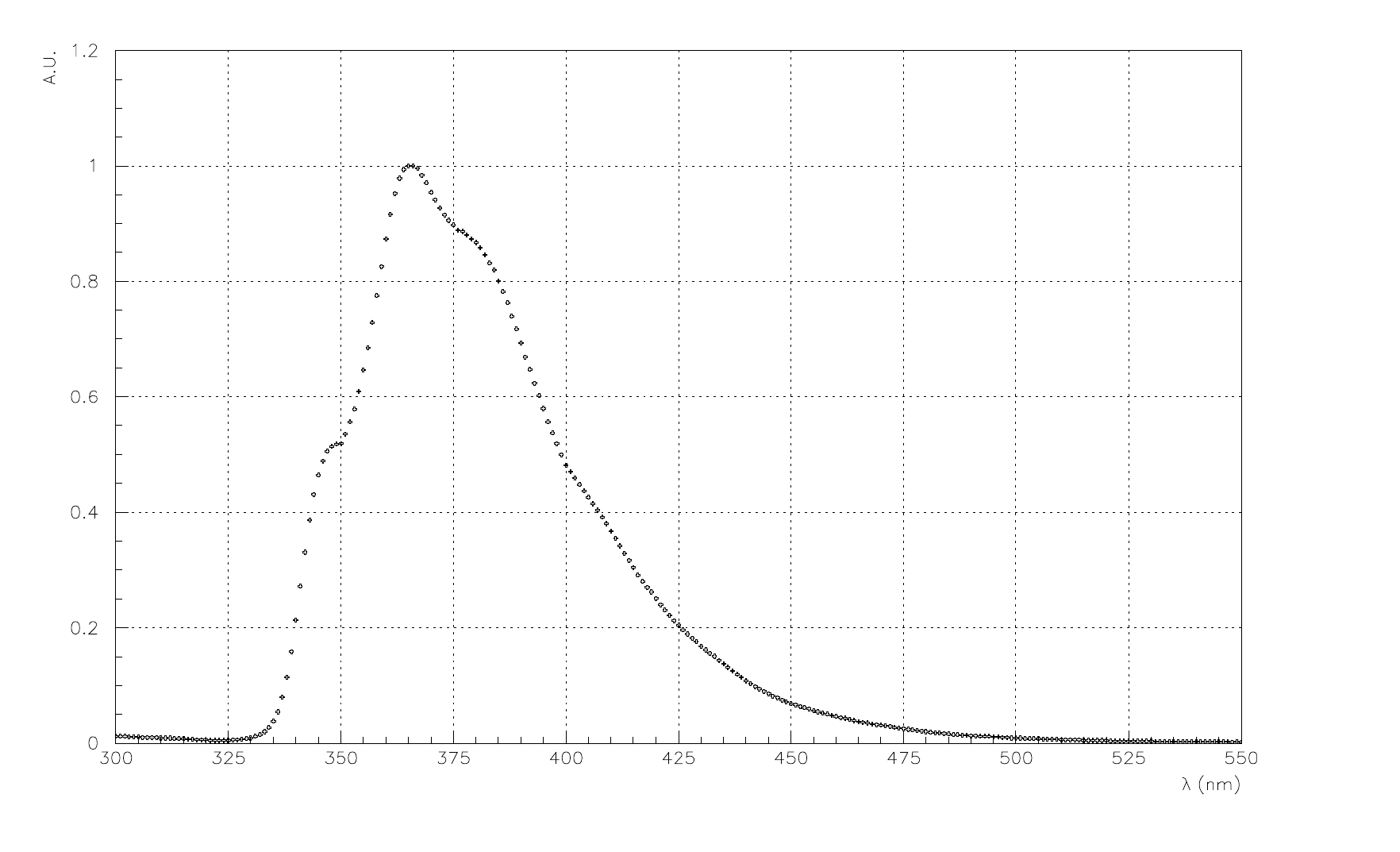}
\caption{Emission spectrum of the PC+PPO mixture used in Borexino}
\label{fig:emis}
\end{figure}
\end{center}

\begin{center}
\begin{figure}
\includegraphics [angle=-90, width=0.53\textwidth,clip=]{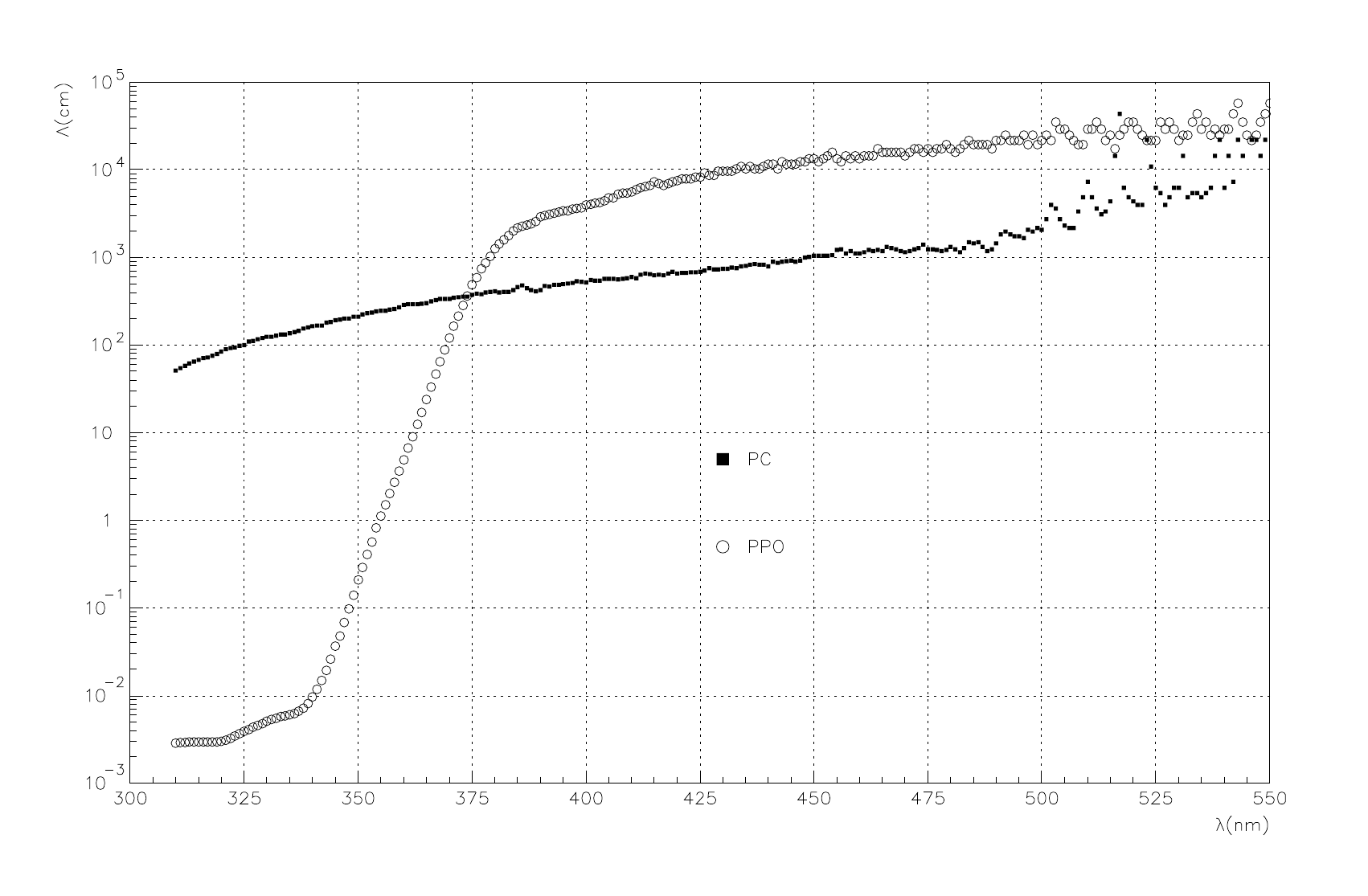}
\caption{Attenuation length of PC (full squares) and PPO (empty circles).}
\label{fig:abs}
\end{figure}
\end{center}

\begin{center}
\begin{figure}
\includegraphics[angle=-90, width=0.53\textwidth,clip=]{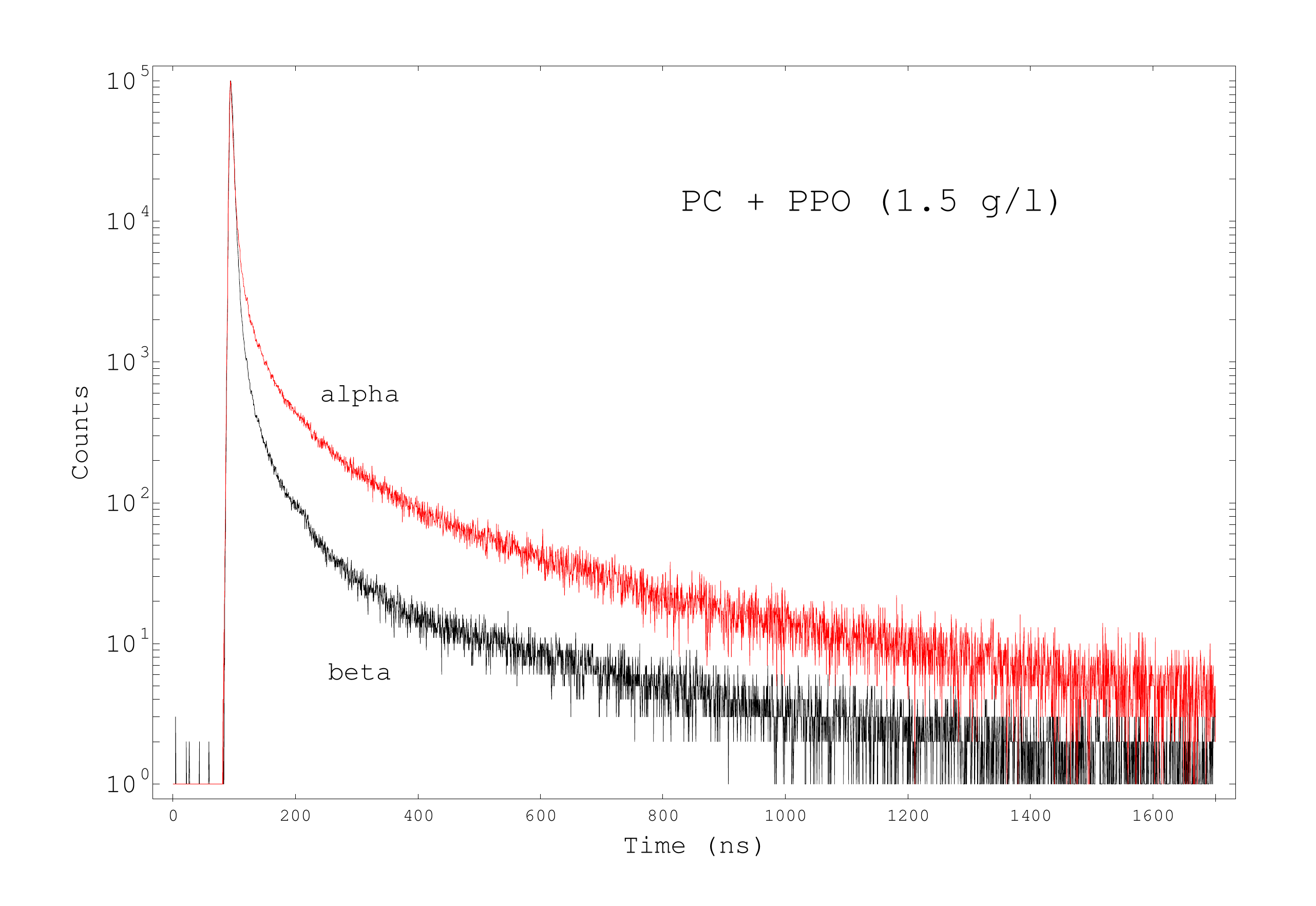}
\caption{Time response of the Borexino scintillator mixture for $\alpha$ and $\beta$ particles.}
\label{fig:alphabeta}
\end{figure}
\end{center}

The final test of a large scale sample of the selected scintillator mixture was performed  
in the Borexino prototype CTF (4 tons of PC+PPO
in a spherical vessel viewed by 100 photomultipliers). Due to  the large size and
4$\pi$ sensitivity of the apparatus, the CTF test made it possible
to single out and quantify the effect of processes like absorption and re-emission
off the fluor (PPO) and scattering on the solvent (pseudocumene). These phenomena were
found to have a significant impact on the overall performance of the detector. In
particular, the energy and position resolution were most severely affected \cite{bib:NIM_scint}.  
The energy resolution critically relies on the high light yield and low auto-absorption 
of the scintillator, as well as having an optimal match between the emitted
light spectrum and the photomultiplier Quantum Efficiency.
The PC+PPO mixture has a light yield of $\simeq$ 10000 photons/MeV 
and its  emission spectrum  peaks at 360 nm (see Fig. \ref{fig:emis}) thus 
matching well the phototube peak efficiency.  This fact, together
with the low absorption of the scintillator\footnote{It should also be noted that 
whenever a photon is absorbed by PPO
it has a high probability (80\%) to be re-emitted with a random direction and with the typical PPO emission spectrum and decay time (1.6 ns). Therefore, in most cases the photon
is not lost, but effectively "slowed down".} and buffer fluid (see Fig. \ref{fig:abs}) leads to an 
overall light collection of $\simeq$ 500 photons per MeV of deposited energy, 
corresponding to a resolution (1 $\sigma$) of $\simeq$ 5\% at 1~MeV.
The time response of the scintillator is also critical: in particular, a fast response 
is needed to achieve  good position resolution, while $\alpha/\beta$ discrimination relies
on the different shapes of the photon time distribution for $\alpha$ and $\beta$ events. 
The time response of the PC+PPO mixture chosen for Borexino was measured on a small 
sample of scintillator and is shown in Fig. \ref{fig:alphabeta}
for both $\alpha$ and $\beta$ particles: the curves can be phenomenologically described 
as the sum of several exponentials having time constants $\tau_i$. 
The first exponential (corresponding to
the fastest time decay) accounts for most of the emitted light (90\% in the case of $\beta$s
and 65\% in the case of $\alpha$s) and has $\tau=3.5$~ns (only a factor 2 larger
than the intrinsic PPO time decay). The difference in the
amount of light emitted at longer
times (10\% in the case of $\beta$s and 35\% in the case of $\alpha$s) is crucial for the effectiveness of
pulse-shape discrimination techniques. See \cite{bib:RanucciAB} and 
\cite{bib:DManuzioThesis}  for details. 


\section{The Nylon Vessels}
\label{sec:vessels}

The design of the Borexino scintillator containment system is a coherent part 
of the general strategy to keep a large volume of scintillator in place while minimizing 
the radioactive background from the materials in contact with it.  
As described in section \ref{sec:Borex}, Borexino exploits a heavy mechanical component, 
the 6.85-m radius Stainless Steel Sphere (SSS), both for the support of large buoyant 
forces due to the need for an external water shield, and also to act as a support 
structure for the PMTs.  (The dual use of the SSS for scintillator containment and 
for PMT support is made possible by the steel having lower radioactivity than the PMTs.)  

In the Borexino design, the SSS and the PMTs are kept as far as possible from the active 
scintillator by an intermediate volume of non-scintillating buffer fluid, in order to 
reduce the background from gamma rays.  With the SSS carrying the heavy structural loads, 
the vessel that separates the scintillator and the inert buffer can be a low-mass, 
relatively delicate membrane made from a material chosen primarily for its properties 
of optical clarity, chemical compatibility with PC and low intrinsic radioactivity.   

The design chosen for the scintillator containment vessel is a thin-walled (0.125~mm thick), 
transparent nylon balloon, 4.25~m in radius, held in place at the center of the SSS.  
Nylon has approximately the same index of refraction (1.53) as PC, and thin nylon sheets 
can be made to be very transparent.  It is chemically compatible with both the scintillator (PC + PPO) 
within this vessel, and the passive buffer fluid (PC + DMP) outside it.  
Nylon also retains its material strength even while immersed in cold water (as during 
the water filling stage of the detector commissioning) for many months.  
In order to prevent the quenching compound (DMP) from leaking into the active scintillator, 
or the fluor (PPO) from leaking into the passive buffer volume, the scintillator 
containment vessel must also be sufficiently leak-tight, to better than 0.01~cm$^3$/s 
liquid leak at a pressure difference of 1~mbar. This should be compared with a measured value
of  0.005~cm$^3$/s \cite{bib:nylon}.

To prevent contamination from the long-lived $^{210}$Pb radon daughter, 
the vessel was constructed inside the Princeton University clean room, which was kept 
low in radon by a dedicated pressure-swing adsorption system 
\cite{bib:nylon2}.  
The ability to assemble the vessel off-site, under very stringent clean-room conditions, 
is an additional advantage of the thin nylon membrane design.
 The Nylon-6 used in the vessel is a blend of Sniamid ADS40T and Ultramid B4, 
 specially extruded at the mf-folien plant in Germany.  
 It is sufficiently radiopure, having a $^{226}$Ra activity of $<$~21~mBq/kg 
 \cite{bib:nylon3}, that it was expected to contribute fewer 
 than seven decays per day of $^{222}$Rn within the Borexino fiducial volume 
 \cite{bib:nylon4}, and in fact contributes fewer than two 
 \cite{bib:be7paper}.  
 The rate of gamma rays within the fiducial volume due to the nylon film of the vessel 
 is thought to be much less than a single event per day \cite{bib:nylon}. 

The scintillator containment vessel (Inner Vessel, IV) is completely surrounded by a second, 
larger Nylon-6 membrane (Outer Vessel, OV) in order to square the effectiveness of the 
system as a barrier against radon atoms diffusing inward from outer parts of the detector.  
The two vessels were pre-assembled in this nested condition even before being shipped to 
Gran Sasso for installation and inflation to their final spherical shapes.  
The OV, 5.5~m in radius, divides the passive buffer volume into two concentric parts.  
As it is not in close contact with the scintillator fluid, its requirements for 
radiopurity and leak tightness are  lower; the measured values of 0.21~$\pm$~0.3~mBq/kg 
of $^{226}$Ra, and 0.1~cm$^3$/s (liquid leak) at 1~mbar $\Delta$P, respectively, 
are more than adequate \cite{bib:nylon3}, \cite{bib:nylon}.

The nylon membranes of both vessels have the material strength (a yield point of 20-70~MPa, 
depending on ambient water content) to survive the forces due to the 0.4\% density 
difference between the active scintillator and passive buffer fluids.  
They were even designed to withstand a 5$^{\circ}$C temperature gradient (which would cause 
an additional 0.5\% density difference) for short time periods.  
However, because their material is flexible, the vessels must be held in place by a 
support structure.  Each nylon sphere is attached to a vertically oriented cylindrical 
"end region" at both top and bottom.  For the IV end regions, very radiopure materials 
(cast nylon and copper) were chosen at the expense of some material strength, 
while stainless steel was used for the OV end regions; these contribute the largest 
fraction of the roughly two gamma events per day in the fiducial volume coming from 
the vessel support structures and instrumentation.  Each IV end region is bolted 
to the inner end of the corresponding OV end region, which in turn is fixed rigidly 
in place to top or bottom of the SSS.  To keep the vessels constrained in roughly 
spherical shapes, more than just two fixed attachment points are needed: each vessel is 
also held in place by two sets of ropes (one set attaching at each end region) that 
loop vertically over and under them.  A third set of ropes goes around each vessel 
horizontally, completing a coarse mesh.  All ropes are made of Tensylon, an ultra high 
density polyethylene (UHDPE).  Though Tensylon is not perfect from a material strength 
standpoint, it is essentially free of potassium (including $^{40}$K, a naturally occurring 
gamma ray emitter), unlike most common rope materials.

The end regions are outfitted with sets of instrumentation for monitoring the detector.  
These include strain gauges that hold the rope ends to the end regions, for monitoring 
the vessels' buoyancy, temperature sensors in the buffer volumes, differential pressure gauges,
 and a set of optical fibers.  The fibers are routed to small PTFE diffuser bulbs 
 fixed at various points on the vessel outer surfaces \cite{bib:pocar}.  
 These bulbs may be lit up with laser light for use in monitoring the shapes and positions 
 of the vessels with the internal camera system, as well as for calibrating the position 
 reconstruction algorithms used in analyzing experimental data acquired with the detector 
 electronics.  Each end region also permits the passage of fluid into the corresponding 
 detector volume for filling and drainage operations.

More detailed information on the Borexino nylon vessels, their fabrication and design, 
and the associated support structures and instrumentation, is available in references 
\cite{bib:nylon}, \cite{bib:cadonati}, \cite{bib:pocar}, \cite{bib:mccarty}.

\section{The Inner Detector}
\label{sec:InnerDet}

As we summarized in section \ref{sec:Borex}, there are two main
detection systems in Borexino: the inner detector, composed of 2212 photomultipliers
collecting the scintillation light inside the SSS and the outer
detector, composed of 208 photomultipliers that detect the \che\ light produced by
muons in water. In this section we describe the inner detector,
focussing on photomultipliers design, front-end and read-out electronics.
Other details concerning the mechanical structure of
the Borexino inner detector can be found elsewhere
\cite{bib:Borex1}, \cite{bib:cadonati}, \cite{bib:pocar}, \cite{bib:nylon}, \cite{bib:pmts1}, 
\cite{bib:pmts2}, \cite{bib:cones}. 
The section is structured as follows: in \ref{sec:pmts_in} we
report the main characteristics of the inner detector photomultipliers and their
PC-proof sealing, while \ref{sec:electronics} describes the inner
detector front-end and read-out electronics.

\begin{table}
\begin{tabular}{|l|ll|}
\hline
 Voltage for gain $10^7$ \dotfill                        	& ~1650~V 		& typical \\
 Maximum voltage \dotfill                                	& ~2200~V 		& max \\
 Maximum cathode-first anode $\Delta$V\dotfill           & ~900~V 		& max \\
 Rise time \dotfill                                      		& ~4/6~ns 			& typical \\
 FWHM \dotfill                                           		& ~7/10~ns 		& typical \\
 Fall time \dotfill                                      		& ~8/12~ns 		& typical \\
 Linearity on peak current (gain $10^6)$ \dotfill          & ~8~mA 		& typical \\
 Linearity on peak current (gain $10^7)$ \dotfill          & ~10~mA 		& typical \\
 Linearity on charge (gain $10^6)$\dotfill 			& ~80/120~pC 	& typical \\
 Linearity on charge (gain $10^7)$\dotfill    		& ~100/150~pC 	& typical \\
 SPE Peak-to-Valley ratio \dotfill                       		& ~2.5 		& typical \\
 Photocathode sensitivity at $420$ nm \dotfill            	& ~26.5 \% 		& typical \\
 Transit time spread fwhm \dotfill                       		& ~2.8~ns & typical \\
 Pre-pulsing ($2\sigma - 20\sigma$) \dotfill             	& ~3/6~\% & max \\
 After-pulsing ($0.05\div12.4\; \mu$ s)\dotfill          	& ~2.5 & typical \\
 Dark current at gain $10^7$ \dotfill                    		& ~25~nA & typical \\
 Dark counts at gain $10^7$ \dotfill                     		& ~3000 & typical \\
 \hline
 \end{tabular}
\caption{Manufacturer specifications for 8'' ETL-9351 photomultiplier tubes.} 
\label{tab:pmts}
\end{table}

\subsection{The Inner Detector Photomultipliers}
\label{sec:pmts_in} \hspace{5mm} 

The necessity to measure the energy of each scintillation event 
and to reconstruct its position by means of a time of flight 
technique put severe constraints on the selection of the best 
photomultiplier (PMT) for Borexino.

Monte Carlo simulations showed that the mean number of photoelectrons 
(p.e.) detected by one PMT are $0.02-2.0$ for events with energy 
between 250-800 keV. It is therefore important that the PMTs have 
a good single electron charge resolution in order to get an 
accurate reconstruction of the event energy. 
The interaction point in the detector is reconstructed using the time
information from the PMTs. The position resolution depends therefore on
the precision of the measurement of the arrival time of a single 
photoelectron.
Hence, the PMTs should demonstrate a good single
electron performance both for the amplitude and the timing response.
Furthermore, in order to minimize the probability of random hits
during the acquisition, the PMTs should feature a low dark rate. Another
parameter to be kept under control is the probability of the delayed
trigger of the system which depends on the PMT after-pulsing rate. 

After several tests (see \cite{bib:Oleg1}, \cite{bib:Oleg2}, \cite{bib:Oleg3}, \cite{bib:Oleg4})   
on the large area tubes available on the market, the 8" E.T.L. 9351 phototube,
formerly Thorn EMI, was chosen for the Borexino inner detector.

The 9351 tube has a hemispherical photocathode  with a 
curvature radius equal to about 11~cm, resulting in a minimum projected
area of 366~cm$^2$. The cathode coating is made of a layer of CsKSb
and the multiplier structure consists of 12 linear focused dynodes
(BeCu). The main nominal characteristics of the tubes are summarized
in Table \ref{tab:pmts}.

The bulb of the tube is made of a low radioactive Schott 8246 glass
that has been carefully studied and characterized underground at Gran Sasso
by means of Ge detectors. Results of these measurements are
published in \cite{bib:Borex2}. We report here just
the main results: $^{238}U$: 6.6 $\pm$ 1.9 $10^{-8}$ g/g,  
 $^{232}Th$: 3.2 $\pm$ 0.3 $10^{-8}$ g/g,  $K_{nat}$: 1.6 $\pm$ 0.4 $10^{-5}$ g/g.
  
The E.T.L. 9351 model was chosen taking into account the impact of the main
technical features of the candidate tubes on the overall detector
performance. For this purpose, the impact of the various PMT parameters
and of their characteristics on the final detector performance was 
analyzed by means of Monte Carlo simulations.

The results of our measurements and simulations, together with the 
information provided by the
manufacturer, were used to define the high priority
specification parameters that are shown in Table
\ref{table:pmt_specs}. These specification parameters meet the tight
scientific requirements of Borexino and, at the same time, are
reasonable enough to avoid a large rejection factor that would have
had an unacceptable impact on the total cost of the system.
\begin{table}[t]
\begin{center}
\begin{tabular}{|c|} \hline
High Priority Parameters \\ \hline
\begin{tabular}{l|c}
Photocathode quantum efficiency & $>$ 21\%  \\
After pulses & $<$ 5\%  \\
Single p.e. transit time spread &  $<$ 1.3~ns  \\
Late-pulsing  &  $<$ 4\%  \\
Dark count rate  &  $<$ 2. $10^{4}$~cps  \\
Single p.e. peak to valley ratio & $>$ 1.5
\end{tabular}\\ \hline
\end{tabular} \\
\end{center}
\caption{\small{High priority parameters and the threshold values 
used to accept or reject a photomultiplier for Borexino. All listed 
conditions were required for a photomultiplier to be accepted.}}
\label{:pmt_specs}
\end{table}

\begin{figure}
\begin{center}
\includegraphics[width=0.5\textwidth]{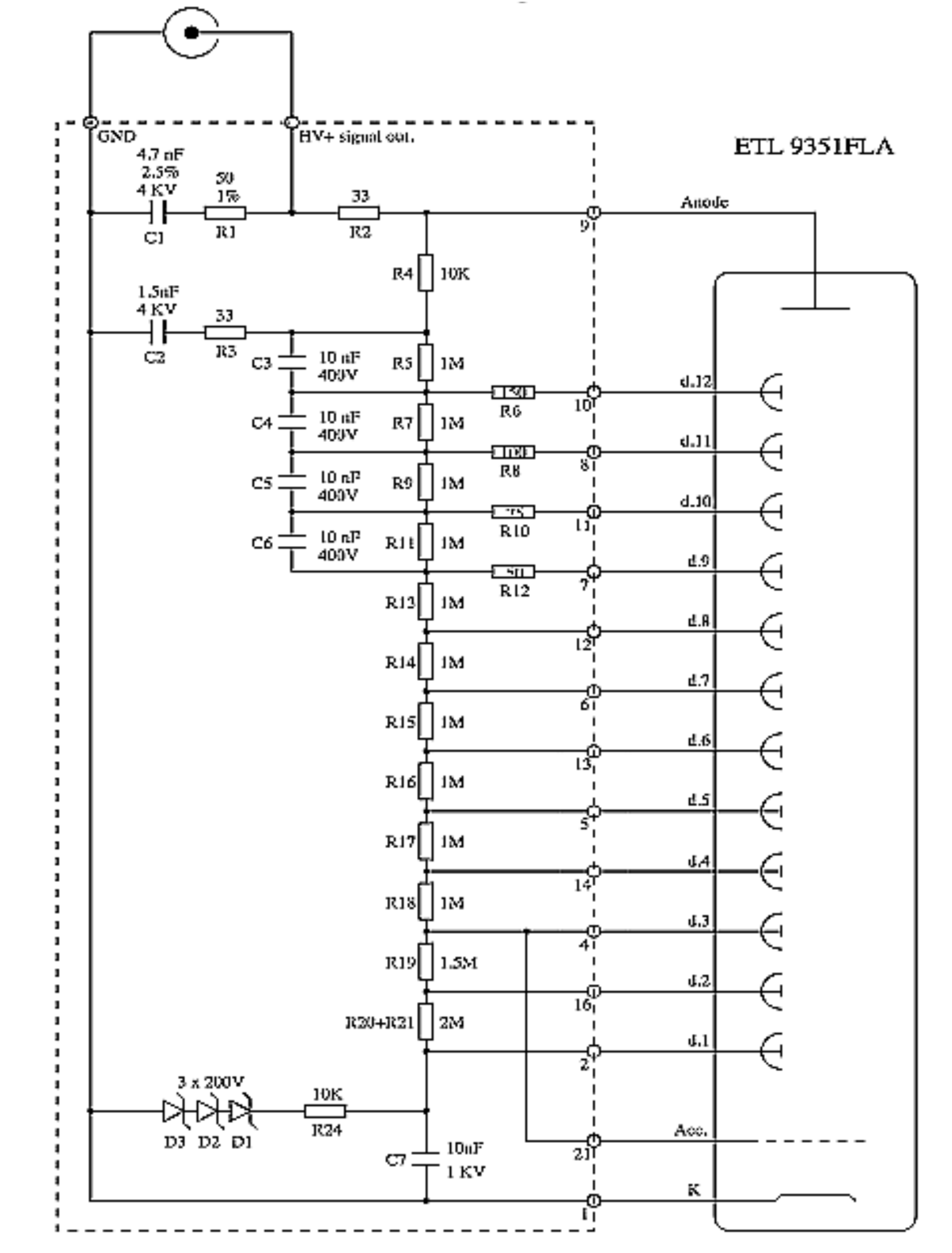}
\caption{The photomultiplier voltage divider.} 
\label{fig:voltagedivider}
\end{center}
\end{figure}

\subsubsection{The Voltage Divider}
\hspace{5mm} The main guideline leading the voltage divider design was to
maximize the photomultiplier efficiency by shaping the output analog signal.
The complete schematic diagram of the voltage divider is
shown in Fig. \ref{fig:voltagedivider}.  The common linear resistor chain
has been chosen, only modifying the resistor values
between the first and second dynode, equal to 2 times the common
value R, and of that between the second and third dynodes, equal to 1.5 R.
As R we chose the value of 1~M$\Omega$, in order to minimize the power
dissipation in the divider itself.  Signal and HV are
coupled/decoupled at the input of the front-end electronics on one
side, and at the divider on the other.  On the divider side this is
accomplished by resistors R2, R3, R4 and C2. The two resistors R2 and R3 have
been added to the circuit to optimize the signal shape.  The dumping
resistors on the last four dynodes (R6, R8, R10, and R12) suppress
the ringing associated with the pulse. The 10~nF capacitors in
parallel with these resistors keep the voltage constant between the
dynodes during the development of the pulse.  An "AC-coupled
resistor" network (C1/R1, 50~$\Omega$ in series with 4.7~nF) has been added
to provide some level of cable back termination in order to minimize
possible signal reflections on the transmission line. The effect of
the capacitor used to ground the 50 $\Omega$ resistor is that of adding a
negative tail. This compensates, almost perfectly, the positive one,
producing a net result of  a pulse with a negligible residual tail. The
dividers are supplied with positive high voltage in the range of
1100-2000 V. The voltage across the cathode and the first dynode
is fixed - by means of the Zener diodes D1 to D3 - to the
value (600 V) suggested by the manufacturer for the optimum photomultiplier
performance.  A simple R-C  low pass filter network (R24/C7) is used
here to minimize the wide band noise that might be introduced by the
Zeners.

\subsubsection{Photomultiplier encapsulation, sealing and mounting structure}

\begin{figure}
\begin{center}
\includegraphics[width=0.52\textwidth]{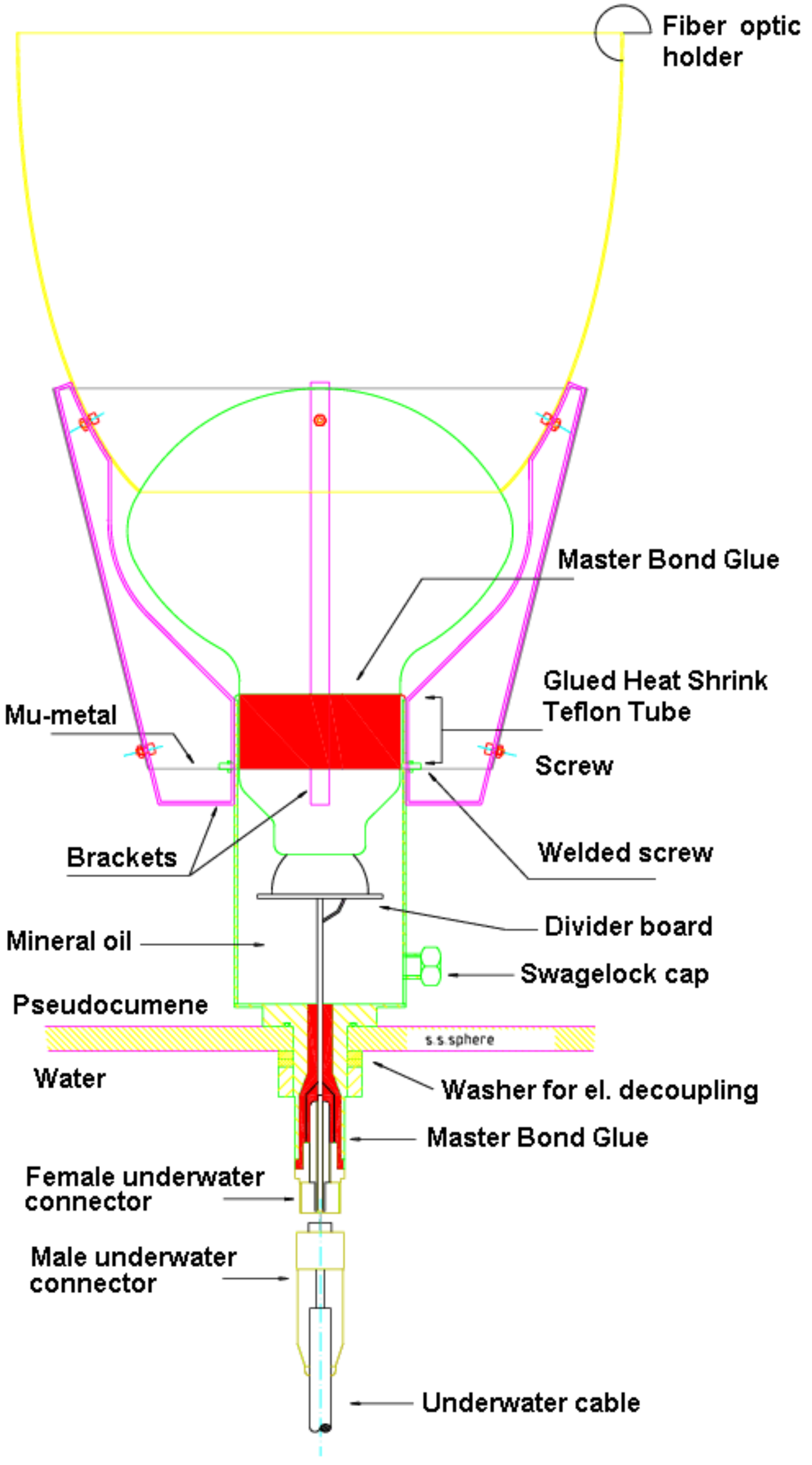}
\caption{Design of the Borexino photomultiplier tubes.} 
\label{fig:housing}
\end{center}
\end{figure}

The 2212 Borexino photomultipliers are mounted in the detector through 
equally spaced holes located on the 13.7 m diameter stainless steel sphere 
already described in section \ref{sec:Borex}. The sphere is mounted inside 
a cylindrical tank  and has been filled with scintillator while the tank has  
been filled with water to act as a final shield against the radioactivity from the
surrounding rocks. To assure a reliable operation of the PMTs in
such a complex environment, it was necessary to study and develop an
encapsulation of the neck of the bulb and the divider. The design
has been based upon the broad experience gained in the research and
development of the CTF experiment.  The base and the neck of the tube 
are enclosed in a cylindrical stainless steel housing whose external diameter 
is equal to 90~mm, as shown in Fig. \ref{fig:housing}. 
This housing is fixed to the glass of the tube neck through the PC proof 
EP45HT epoxy resin from Master Bond, which acts as a structural adhesive as well as a
protective barrier against PC ingress. The end-cap of the
cylindrical housing, welded and helium leak tested 
(sensitivity 10$^{-9}$ scc/s), carries the feed-through which is 
inserted into a hole on the sphere surface and then secured through a rear nut. 
In this way the front part of the photomultiplier is
immersed in PC and the rear part in water. 
Phenolic resin is used as an insulating material that electrically 
decouples the
device to avoid ground loops and also acts as a groove for the viton O-ring
assuring the tightness
between the tank and the sphere. All 2212 feedthroughs have
been tested with a custom made spectrometer, using argon as a
trace gas to evaluate the leak rate.

It should be reminded that the feed-through is also designed to
accommodate the underwater jam-nut connector from the company Framatome.
The connector is screwed into the feed-through until its O-ring is
properly compressed; a further potting, with the same epoxy resin
mentioned before, is also made to have a second barrier against
water infiltration.  The space inside the cylinder is filled with an inert organic oil inserted inside the can
through a 10~mm pipe port sealed afterwards with a Swagelok cap (see Fig. \ref{fig:housing}).
This mineral oil prevents water from condensing on the divider without stressing the
very delicate joints between the metal pins and the glass.  The last
barrier against PC is provided by a heat shrink Teflon tube glued
with epoxy resin between the glass neck of the photomultiplier and the steel
can. In order to fully benefit  from this barrier, we exploited a patented
technology of the Gore company which implies a preliminary surface etching of the
Teflon film and allows a strong Teflon adhesion on other surfaces.

\subsubsection{Optical Concentrators and $\mu$-metal}

The intrinsic radioactivity of the photomultipliers used in Borexino
does not allow them to be placed close to the center of the detector. In fact, 
although the PMTs were made from very pure glass, the typical contamination
of the glass itself and of the mechanical structure of the PMTs is of the order of 1 Bq/kg,
mostly due to Th and K contaminations.

In order to enhance the photon detection efficiency,
1800 photomultipliers out of the total 2212 have been equipped with optical
concentrators (OC). The OC is designed to collect scintillation
photons emitted inside the Inner Vessel with high efficiency and its
shape is designed to reflect photons to the curved photocathode of
the photomultiplier if the incidence angle is below 32.5 degrees.  The OC
consists of high purity, soft aluminum which is spinning processed
to get its final shape (surface of a body of revolution). The surface
was anodized to guarantee chemical stability both in water and in PC. 
In order to increase specular reflectivity (90\% in the
wavelength region between 370 and 450~nm) the surface was then 
manually polished and cleaned. 
The aging tests confirmed a good resistivity against water 
corrosion and PC compatibility.  

The E.T.L. 9351 photomultiplier, due to its large size, is very sensitive to  the
Earth's magnetic field. Thus it was necessary to shield it with a 
conic $\mu$-metal foil
(0.5~mm thick) placed around the cathode and the base region.
Accelerated aging tests showed that this material, in contact with
pseudocumene, strongly catalyzed the scintillator oxidation. It
was mandatory to protect the $\mu$-metal with a lining (20~$\mu$m thick) of
clear phenolic paint from Morton. The support of the $\mu$-metal and
the light concentrators is sketched in Fig. \ref{fig:housing}:
it is mounted
with four holders at 90$^{\circ}$, connected to the housing 
of the tube with some screws welded onto it.

\subsubsection{Cables and Connectors}

The Borexino cables and connectors will be completely immersed
in high purity water for a period of many years. This has required
solutions and materials explicitly developed for submarine
applications. The fundamental requirements are material
compatibility and electrical performance. Plastics must have a very 
low water absorption coefficient to withstand long term exposure in water
and the design must include multiple barriers against
radial and axial diffusion. It is also necessary to avoid effects
caused by the concurrent presence of different materials together,
such as the development of galvanic couples and of localized corrosion,
as well as the release of impurities into the water. Obviously, also the
level of radioactivity of the materials is of major concern.  The transmission 
line conducts both the signal and the
high voltage (HV), while assuring the connection to a 50~$\Omega$ 
front-end
where the decoupling of the HV and the signal is accomplished.  The
cables that connect the photomultipliers directly to the
electronics, are  custom made RG 213 coaxial 50~$\Omega$ cables. The outer
jacket is made of solid extruded high density polyethylene while a
second barrier is made with a laminated copper foil,
bound to the braid with a copolymer coating. All cables have
an electrical length of 282.1~$\pm$~0.25~ns ($\simeq$~57~m) most of which
operates under water.  Underwater connectors  work in a non
critical condition regarding pressure (the maximum pressure is that of a column
of water of 17 m, while these cables are designed for submarine operations), 
but an immersion time of many years without maintenance requires a very high confidence and
experience in long term applications in the submarine field. 
The company Framatome provided suitably strong stainless steel
connectors with viton O-rings.
Special attention was also dedicated to the
connector cabling to optimize the overall electrical response
of the line; dedicated tests with a dual port network analyzer were
performed both at the company and at the experiment site.

\subsubsection{Design Qualification}

Various testing stages have been foreseen in order to assess the
quality and long-term reliability of the photomultiplier and its encapsulation.
As a first step, the device was immersed in its final configuration
in deionized water inside a steel tank, to perform an
accelerated aging test under pressure and temperature. At the same
time, other prototypes were thermally cycled several times in order
to identify any possible weakness in the glass-to-metal joint. The
results of the tests were compared with a simulation using a
finite element analysis.  After this first campaign, a more
realistic, long term and large scale test set-up was realized at
Gran Sasso. A set of 48 photomultipliers was immersed, at the end of 1999, partially in
pseudocumene and partially in water in a two-liquid-tank explicitly
designed to reproduce the main geometry, mechanics and environmental
conditions of the experiment. The test was successful: after 3 years of testing, 
all photomultipliers were working properly and all the main electrical, 
mechanical and chemical parameters were within the specifications.
See section \ref{sec:pmts_results} for PMT performance after Borexino 
filling.

\begin{figure}[!tb]
\begin{center}
\includegraphics[width=0.5\textwidth,clip=true]{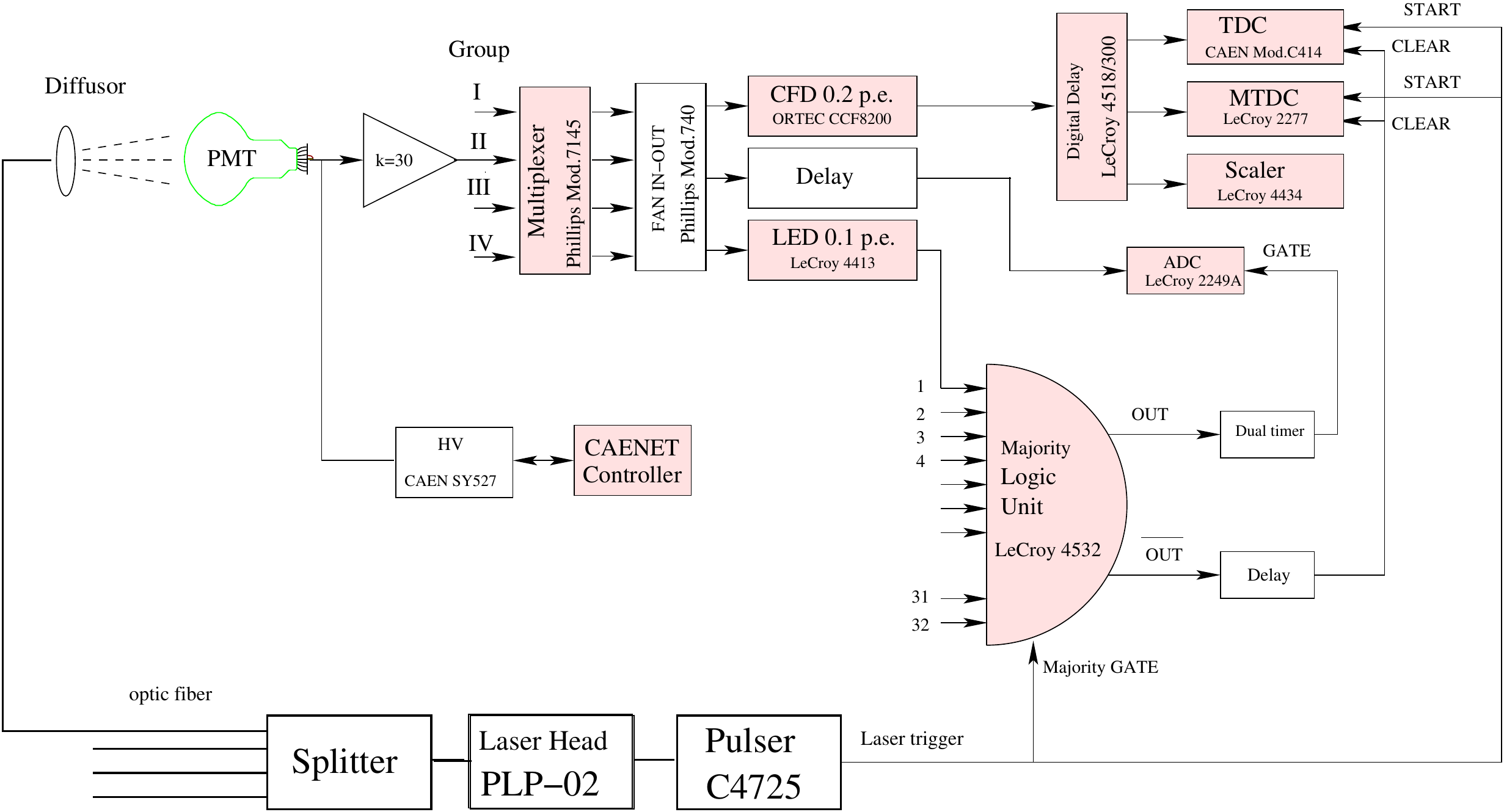}
\caption{One channel of the electronics for the PMT test facility.} 
\label{Electronics}
\end{center}
\end{figure}

\subsubsection{Test System}

The photomultipliers delivered from the supplier (ETL) are factory tested
and are provided with the data sheets reporting 
the operating voltage and other 
performance parameters. However, due to the
difference in the dividers used by ETL and the ones used in
Borexino, several of the parameters were measured again.  For this
purpose a test set-up has been realized at the Gran Sasso
Laboratories.  

The Borexino PMT test facility
is placed in two adjacent rooms. In one room the electronics is mounted,
while the other one is a dark room with 4 wooden tables designed to
hold up to 64 PMTs. The tables are separated from each other by black
shrouds, which shield the light reflected from the PMT's photocathode.
Together with the acceptance tests, the performance of the PMTs while immersed
in water was also tested, in order to study the possible mechanical problems
for the PMTs to be mounted at the bottom of the Borexino detector.
For this purpose, three pressurized Water Tanks (WT) were installed, each
of the tanks can hold up to 20 PMTs. For the purpose of the long-term
testing of the PMTs in conditions close to the ones of the Borexino
detector a Two-Liquid Test Tank (TLTT) was designed and installed
in the laboratory \cite{bib:laser_calibr}. 48 PMTs are installed in the TLTT,
with the base being in water and the bulb immersed in liquid scintillator.
The overall test facility is able to measure the characteristics of
172 PMTs if completely loaded.
The dark-room is equipped with an Earth's magnetic field
compensating system. The non-uniformity of the compensated field in
the plane of the tables is no more than $10\%$. In the WTs and the
TLTT the Earth's Magnetic Field is not compensated.
 All the photomultipliers under test are
illuminated with a picosecond solid state laser (Hamamatsu). The light
of this laser is uniformly distributed and attenuated to reach the single
photoelectron condition.  The photomultipliers are connected through a patch
panel to the electronics located in two racks; there are 32
independent channels, quad-multiplexed, to obtain a total of
128 input ports.   The system uses the modular CAMAC standard
electronics and is connected to a personal computer by the CAEN C111
interface. The majority logic unit LeCroy 4532 is able to memorize
the pattern of the hit channels and to activate the reading as soon
as one of the signals on the inputs is inside the external GATE
on the majority unit. Every laser pulse is followed by an internal
trigger used as the majority external gate. One channel of the dark room 
electronics is shown in Fig. \ref{Electronics}.  The system performs a
preliminary, but very accurate, high voltage tuning \cite{bib:Oleg5} to set the real
gain of each photomultiplier at a level of $2 \cdot 10^7$; 
several parameters are then acquired during a few hour test run and displayed as histograms
on the screen in real time.  The most important are
the charge spectrum, the transit time spread, the after-pulses, and
the relative sensitivity of the corresponding tube.
Using the test facility described above, 2212  phototubes, out of the 2350 delivered 
by the manufacturer, were selected with the optimal characteristics for Borexino.
The precision study of the amplitude and timing response of the
E.T.L. 9351 series PMTs has been performed using the
test facility, results have been reported in \cite{bib:Oleg6} and \cite{bib:Oleg7}.

\subsection{The Inner Detector Electronics}
\label{sec:electronics}
\begin{figure}[!ht]
\begin{center}
\includegraphics[height=19 cm]{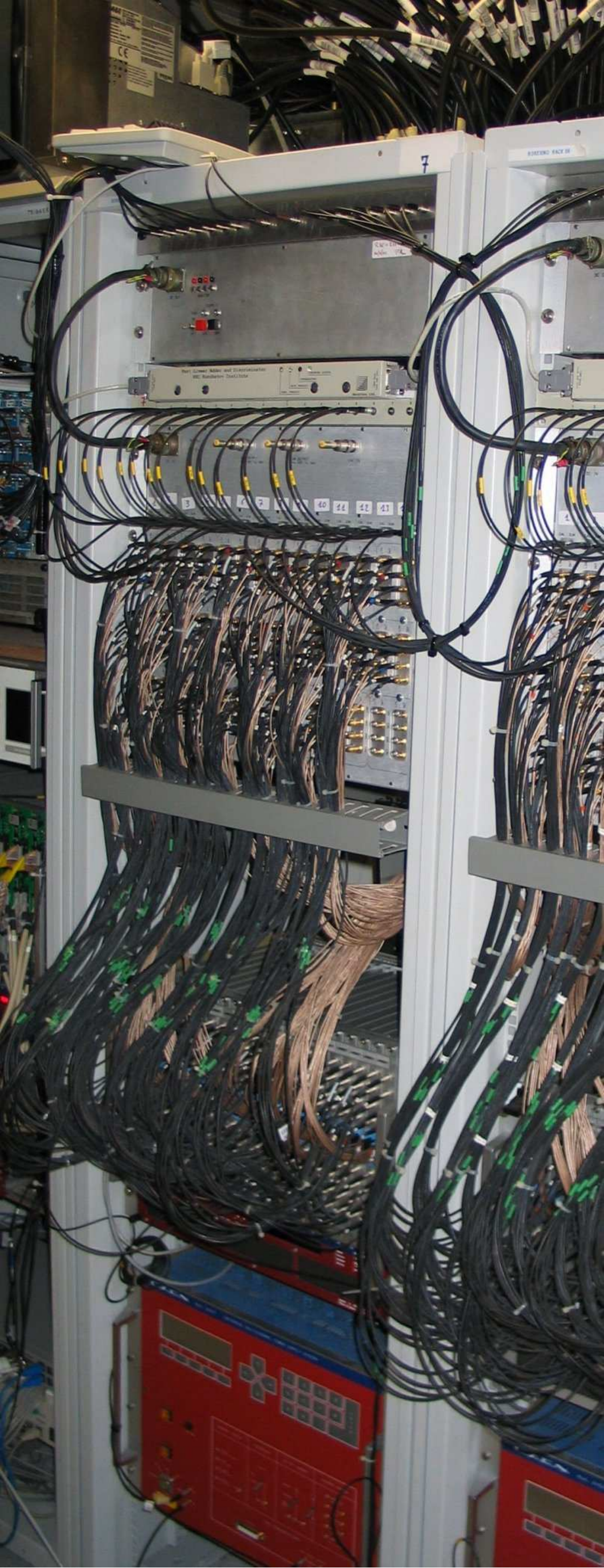}
\caption{Picture of the inner detector electronics rack. From the bottom to the top
the picture shows the high voltage mainframes, the digital VME electronics, 
the front end electronics, the analogue adders and the low voltage power supplies.}
\label{fig:rack}
\end{center}
\end{figure}

The inner detector electronics is organized in 14 identical racks, each of them
handling 160 PMTs. 
Fig. \ref{fig:rack} shows a picture of one of these racks. From the bottom to
the top, the figure shows the high voltage mainframes, the digital VME electronics, 
the front end electronics, the analogue adders and the low voltage power supplies. 
The following sections describe the main features of these devices.

\subsubsection{The High Voltage system}
\label{sec:hv}
The high voltage power supply system is based on SYS527 mainframes
and A932AP boards, both produced by CAEN\footnote{www.caen.it} s.p.a. 
(see the bottom of Fig. \ref{fig:rack}).
Each mainframe can house up to 10 boards for a total of 240 channels. 
However, in order to keep each electronics rack independent from the others and simplify 
the cable layout in counting room, for the internal detector we have used 
only 7 boards per mainframe, for a total of 168 channels per rack. Only one
mainframe has 9 boards and it is used to power the PMTs in the Water Tank.

The A932AP board can provide 24 independent HV channels. The 24 channels share a 
common high voltage source, and only the total current drawn by the 24 channels
is measured.

The mainframes have a CAENET serial interface that is used to set parameters,
control the high voltage and monitor high voltage values and currents during
data taking. The system is continually monitored by a dedicated slow control
server that periodically store the system status into the experiment's data base.
 
\subsubsection{Inner Detector Front-End Electronics}
\label{sec:FrontEnd}
In order to determine the energy and the position of a scintillation 
event, the amount of emitted light and the time
distribution of the photoelectrons ({\it p.e.}) must be measured. 
For these reasons, the front-end circuit
connected to each photomultiplier produces two signals (one for the
energy and one for the time measurements) that are processed
by the digital boards described in section \ref{sec:Laben}.

The number of photoelectrons collected by the Borexino detector 
is approximately 500~p.e./MeV and the interesting energy 
range extends from a few tens of keV up to a few MeV. 
Therefore, Borexino PMTs work mostly in 
single photoelectron regime, although the probability of having more
than one photoelectron in a given PMT is not completely negligible,
even at relatively low energies.

The signal due to a single photoelectron at the front-end input 
is a pulse having an amplitude of about 12-15~mV and a total width of 15~ns. 
The multiple hit probability is of the order of 10\% for a 1~MeV 
energy deposit in the detector center. Obviously,  this probability
increases for off center
scintillation events, and at higher energies. The shape of the signal for
multiple hit events depends on the scintillation light decay time, 
the photomultiplier transit time jitter, and the 
spread of the arrival time of photons
on the photomultipliers. For these reasons, the information about
the energy deposit associated with the scintillation event is
contained in the total charge collected at the photomultiplier's
output within a proper time interval (80-300~ns for single beta or
gamma events). Time correlated events, {\it i.e.} two isotopes decaying
in sequence with the second one having  a short lifetime (hundreds 
of $\mu$s or less), are very important for tagging some
types of radioactive backgrounds, like U, Th and Rn. 
The detection of these events with high efficiency requires circuits
with dead time in the range of tens of ns or less. 

\begin{figure}[!tb]
\begin{center}
\includegraphics [width=0.5\textwidth,clip=true]{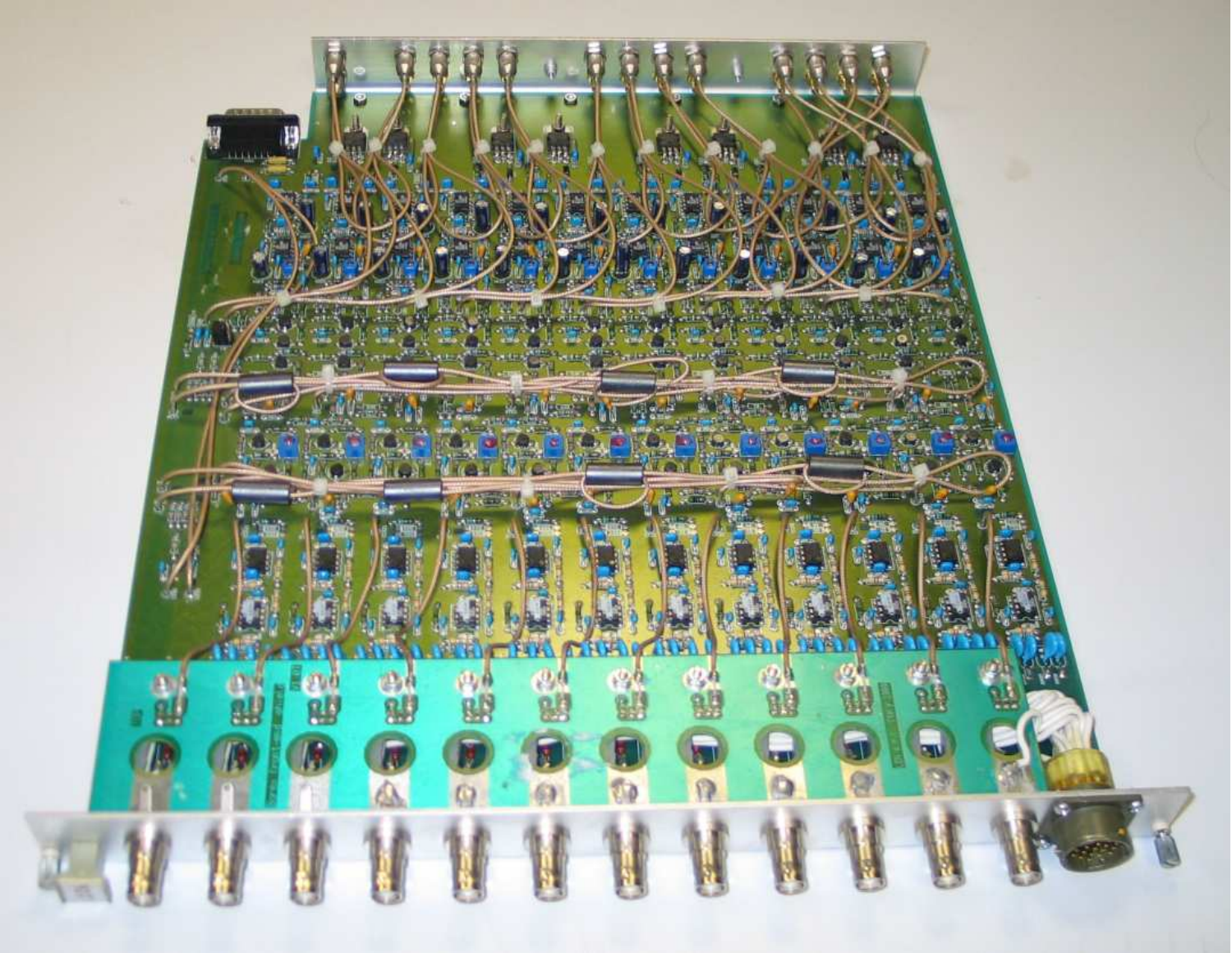}
\caption{A picture of the custom made analogue front end board.} 
\label{fig:FEboard}
\end{center}
\end{figure}

A special analogue
gateless charge integrator has been designed to fulfill the
energy reconstruction requirements. The circuit works as an
ideal integrator always integrating the signal present on the input, without any
external gate. 
The key factor of the integrator design is the AC-coupling with the
photomultiplier signal due to the presence of a single cable carrying both the
high voltage and the photomultiplier signal. 
This AC coupling makes the total charge associated with each pulse 
(single or multiple) equal to zero. 
The integrator output rises at the pulse arrival time and
then decays within a time related to $\tau_{int}=RC$ where C is
the capacitor filtering the high voltage (C = 4.7 nF) and R is
the 50 $\Omega$ cable impedance 
(for the relatively high frequency PMT signal). This integrator
automatically resets itself after each pulse without the use of
switches, integration gate signals, baseline restoration networks
and it has no dead time. The RC value (500~ns) sets the
limit for the maximum frequency of the signals that can be
measured. A detailed description of this gateless charge integrator
and of the whole front-end circuit  is reported in \cite{bib:fepap}.

By sampling the integrator output V at the time $t_0$ and at the time
$t_0+ \Delta t$ ($\Delta t=80$ ns) the charge Q due to the
photoelectrons arriving within the interval $\Delta t$ is easily
obtained using
$$Q= G (V(t_0+\Delta t)-V_{off} )- (V(t_0)-V_{off}) exp^{-\Delta t /\tau_{int} } $$

where $G=129$~mV/pC is the integrator gain,  $V_{off}$ is a DC
offset value. The offset value is intentionally  added to the output
to maintain a positive output polarity in presence of low frequency
fluctuations of the output baseline. These are present due to the
high gain (close to  $10^3$ from a few Hz up to about several kHz)
of the integrator at low frequency. In this way, the low-frequency 
fluctuations do not influence the
precision of the charge measurement due to the double sampling
procedure. High frequency noise components introduce distortions to
the integrator output not compensated by this double sampling procedure.
The rms value of a fixed charge measurement due to this effect is 3~mV/pC 
to be compared with the integrator gain of 129~mV/pC.

In addition to the integrator signal, the front-end
circuit outputs, for each channel, the photomultiplier signal 
amplified by two low noise amplifiers with fast transistors. 
This timing output is fed as an input to a discriminator  in the digital board  to get the
photoelectron arrival time. The typical timing signal
amplitude  is 230~mV corresponding to one photoelectron. The distribution of
the timing output  noise level (measured on the 2212 channels) has a
mean value of 0.8~mV with 0.4~mV rms with a discriminator
threshold corresponding to 1/100 of the single photoelectron
amplitude. 

The front-end electronics is organized in boards
corresponding to high standard rack units. Each board (see Fig. \ref{fig:FEboard}) contains  12
single electronics chains  and a circuit
providing the resistive sum $S_{12}$ of the 12 photomultiplier signals.
This sum is used as input to the FADC system described 
in section \ref{sec:Flash}.

The sum  chain is linear up to 35 contemporary photoelectrons/channel. 
Higher numbers of synchronous photoelectrons leads to saturation of the 
amplifier that is shared by the sum and the timing circuit. Due to the time spread 
in the distribution of the number of photoelectrons belonging to the 
same event the true sum dynamic range is higher. In addition, a 
calibration signal can be distributed to all 12 channels. 

\begin{figure}[!tb]
\begin{center}
\includegraphics [width=0.5\textwidth,clip=true]{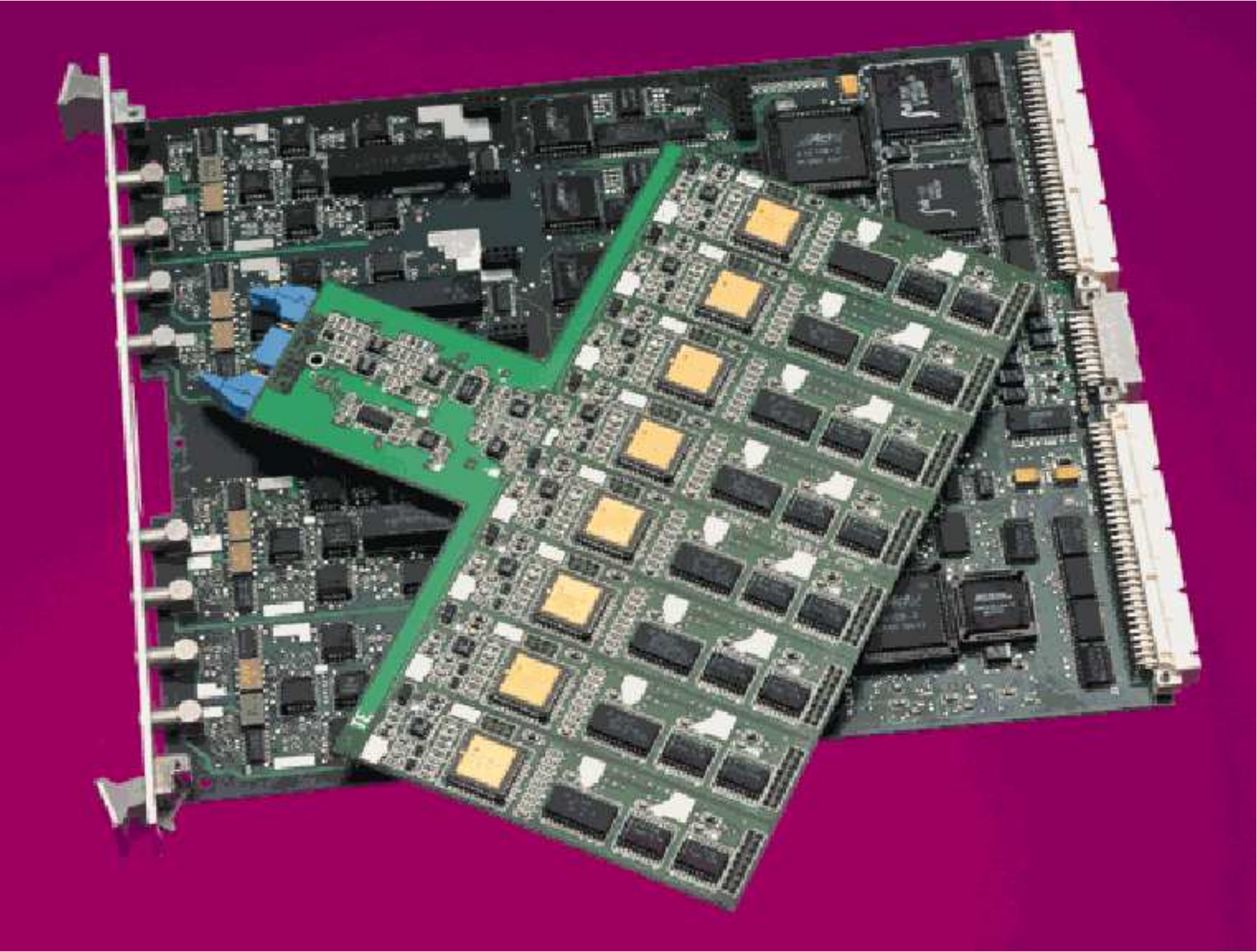}
\caption{A picture of the custom made VME data acquisition board.} 
\label{fig:labenboard}
\end{center}
\end{figure}

\subsubsection{Inner Detector Read-out Electronics}
\label{sec:Laben}

The inner detector digital electronics provides  amplitude and
time-to-digital conversion for each front-end channel. Each board (see Fig. \ref{fig:labenboard})
houses electronics for 8 front-end channels and up to 20 boards can
be plugged into a VME crate, for 160 total channels in one
electronic rack. Each VME crate is also equipped with a backplane board that
acts as an interface to the trigger system. These boards were designed and built in
collaboration with Laben s.p.a.

The board architecture is composed of two main functional blocks:
the single acquisition channel, replicated eight times and the main control block. 
The single acquisition channel
acquires, asynchronously and continuously, the hits
coming from the photomultipliers. Each channel has two inputs from
the front-end stage: an inverted and amplified copy of the fast photomultiplier
signal and an integrated output. The fast signal provides the
arrival time of the light pulse, while the associated charge
(energy) is obtained from the integrated output.

The timing signal is sent to a programmable dual threshold
discriminator. The choice of a dual threshold discriminator was done
to find the best compromise between a small amplitude {\it walk effect} and 
{\it time jitter} (at the
level of a fraction of a nanosecond) and a good rejection of the non
negligible dark rate of photomultiplier pulses smaller than a
single photoelectron\footnote{The first point would require to set
thresholds as low as possible, the second one would instead require higher
values.}. This discriminator is built with two
comparators in AND condition: the channel is fired only if the
photomultiplier signal crosses the higher threshold but the timing
information is related to the lower crossing. The higher threshold is
programmable between 0 and 500~mV; a single photoelectron signal is
about 220~mV and the discriminator threshold currently used in the Borexino runs 
is about 40~mV, corresponding to about 1/5 of a photoelectron. The detection efficiency
depends slightly on the PMT pulse shape and is of the order of 95\%.

If the discriminator fires, it is automatically disabled for about 140~ns.
This dead time was introduced to prevent data jamming during the
sampling and the writing of the data on the internal memory buffers.
This is a real DAQ dead time only for multiple hits with time delays
between 80 and 140~ns. The impact of this dead time on the number of
detected hits is 
of the order of few percent for the typical neutrino events in the
fiducial volume. However, whenever two photons hit the same photomultiplier in
less than 80~ns, only the timing information of the second one is lost,
while the charge sampling is able to keep track of the presence of
this second hit. 

The discriminator output
is  used to fire a double pulse generator. Two 30~ns-wide, 
80~ns-apart pulses are generated and sent as triggers to a dual
input FADC, with 8 bit resolution. The first of the two
pulses fires an on-board coincidence unit, that computes the
number of activated channels in an adjustable time window of about
50~ns, acting as a first level trigger (see section \ref{sec:Trigger} for details).

The first FADC channel samples a 10 MHz triangular
wave twice\footnote{Generated on the board by a triangular wave generator
synchronous with the 20 MHz base clock.}. At the same time, the first
of the two pulses latches the content of a 16 bit
counter (Gray counter) driven at 20 MHz in a buffer register.

The value of the Gray counter plus the knowledge of the ADC sampling 
of the triangular wave is used to reconstruct the timing information
of the hit with a maximum span of 3.2~ms and a resolution  better
than 500~ps. In order to obtain a good ramp linearity and low
noise, special care has been taken in designing the TDC
hardware.

The second FADC samples the integrated charge signal. Thanks to a
well calibrated delay line, the two samples fall exactly on the
baseline and on the peak of the charge pulse.

The ADC and TDC data temporarily stored in an internal memory (local
FIFO buffer) are then managed by the main control block, common to
all channels of a board. Almost all the functions of this block are
executed by a DSP (Digital Signal Processor). The DSP
controls the status of the eight channels' FIFO and waits for
triggers. 

Old data are suppressed by an automatic procedure which every 6.4 $\mu$s
drops an hit from the FIFO for every channel. A global Daq Gate Signal (DGS) generated by
the trigger system and distributed to all boards inhibits this procedure, leading 
to a programmable read-out time width. Data have been taken with a DGS 7.2 $\mu$s
wide from May 2007 until December 2007. This value was increased up to 
16 $\mu$s since then.

When a trigger condition occurs, the Borexino Trigger Board (see
section \ref{sec:Trigger}) 
delivers to the DSPs a master trigger signal, the global gate signal mentioned above, 
and the 16 bit trigger number identifier. The master trigger signals acts as 
Common Stop to all the channel FIFOs and interrupts the DSP. 

Under this interrupt the DSP starts the procedure to build the data record: all the hits
registered in the digital FIFOs are collected and stored in a 8~kbyte  
Dual Port Ram (DPR), together with a header containing the trigger
number. The depth of this memory allows it to store hundreds of
typical size Borexino events. The DSP processing time under a trigger
condition is 30~$\mu$s plus 6.4 $\mu$s for each hit. Since the fast coincidences with time
delays less than 80~$\mu$s are very important for Borexino, 
the DSP architecture guaranties that the board is able to recover the
data of subsequent events occurred during the busy state of the DSP 
(up to a maximum of 10 events or up to the saturation of the internal FIFOs); 
this is necessary to acquire the fast coincidences in Borexino.

The DPR is accessible both from the DSP and from a VME bus. A dedicated 
process running on a VME Power PC computer (PPC) 
waits for an interrupt to read these data, store them in a
single data structure, and send them to the DAQ processes. For more details about
the boards and the trigger system see \cite{bib:laben1}.

\subsubsection{The Fast Waveform Digitizer system}
\label{sec:Flash}

The main acquisition system of Borexino has been designed and optimized for
the detection of low energy solar neutrinos in the sub-MeV range.
A reasonable reconstruction of higher energy events is possible 
by applying a correction to take into account the saturated channels. However,
given the importance of the physics which could be explored by Borexino
in the energy range between 3 and 20~MeV (Supernova neutrinos, reactor and 
Earth anti-neutrinos...), a separate system was designed which is explicitly 
dedicated to the higher energy range. 

This system is based on the idea that at high energies
it is possible to retain precision while not having to record each
of the 2212 channels individually. The photomultiplier signals are therefore
grouped by solid angle sectors, thus reducing the number of acquisition
channels. Each of these sums is recorded by a 400~MHz waveform digitizer.

As described in section \ref{sec:FrontEnd}, each module of the
front-end electronics provides a sum of the fast signals of its 12
channels, obtained by a resistor network. A set of analog adders,
active and passive, is used to build, from these sums, 98 groups of
up to 24 phototubes, plus the total sum of all channels. These 99
channels are all recorded by fast waveform digitizers.

There are 34 digitizer modules with 3 input channels each, located
in 4 VME crates, among which 33 account for the 99 analog signals
and the last serves to record logical signals such as the trigger.
The 4 VME crates are interconnected by a PVIC\footnote{The PVIC is a
product of \emph{Creative Electronics Systems}} bus and read by a
single VME master MVME-2302\footnote{The MVME-2302 is a product of
\emph{Motorola}}.

The device used was originally designed at APC, Paris, then
developed in collaboration with CAEN and produced  as
\emph{model V896}. It is a VME module housing 3 analog channels,
with internal or external clock, a multi-event capability and zero
dead time read-out, as long as the read-out occurs faster than the triggering
rate. The dynamic range is 8-bits and the sampling period is 2.5~ns.

The V896 is based on a 100~MHz 8-bit flash ADC with a 475~MHz analog
bandwidth. Four such flash ADCs, strobed by four 100~MHz clocks,
contribute to the digitization of each analog channel. These four clocks, with
$\pi /2$ phase shifts are generated internally from a 50~MHz clock.

The waveform digitizers operate in stop mode; they digitize
continuously, writing into a circular buffer which we call a page.
The storage capacity is 655.36~$\mu$s, divided into a configurable
number of pages. In Borexino, they are configured as 64 pages of
10.24~$\mu$s. On receipt of a trigger they simply leave
that page with all the data it contains and start writing to the
next page. The written pages can be read anytime by a VME master.
This VME master frees the pages it no longer needs, thus allowing
the V896 to rewrite them.

The V896 features a 16-bit TTL front panel input that is 
read when a trigger occurs and is used to associate each page 
with the event number generated by the main trigger logic.

The 34 V896 modules are synchronized by an external 50~MHz clock;
the trigger issued by the main trigger logic is detected and
synchronized with this clock by NIM logic and distributed to all the
V896 modules. The purpose of synchronizing it with the clock is to
guarantee that all waveform digitizers detect the trigger on the same
clock edge.

A second level of triggering, equivalent to an energy cut, is
performed online by the acquisition process to reduce the
amount of data. This second level trigger is done by analyzing the
total sum signal. When the amplitude exceeds a programmable
threshold, the event is marked to be read, together with other
events in coincidence with it, before or after, within
a programmable delay.

A further reduction of the data, by a factor of around 20, is performed
by keeping only part of the total 10.24~$\mu$s time
interval of the event: time windows are delimited where the total
sum signal exceeds a second programmable threshold. Finally, only
the relevant part of the data is read from the VME, formatted,
and sent to the event builder. The level of the energy cut is
tuned to set the waveform digitizers' data flow to about 50\%
of the total.

The gain of the active adders is adjusted to reach a mean
amplitude of 5 flash-ADC counts per photoelectron. This
would lead to a maximum dynamic range of about 50 photoelectrons if
they were all synchronous, leading to a maximum of 12.5~MeV at the
center of the detector and much less at the limit of the fiducial
volume.
Actually, due to the dispersion in the emission time of the photons,
the simulation shows that, for the whole fiducial volume, the mean
value of the energy loss due to saturation reaches about 1\% at 10~MeV. 
Since saturation occurs, by definition, with high amplitude
signals and high amplitude signals are not subject to large
statistical fluctuations of their shape, the lost energy is
recoverable by fitting the expected pulse shape on the unsaturated
part of the signal.

\section{The Outer detector}
\label{sec:OuterDetector}

Even though the rock shielding reduces the muon flux by a factor $10^6$, 
high energy muons originating within cosmic ray interactions in the 
atmosphere are still able to penetrate to the depth of the Gran Sasso 
laboratory.
In this context they constitute a relevant source of background for the 
experiment and must be tagged with high efficiency for the success of 
the $^7 Be$ neutrino flux measurement as well as of the rest of the physics 
program.

Among the tagging methods, a key role is played by the Outer Detector (OD), 
a water \che\ detector composed of 208 additional 
photomultipliers installed in 
the volume of the Water Tank (WT), where an ultra-clean water buffer is 
present for shielding purposes.

The photomultipliers are of the same model of those used for the Inner Detector 
(8" encapsulated E.T.L. 9351) and are arranged on the outer side of the SSS 
and on the floor of the Water Tank. The photomultiplier's 
parameters are summarized
in Table \ref{tab:pmts}.
The photomultipliers must work in water and under almost 2 atmospheres of 
pressure at the WT bottom. Therefore, their mechanical arrangement, 
the connectors and the cables are designed and chosen
to be reliable in these conditions.

\begin{center}
\begin{figure}
\includegraphics[width=0.52\textwidth]{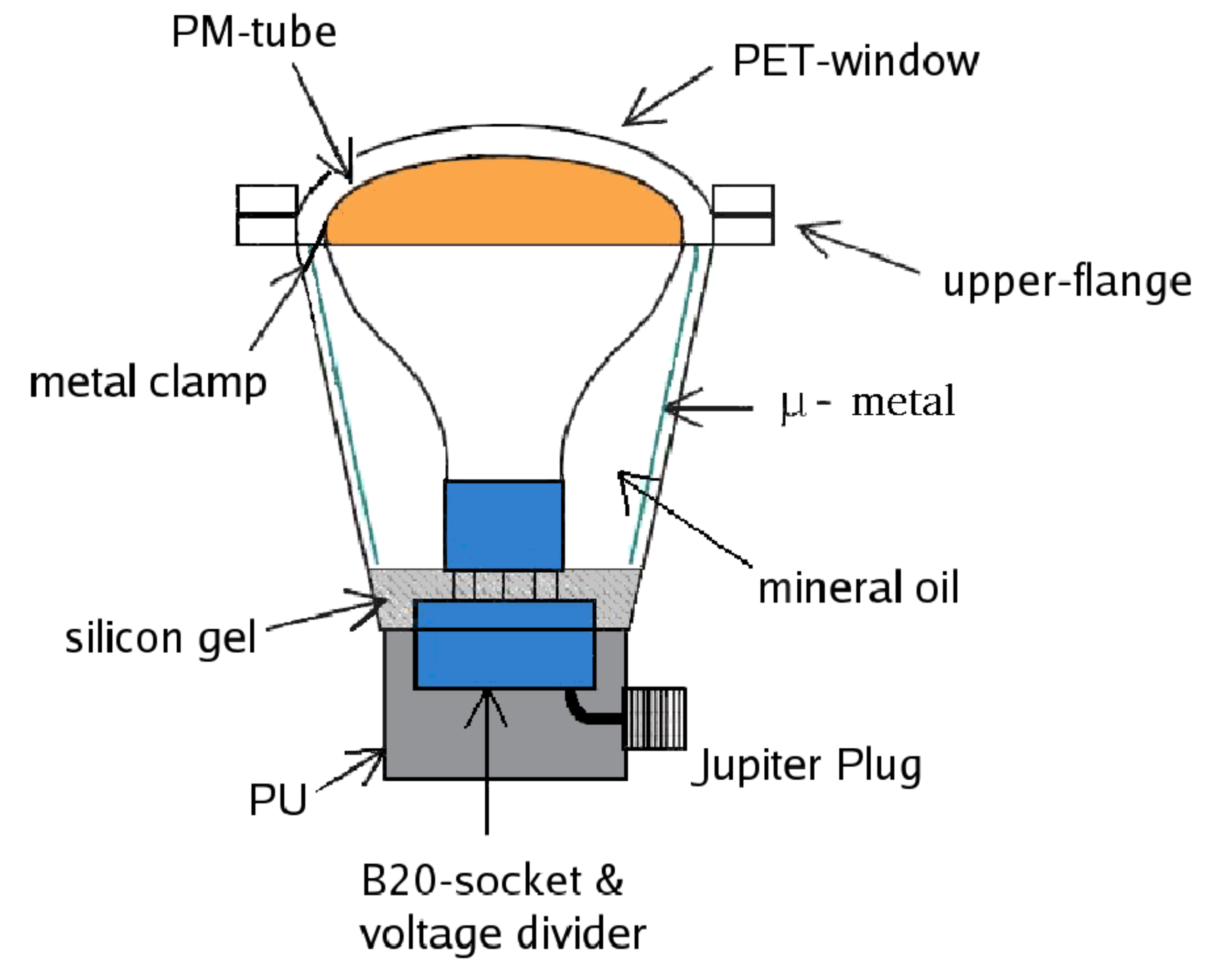} \\
\includegraphics[width=0.5\textwidth]{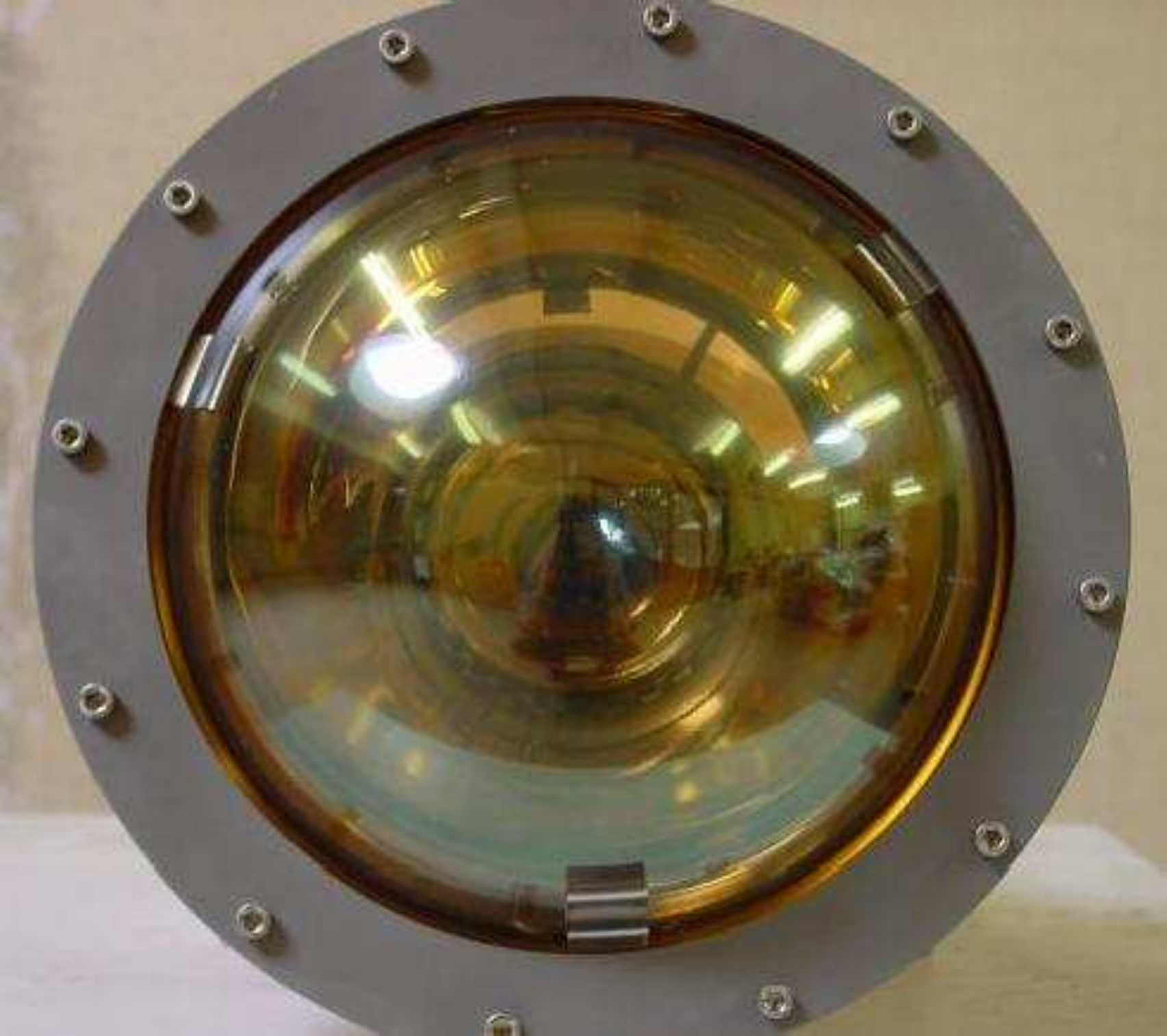}
\caption{{Sketch of the encapsulation of the outer detector PMTs (top) and a front picture of an encapsulated PMT (bottom).}}
\label{fig:muencapsulation}
\end{figure}
\end{center}

Unlike those mounted inside the sphere, OD photomultipliers are equipped 
with a full encapsulation (see Fig.~\ref{fig:muencapsulation}).
This is a stainless steel cone-shaped case, housing the photomultiplier, 
the voltage divider, 
the $\mu$-metal shielding and a female HV connector. 
The $\mu$-metal shielding is a thin metal foil required to protect the 
photomultiplier against the Earth's magnetic field that could
spoil the time resolution.
The HV connectors mate with 55~m submarine cables identical to those used for 
Inner Detector photomultipliers.
These cables carry both High Voltage and the signal connecting 
the photomultiplier to the electronics chain.
The sensitive photocathode is covered only by a transparent PET foil 
(the transparency is 90-92\% in  the frequency range of interest), so the photomultiplier retains 
an acceptance angle close to $180^{\circ}$.
The space between the case (including the PET foil on the front) and the 
photomultiplier is filled with mineral oil in order to minimize the 
index of refraction discontinuities in the light path to the photocathode.
A detailed description of the encapsulation design and its pressure 
tests can be found in \cite{bib:Len05D}.

%

In order to increase the light detection efficiency, most of 
the WT and SSS surfaces ($\approx$95\%) are covered  with sheets of Tyvek, 
a white paper-like material, $\sim$ 200 $\mu$m thick, made of pressed 
polyethylene fibers.


In the original detector's layout the 208 photomultipliers 
were arranged in 12 horizontal rings on the outer SSS surface, 
every photomultiplier looking radially outward. 
In this way the $590$~m$^2$ surface of the SSS was populated with one 
photomultiplier every $2.8$~m$^2$ ($1.7$~m linear average distance) 
and, according to simulations, the 99\% detection efficiency required 
to suppress the muon flux was met. 

However, though not strictly required for the \ber\ primary physics goal 
of Borexino, the additional possibility of reconstructing also the track of 
the through-going muon is of fundamental help in the reduction of 
$^{11}$C background for $p e p $ and CNO measurements. This ambitious 
goal requires the highest possible efficiency of direct light
detection, reflected light  being of hinderance in this respect. In the new design, 
the lowest quarter of the photomultipliers are re-positioned 
on the Water Tank floor looking upward in four concentric rings, plus 
one ring situated on the ``slope''.
This is a structural feature of the Water Tank: along the circular perimeter 
of the floor, the volume up to a height of $\sim1.5~$m 
and degrading inward for about the same length is filled with massive 
steel for engineering reasons.
This offers a comfortable surface with a $\sim45^{\circ}$ inclination 
where installed photomultipliers look roughly toward the sphere. 
Monte Carlo simulations have shown that this configuration 
reduces the amount of reflected light detected by the photomultipliers, 
without spoiling the muon detection efficiency.

As said earlier, each photomultiplier is connected to the electronics 
via a single cable that brings both high voltage and the signal. 
Unlike the ID electronic chain, the high voltage decoupling 
is not performed within the front-end 
electronics, but is performed externally in custom decoupling boxes (HVDs) 
with a high pass filter circuit. 
The signal goes from the HVD boxes to the front-end electronics.


The front-end electronics of the OD is made of 14 
\emph{Charge-to-Time Converters} (QTCs).

The board 
is a 9U single VME unit which takes 
only power from the VME backplane with no connection to the data-way 
bus. One QTC board is made of 16 channels.

Each QTC converts the analog signal coming from the 
photomultiplier into a differential digital
signal whose time duration (distance between the trailing and 
the leading edges) is proportional to the total charge of the pulse.
The output of the QTC is connected to a commercial double-edge TDC (CAEN v673).
The TDC measures the time of the two signal edges with respect to the common stop
provided by the trigger system. The first rising edge yields the time of the PMT pulse,
while the time distance to the second falling edge yields the charge.  

Besides this main function, each QTC provides two additional signals that
are used for triggering purposes, the so called secondary and tertiary outputs.
The secondary output is a one-per-board analog step function pulse whose 
height, in steps, represents the number of channels firing in coincidence.
Both height and width of the steps can be adjusted. 
The width defines the coincidence window. All secondary outputs (14, one per
QTC board) are sent to the Muon Trigger Board (MTB) for the Outer Muon Trigger 
(OMT) formation.
The Tertiary Output (TO) is a one-per-board analog pulse given by the sum of 
the input signals with a built-in fixed amplification factor of about 2. 
These outputs are sent to an additional analog trigger formation system. 

\section{The trigger system}
\label{sec:Trigger} 
The main requirement for the Borexino triggering system 
is to be able to identify, quickly and efficiently, scintillation events that are
detected by the quasi simultaneous occurrence of several photomultiplier hits.  
Due to the large number of photomultipliers, and the relatively high 
total dark current, we designed a triggering system using purely digital logic. 

The trigger should fire when a programmable number of photomultipliers
(typically a few tens, see below) are hit within a short trigger time window (TTW). The TTW 
must be set to be larger than the maximum arrival time spread of photons 
at the photomultipliers. Being the total transit time of photons throughout
the SSS at most about 50 ns, the system allow the TTW to be set from a minimum
of 48 ns up to a maximum value of 99 ns. The typical value used during data taking
was 60 ns (see section \ref{sec:per}). 
The energy threshold is set by the 156~keV
$^{14}C~\beta$ decay end--point energy. Although this background cannot be removed, 
it is very important to collect at least the last part of the $^{14}$C~$\beta$ spectrum 
for calibration and monitoring purposes. 
With a nominal effective yield of about 500~photoelectrons/MeV, this 
requirement brings the triggering threshold  down to about 40 photons. 

\begin{center}
\begin{figure}
\includegraphics[width=0.52\textwidth]{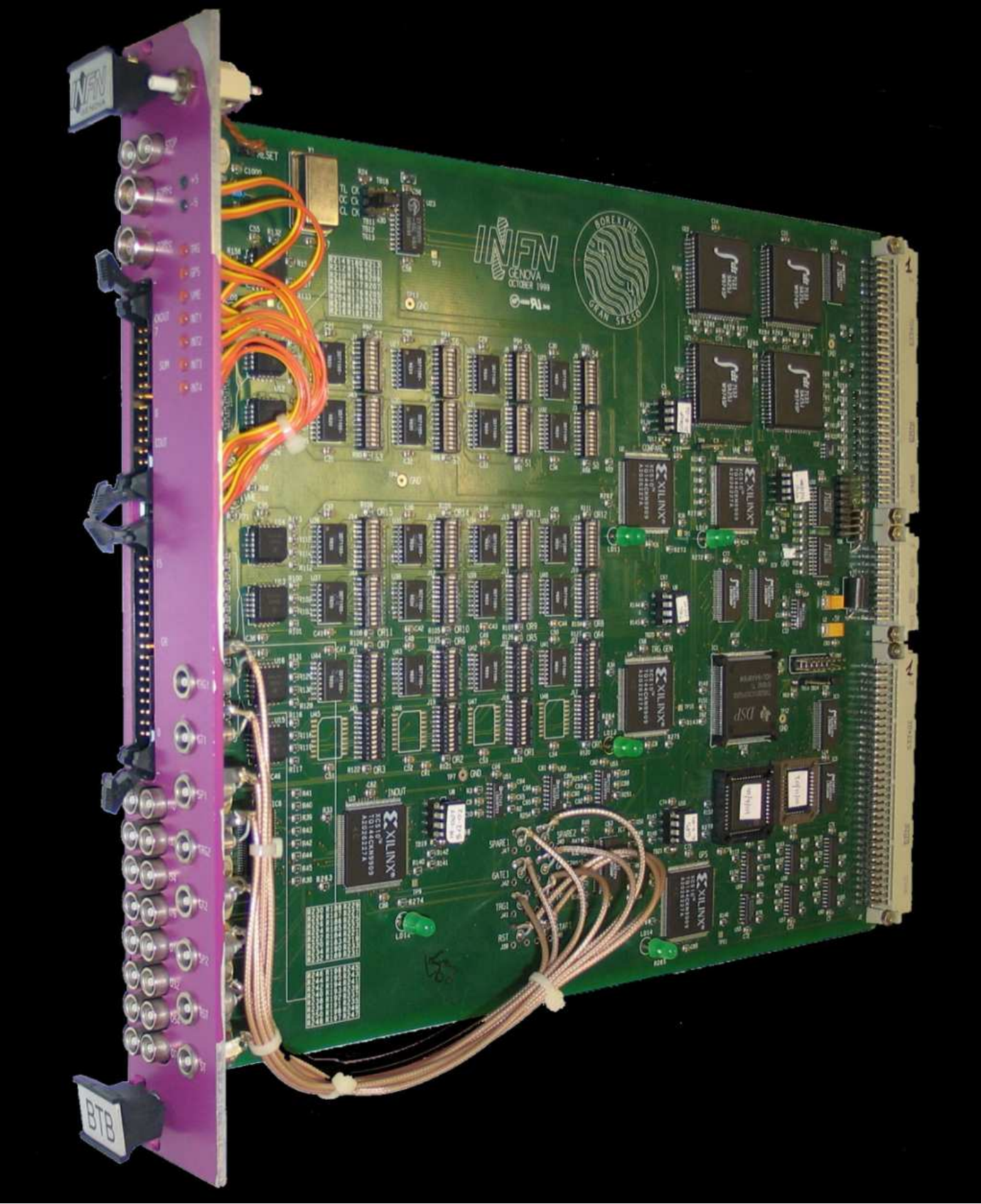}
\caption{A picture of the Borexino Trigger Board.}
\label{fig:btb}
\end{figure}
\end{center}

In order not to miss the  double coincidences from fast radioactive chains, 
and particularly the \Bipo\ coincidence of the $^{238}$U chain 
whose mean decay time is 236~$\mu$s, the system must have a 
short recovery time. Therefore, the whole system is implemented using
combinatorial logic (implemented in a set of Field Programmable Gate Arrays, FPGAs) 
and a Digital Signal Processor (DSP) that supervises and controls the logic. 
The DSP and all FPGAs are housed in a custom made VME board (see Fig. \ref{fig:btb}) 
that is controlled and read by a dedicated Power PC of the same type used for data taking. 

The system works as follows: as described in section
\ref{sec:Laben}, each digital crate is equipped with a backplane
board that is capable of counting the number of hits occurring
within a TTW. Particularly, every 16~ns-33~ns (one period
of the common 30-60~MHz programmable clock that is distributed throughout the
system) each digital crate counts the number of channels in which
the photomultiplier signal exceeds the analog threshold\footnote{This threshold
is set to 1/5 of the photoelectron signal amplitude. After front-end
amplification, the single photoelectron signal amplitude is about
230 mV; the threshold we set is 40 mV}. At each clock edge, this
number is the total number of counts that have occurred in the previous 3
clock cycles, making the 48-99 ~ns TTW range stated above. 
This 8--bit number is synchronously sent
to a set of Trigger Adder Board (TAB) via a flat cable with differential
Positive ECL (PECL) signals every 16-33~ns. There are 5 TAB boards in the system. Four
of them perform the sum of a group of 4 crates (there are 14 crates)
and the fifth one perform the sum of the output of the other four
TABs. At the end of this chain, every 16-33~ns the output of the fifth
TAB yields the total number of hits detected in the previous 48 - 99 ~ns.
This number feeds the input of the Borexino Trigger Board (BTB), a
custom made VME readable board that implements all triggering logic
by means of programmable FPGAs and a Texas Instruments TMS320C50 DSP.
When the total number of hits exceeds the programmed threshold, the
DSP is interrupted, and the trigger sequence begins: the triggering
signals are generated, the absolute time is read from a GPS clock
and a trigger record is written in the VME readable memory. In this
way each event is labeled with its unique 16 bits event number and its absolute
time is known with about 100 ns accuracy.

The BTB handles several additional triggers, besides the main one
described above: every 0.5~s, a random trigger, an electronic pulse
trigger and a timing laser trigger are fired for monitoring purposes.
In a random trigger, an event is read regardless of the detector status;
in an electronics pulse trigger, a pulse from a pulse generator is sent to 
the test input of each front-end channel; in the timing laser trigger,
the laser system and the Outer Detector LEDs described in 
section \ref{sec:CalibHardware} are fired. For the purpose of 
sub-ns time allignment, the signal sent to the laser and LEDs systems
is also copied and sent to a set of spare data acquisition channels. 
By means of these 3 triggers we can identify the number of 
valid channels in each run and disentangle the reasons for failure. This is of
particular relevance for Borexino, because the stability of the energy
calibration is a crucial feature for observing the seasonal 
variation of the neutrino flux.

A special triggering mode was introduced a few months after the beginning
of data taking to cope with the necessity to acquire the spallation neutrons
produced by muons crossing the scintillator. In fact, we have observed from
the data that $\approx$70 times per day a muon produces one or more 
spallation neutrons that propagate in the scintillator and are finally 
captured by a proton with the emission of the characteristic 2.26 MeV gamma line.
The capture time is about 250 $\mu$s. For several analyses ($^{11}C$ cosmogenic
background tagging, antineutrino and others) it is very important to detect 
these neutrons efficiently (see also Fig. \ref{fig:neutrons} in section \ref{sec:}). 
The original triggering scheme used from May 2007 until the end of 2007 
did not allow this because of an intrinsic dead time present in case of 
3 or more consecutive triggers in a time window of 1 ms. To solve this problem
we have modified the logic so that every time a muon (identified by the 
outer detector) crosses the SSS, a very long gate of 1.6 ms is issued
instead of the standard gate. In this way all neutrons can be acquired.

A triggering system is also required for the outer muon detector. As 
described in section \ref{sec:OuterDetector}, the QTC boards are
capable of providing an analogue signal whose height in steps is
proportional to the number of channels that fired in coincidence. 
These QTC secondary signals are sent to a custom VME board, the Muon
Trigger Board (MTB), which allows us to set a threshold on the number of PMTs
firing in coincidence, and sends a signal to the BTB. When the MTB signal
arrives, the BTB initiates the normal trigger sequence that is used
for any other trigger. The typical threshold used for physics corresponds to about
6 PMTs.

The whole detector, both internal and external,
is always read--out for any trigger type. For more details about trigger
logic see \cite{bib:DManuzioThesis}, \cite{bib:DDangeloThesis}.

\section{Data acquisition system}
\label{sec:Daq}
All the inner detector digital boards have a VME interface and are organized
in crates hosting 20 boards and a VME Single Board Computer (SBC). 
The Outer Detector TDCs and the Flash ADC boards are also VME based and 
are read by means of the same kind of SBC (1 crate for the Outer Detector readout
and 4 crates for the Flash ADCs).
The data are read crate by crate by the SBC and sent through the network to a 
workstation.
The VME computers (Motorola MVME 230x) are based on the Power PC (603e) CPU and
on the Universe VME/PCI bridge; the operating system running on these machines 
is Debian GNU/Linux with the 2.2.12 kernel. The performance of the system in
a real-time analysis has been studied; the interrupt latency has been found to always be
lower than 100~$\mu$s, with a mean of 20~$\mu$s.
The SBC are diskless and a cross-development setup has been installed on the
i386 based workstations. See \cite{bib:ARazetoThesis} for details.

The read-out is trigger based: once the system is started, the trigger 
signal is propagated to all the racks where the SBCs collect the data 
from the digital boards for network delivery; as soon as all event 
fragments are received by the builder workstation,
the full event is reconstructed and written to disk. 
At the end of the run all the data are sent to a
Raid storage, in a compressed form (the compression algorithm and a 
MD5 hash, assure that the data can not be modified).

The data throughput is dominated by $^{14}$C, the largest source 
of trigger rate; a concentration of about 
$3\cdot 10^{-18}$~g/g corresponds to $250$~Bq of $\beta$~decay in 
the active volume. 
Depending on the threshold, the trigger rate can be up to $100$~cps with 
a mean energy of 150~keV; this produces a data flux of the order of 
$50$~kbyte/s and a total storage of about  4~GB per day.
These numbers are compatible with a simple network design, 
based on 100~ Mb/s ethernet technology with a switched layout.
A full private network setup has been installed (with private 
IP addressing compatible with RFC1597) 
and router/firewall interconnects the Gran Sasso LAN with the Borexino one; 
the system includes all
the facilities commonly found in a full network deployment.

The data acquisition system is controlled 
by a set of web pages which in turn interact 
with the different processes using CORBA. A PostgreSQL database is 
used to maintain the run  information (like the start time, the disabled 
channels list and so on), which are necessary for the
analysis and calibrations. 
All the code has been developed in C, C++ and in Perl scripts.

The maximum data flux that the internal read-out can sustain is 
a 400~cps trigger rate with high energy events (one hit per channel 
which is approximately equivalent to the deposit of a 
5~MeV $\beta$--particle); 
this high rate fulfills all the Borexino physics goals. 
The main use of such a high trigger rate is for laser calibrations.

\section{The Photomultiplier Calibration System}
\label{sec:CalibHardware} 
Borexino relies on the precise
determination of the time of flight of the photons from the location
of the scintillation event in order to reconstruct the event position 
and  to define the Fiducial Volume.
Furthermore, the knowledge of the total charge collected by each
photomultiplier is important for the energy determination of high
energy events. For these reasons, both time and charge calibration of the photomultiplier
system is of utmost importance.  Therefore, a multiplexed
system of optical fibers has been developed for the precise
photomultiplier calibration both in time and in gain.

Precision in the time measurement of single hits affects directly the
position reconstruction precision. Position resolution is limited
by the scintillator fluorescence decay time of 3.5~ns
(effectively increased to 5.5~ns after light propagation
effects~\cite{bib:NIM_scint}), by the photomultiplier transit-time jitter of $\sim
1$~ns and by the inter-photomultiplier time equalization, which should be
maintained at the sub-nanosecond level and is regularly checked. The
precision of the time measurement is also crucial for $\alpha/\beta$
discrimination, based on the different fluorescence time profiles
for $\alpha$ and $\beta$ scintillation events (see 
~\cite{bib:scheda_piero}). An accurate energy determination and
resolution are crucial for the spectral shape recognition of the
neutrino signal: the energy resolution of the detector depends on
good  charge calibration, the energy being determined, through a
proportionality relation, from the number of detected photons or from
the total charge collected by all photomultipliers.

\subsection{System Design}

In the design of the calibration system for the time and charge
response of the Borexino photomultipliers the following requirements have been
taken into account:

\begin{enumerate}

\item accuracy in time equalization: inter-photo\-multi\-pliers equalization
inaccuracy should be lower than the photomultiplier time jitter, 
which is $\sim 1$~ns;

\item accuracy in charge calibration: the system must illuminate
all 2212 photomultipliers at the single photoelectron level in order to measure
the Single Electron Response (SER) parameters of each photomultiplier. If we
require less than 1\% contamination from multi-electron pulses and
from the dark noise accidental coincidences, the illumination level
should be, respectively, lower than a mean value $\mu = 0.05\ p.e.$ (photoelectrons)
and higher than $\mu = 0.01\ p.e.$~\cite{bib:laser_calibr};

\item linearity check: the photomultiplier illumination level should be adjustable,
in order to verify the linearity of the photomultiplier response. The value of
$\sim 8.6\ p.e.$   corresponds to the saturation limit of the
front-end electronics (in an $80\ ns$ interval): higher intensity
calibration sources would allow non-linearity studies;

\item calibration rate: the photomultiplier calibration should be performed at
the fastest rate allowed by the DAQ system (a few hundreds Hz at
an illumination level of $\mu = 0.05\ p.e.$), in order to minimize
the duration of the measurement;

\item operational convenience: the system should allow a high
level of automation for frequent use and must be operable from
outside (a remotely controllable system also meets the requirement
for a synchronization with the main electronics);

\item radioactivity: the contamination induced by the calibration
system materials must not add a significant rate to the dominant
(unavoidable) background generated by the photomultipliers;

\item long-term reliability, for at least 10 years of data taking.

\end{enumerate}

Several alternative solutions were considered, that could suit the
detector geometry consisting of consecutive shielding regions. The
simplest method would be a light diffuser in the center of the
detector (as in the SNO experiment~\cite{bib:sno_dete}), illuminating
all the photomultipliers simultaneously. However, a permanent diffuser would not
be acceptable since it would represent a constant source of
radioactivity in the innermost region, while a removable one would
not allow for the simplicity and safety of operation needed for
frequent calibrations. Permanent light diffusers could be placed in
the buffer region (as in the LSND~\cite{bib:LSND} experiment), or light beams could
be generated by optical fiber couplers in the SSS;  in both of these
solutions, the light would cross several meters of the buffer and
scintillator regions, requiring off-line corrections to the
light-pulse arrival time: the consequent dependence on the
scintillator properties (such as index of refraction and attenuation
length), which could drift in time, would compromise the calibration
accuracy.

\begin{figure}[!tb]
\begin{center}
\includegraphics[width=0.48\textwidth,clip=true]{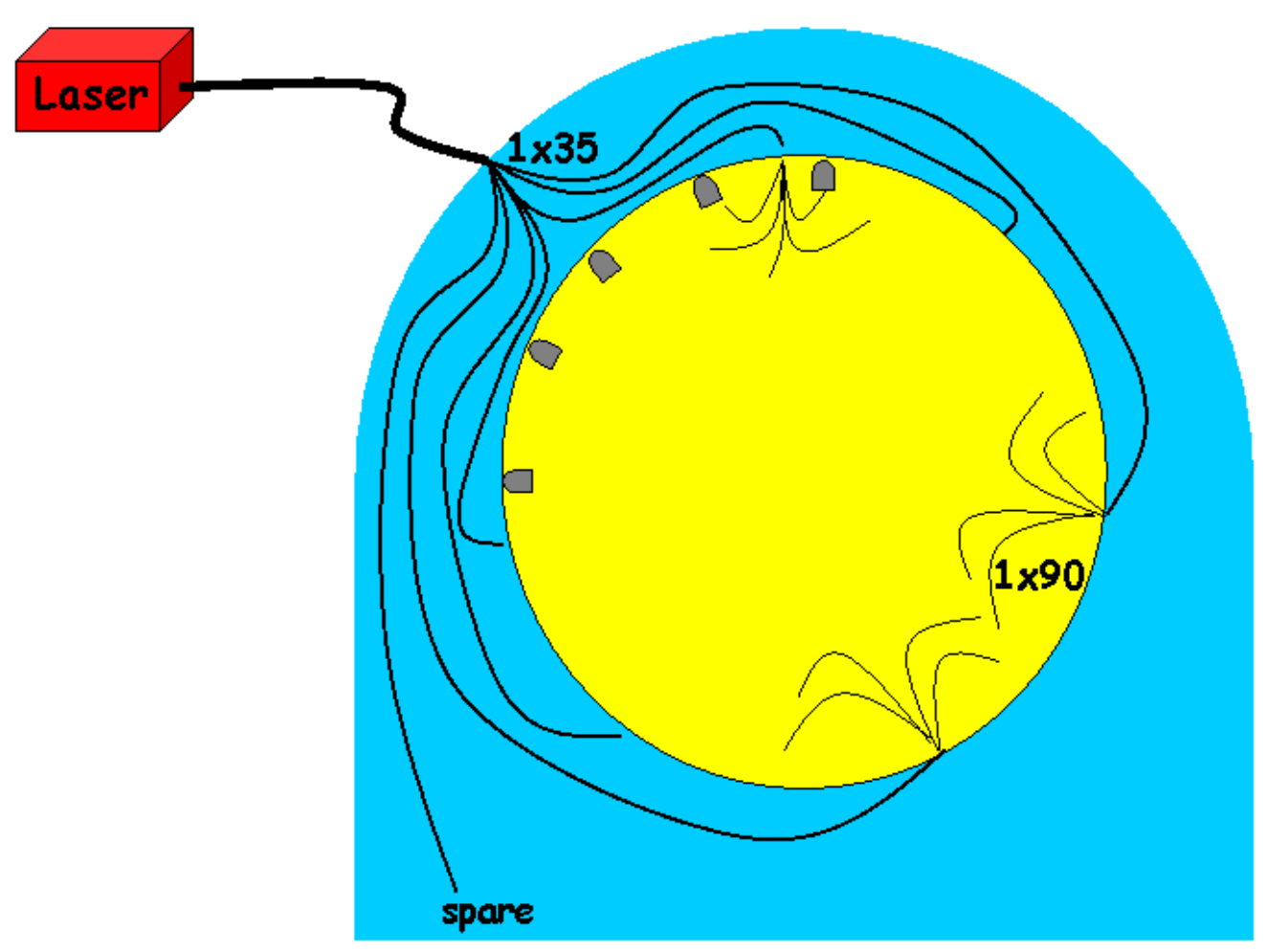}
\caption{Scheme of the multiplexed laser system for photomultiplier calibration.
A single very short laser pulse is distributed to all PMTs through a set of optical
multiplexers and fibers.}
\label{fiber_split}
\end{center}
\end{figure}

The solution adopted for Borexino is based on a different idea:
the light emitted by an external laser source is carried simultaneously 
to each photomultiplier tube by a dedicated system of optical fibers.
As shown in Fig. ~\ref{fiber_split}, the laser light is first distributed to 35 fibers
which reach the Stainless Steel Sphere in 35 different points. Light enters the SSS
via 35 custom designed optical feedthroughs and is again split into 90 fibers
each individually coupled to a photomultiplier. In this way, radioactive contamination is avoided
since no part of the system is immersed in the innermost part of the detector. Furthermore,
light reaches the photocathodes without crossing a large amount
of scintillator and/or buffer liquid, thus avoiding dependency on the medium characteristics.
The entire fiber system design follows a requirement of mechanical decoupling of the
internal/external regions to facilitate the mounting operations:

\begin{itemize}

\item air/water interface: a fast light pulse emitted by an
external laser is focused onto a bundle of 35 external fibers 40~m long,  running 
in the water buffer region;

\item water/SSS interface: the 35 external fibers reach different
locations on the sphere supporting the photomultipliers. Each single external
fiber couples to a proper optical feed-through on the SSS itself;

\item SSS/PC buffer interface: each feed-through is coupled to a
bundle of 90 internal fibers $6\ m$ long, running inside the SSS and
reaching every single photomultiplier.

\end{itemize}

\subsection{Photomultiplier Calibration Light Source}
\label{sec:laser_timing}
The light source is a diode laser ({\it PicoQuant} LDH400), emitting
a fast (50~ps time width) light pulse at a wavelength of 394~nm, 
where the photocathode quantum efficiency is about 27\%. The
maximum peak power is 400~mW, corresponding to $1.7 \times 10^7$
photons/pulse. The laser driver (PDL 800-B) allows a maximum
repetition rate of 40~MHz, well above the maximum rate of the
Borexino DAQ system and can be triggered by an external pulse to
allow synchronization with both main electronics and DAQ; laser
intensity can be selected by means of a potentiometer which, by
design, is only manually adjustable (an upgrade of the controller
allowing remote setting of the intensity is under study).

The optical components associated with the coupling of the laser to
the first segment of the calibration system consists of a series of
neutral density filters for attenuation and a lens to focus the beam
on the surface of a 2.5~mm diameter rigid quartz fiber (cladrod) that then
distributes the light to the 35 fiber bundle. The lens is a
$10\times$ standard microscope lens; focusing is necessary since the
area of the laser spot ($2\times3.5$~mm), is larger than the quartz
fiber core. The rigid quartz fiber is 10~cm long and its
numerical aperture is lower than the 35 fiber bundle area, for
better optical transmission. The optical apparatus is mounted on a
passive anti-vibration platform and enclosed in a light-proof
stainless steel box; the laser controller can be activated from
outside the box.

\subsection{External fibers}

\begin{figure}[!tb]
\begin{center}
\includegraphics [width=0.5\textwidth,clip=true]{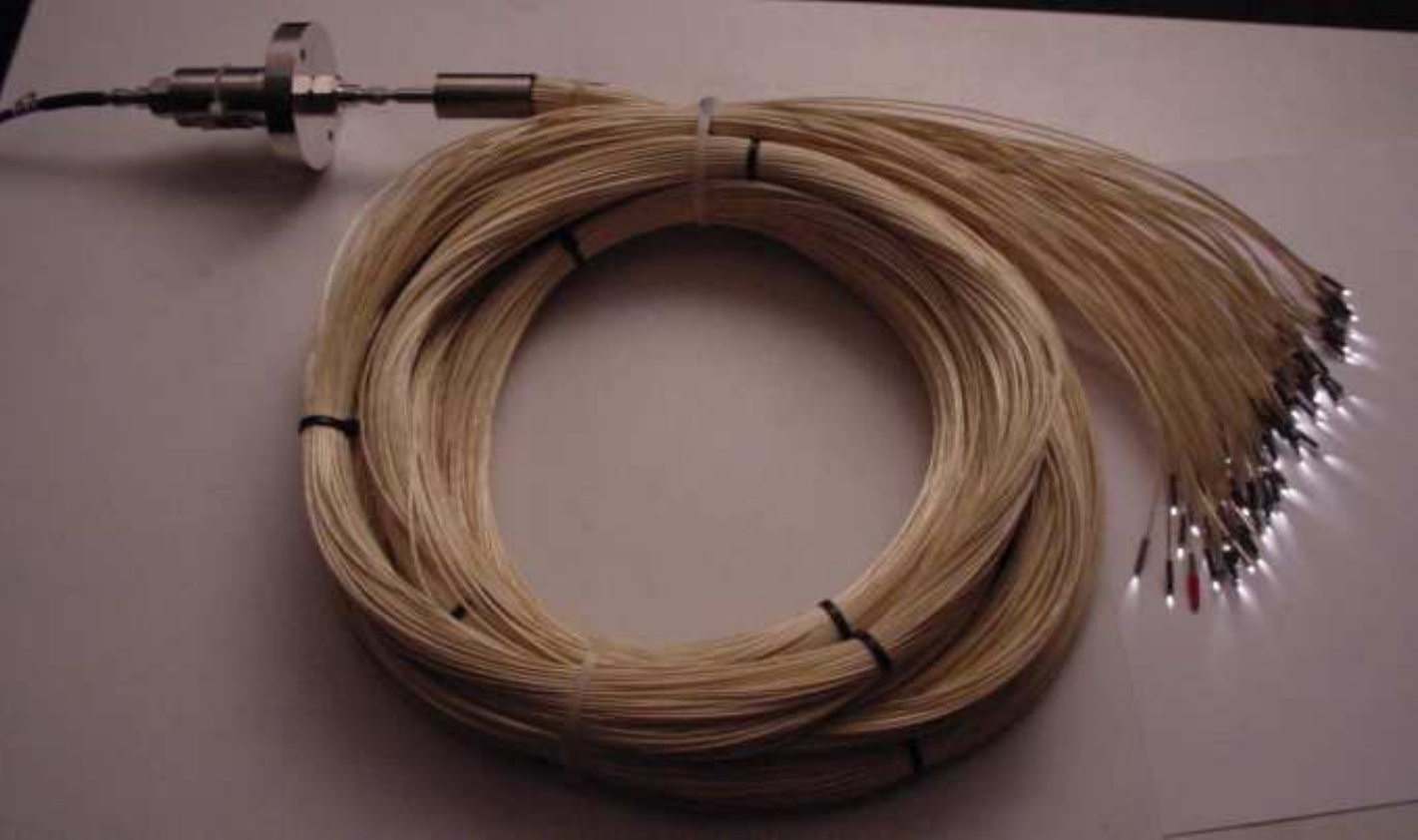}
\caption{A bundle of 90 fibers connected to the feed-through and
illuminated.} \label{fiber_bundle}
\end{center}
\end{figure}

As previously said, the laser beam illuminates the fiber bundle of the first splitting
point through a focusing lens and an optical rod that renders it
uniform and wide enough to cover 35 quartz fibers. The diameters of
the fiber core and cladding are 300 and 325 $\mu$m respectively.
The fibers enter the detector through a single feed-through on the
top of the Water Tank and form separate cables, with
Kevlar strands (for mechanical resistance) and a polyethylene
coating, until they reach their entry points on the SSS. The photomultipliers
are grouped in 28 clusters of about 80 elements. For mounting
reasons, the tubes installed on the 3~m main entrance door
require an extra fiber bundle. The locations on the SSS are chosen
as the central position of each photomultiplier cluster. Six out of the 35
fibers are spare.

\subsection{Light-transmitting feed-through}

A delicate part of the system is the feed-through on the SSS, since
it must distribute the light from 1 input to 90 output fibers as
uniformly as possible, while maintaining a high level of tightness
against liquid transfer between both sides of the SSS. The body and
the connectors are made of electro-polished stainless steel, the
sealing with the stainless steel flange is accomplished with a Viton
O-ring. A quartz fiber, of length 10~cm and diameter 1.5~mm, is placed 
inside the connectors, optically coupling the PC and
the water buffer side. The sealing against PC diffusion by
capillarity near the fiber is assured by filling the inner part of
the feed-through with a PC-proof, quartz-steel adherent epoxy resin.

\subsection{Internal fibers}

The second step of the multiplexed chain is formed by coupling the
SSS feed-through to a bundle of 90 quartz fibers of 110~$\mu$m
core diameter (Fig. \ref{fiber_bundle}). Taking into account the
packing ratio of circles on a plane, 90 closely packed fibers with a
diameter of 120~$\mu$m (cladding) occupy an area of 1.12~mm$^2$.
The cross-sectioned area of the 1.5~mm diameter fiber in the feed-through is
1.76~mm$^2$, thus a tolerance of 50\% for non-optimal packing of
the fiber bundle is achieved.

Both internal and external fibers, as well as the lens and clad rod
used at the beginning of the optical path, have a quartz core, in
order to optimize light transmission efficiency in the ultraviolet
wavelength region. The cladding of the external fibers can be made
of plastic, while the strong chemical reactivity of pseudocumene
determines the choice of quartz for the cladding of the internal
fibers. This is one of the reasons for using Teflon for the coating
of the internal cable; in addition, measurements carried out by the
collaboration showed that Teflon has a very low rate of Radon
emanation. A Teflon support attached to the photomultiplier light concentrator
points the fiber termination in the direction of the photocathode,
at a 20~cm distance.

\begin{figure}[!ht]
\begin{center}
\includegraphics[width=0.5\textwidth,clip=true]{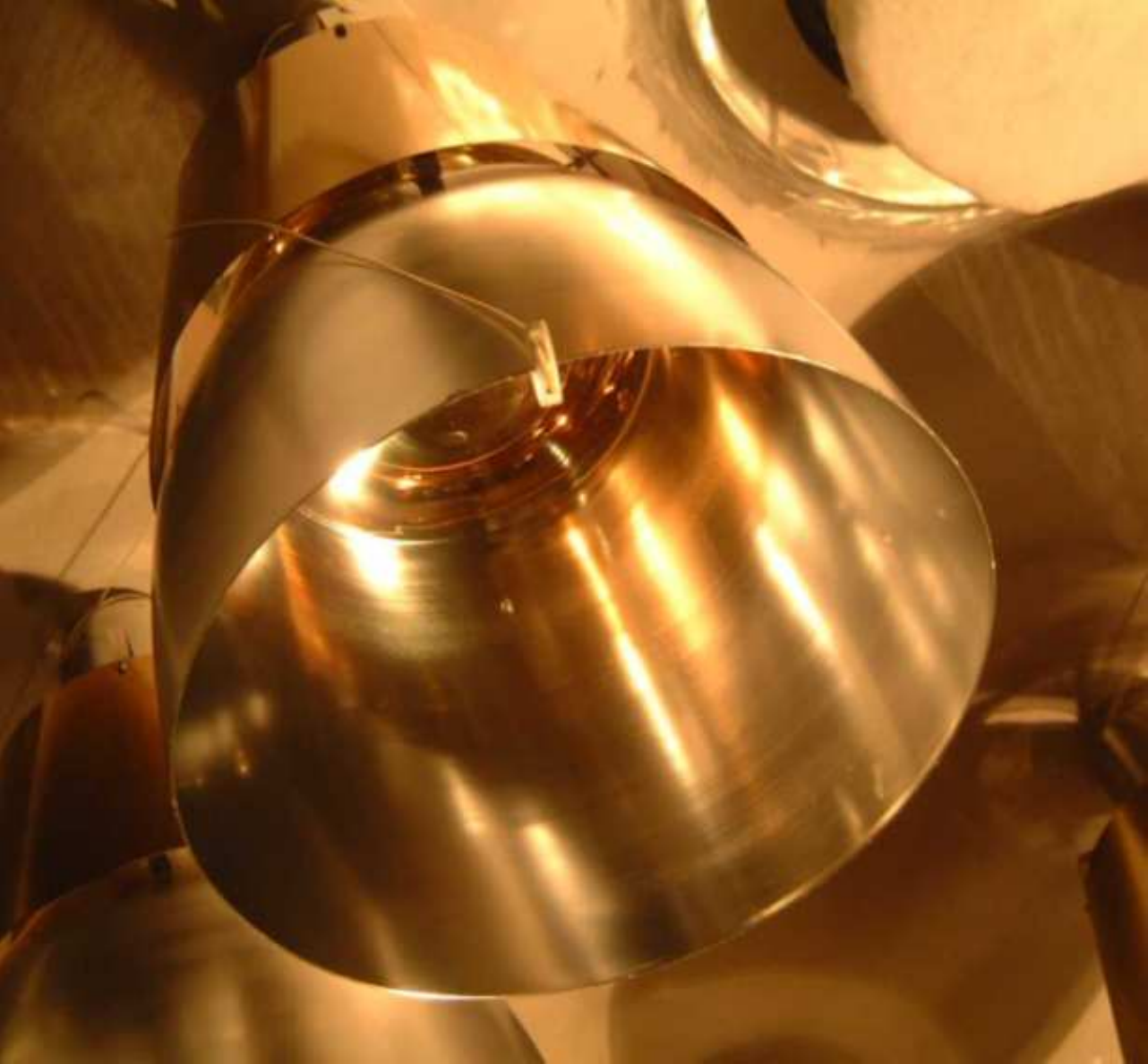}
\caption{A picture of an installed photomultiplier: it shows the optical fiber used for timing and charge
calibration and the Teflon support attached to the light concentrator.}
\label{installed_pmt}
\end{center}
\end{figure}

\subsection{System feasibility tests}

During the design phase, several tests were performed to guarantee
the feasibility of the system for Borexino. 
On the one hand, the
compliance of the system with the general requirements of the
experiment in terms of radioactivity and chemical compatibility had
to be verified and on the other, the total light transmission
expected at the end of the chain had to be measured. The attenuation
due to geometrical effects and fiber transmission can indeed be
safely gauged from the a priori known fiber characteristics, while
the inefficiency caused by losses at  optical couplings can only
be measured in the final experimental conditions. The results of the
system feasibility tests are described in detail
in~\cite{bib:laser_calibr}, along with the light transmission properties
measured on the assembled and installed system.

\subsection{Outer detector calibration system}
\label{sec:CalibOuter}

\begin{center}
\begin{figure}
\includegraphics[width=0.49\textwidth]{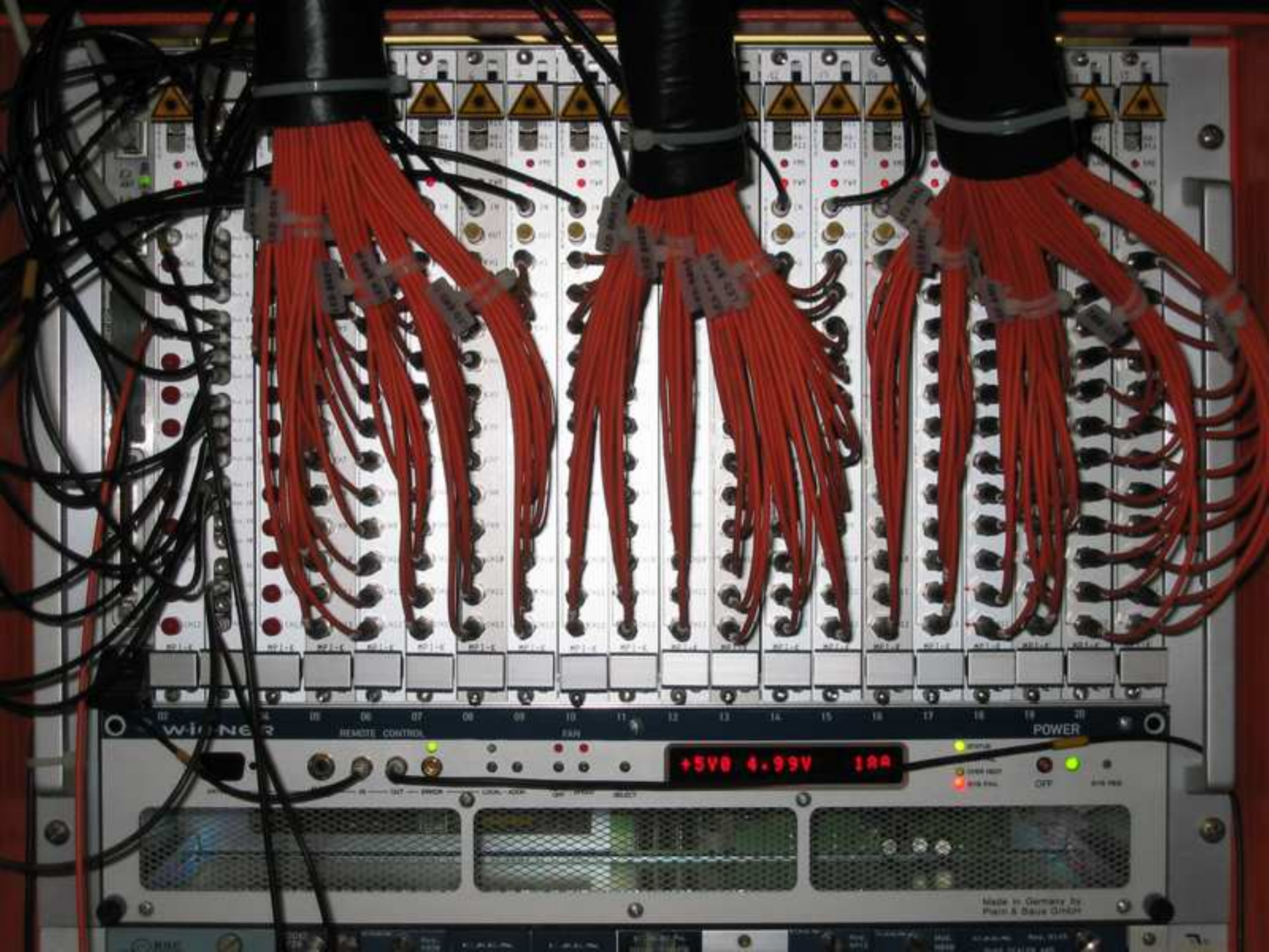}
\caption{The Outer Detector calibration system. 
  LEDs housed in electronic boards deliver light pulses to PMTs in the Water Tank 
  through individual optical fibers (orange).}
\label{fig:led_calib1}
\end{figure}
\end{center}

\begin{center}
\begin{figure}
\includegraphics[width=0.49\textwidth]{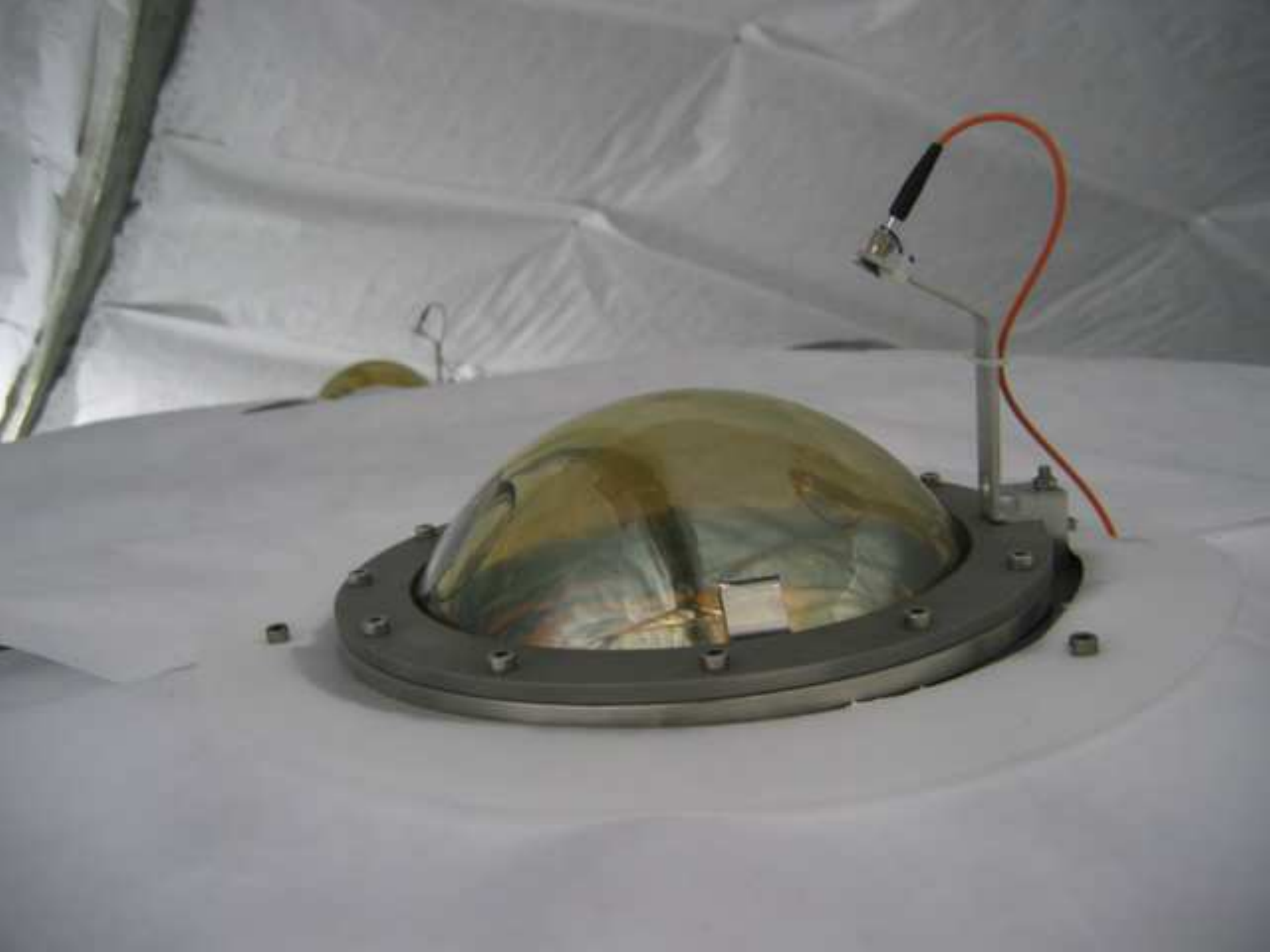}
\caption{The Outer Detector calibration system: the tube 
  enclosing the fiber and its mechanical support are visible above the PMT glass. }
\label{fig:led_calib2}
\end{figure}
\end{center}

The main goal of the Outer Detector (OD) is to tag muons crossing Borexino in order to
exclude them from the samples of events used for the neutrino analysis. Moreover,
since muons  can produce radioactive spallation products inside the scintillation
core, it would be important to reconstruct the muon track in order to spatially
correlate it with a subsequent internal event.
In order to develop refined muon tagging tools and to perform a reliable muon track 
reconstruction it is important to know precisely the relative arrival times 
of each \che\ photon at the phototubes.
However, the transit time of a signal through a photomultiplier tube and 
the electronics  to which it is connected can differ from one channel to another 
by several nanoseconds.
It is therefore mandatory to develop a calibration system to perform the
time alignment of all channels within 1~ns by means of a light pulse
sent simultaneously to all photomultipliers. 
The idea is similar to the one described in the section for the Inner Detector.
The OD photomultipliers are operated in a
wider dynamics than the ID ones: even though it is difficult to quote an average 
hit occupancy for a muon event since 
this strongly depends on the geometry of the track, 
simulations show that photomultipliers can easily detect bursts of 50 or even 100 photoelectrons
 if directly hit by the \che\ cone.
On the other hand, photomultipliers seeing only light diffusively reflected by the Tyvek 
surfaces will yield very few photoelectrons. 
Although the photomultiplier linearity is not in question, a reliable calibration 
should extend as much as possible in the operational dynamic range.
It was therefore decided to develop a  calibration system for 
the outer detector capable of working both in single photoelectron mode, as well as 
in charge regimes up to 40 photoelectrons (and possibly beyond).
The system is made of 208 \emph{Light Emitting Diodes} (LEDs), 
each one delivering the light pulse to one PMT in the Water Tank through an optical fiber.
The optical fiber is mounted on the photomultiplier with a steel support that 
holds it about 8 cm and $30^{\circ}$ off-axis.
Considering the opening angle of the fiber ($15^{\circ}$ semi-aperture) the support 
was designed for 100\% geometrical efficiency and minimal shadowing of 
the photocathode. Fig. \ref{fig:led_calib1} and Fig.  \ref{fig:led_calib2} 
show the layout of the system.

\section{Optical calibration of the scintillator and buffer liquid}
\label{sec:transp}

As already mentioned, the energy and position reconstruction 
in Borexino rely critically on the optical
properties of the scintillator and buffer liquid. The number of emitted photons 
and their time distribution  depend on the scintillator;
the number of detected photons and their arrival time depend on 
the transparency of the media they cross.
Many studies have been performed in the laboratory to measure critical parameters  of
the scintillator  and buffer  liquids, such as the emission and 
absorption spectra, interaction length and so on.  Unfortunately,
the extrapolation of these results to a large--volume 4$\pi$ detector like Borexino may not be 
straightforward (see for example the  Counting Test Facility data~\cite{bib:NIM_scint}) 
and it is therefore necessary to check them $in$ $situ$.  
Furthermore, since Borexino will take data for many years, 
it is important to keep the relevant optical parameters of 
buffer and scintillator under control
during the experiment's lifetime, in order to make sure that no significant variations occur. 
For example, it is known that some mixtures of organic scintillators can change 
their light absorption properties when exposed to a large amount of untreated steel 
for an extended duration of time~\cite{bib:CHOOZ}. 

\begin{center}	
\begin{figure}[!tb]
\includegraphics[width=0.50\textwidth]{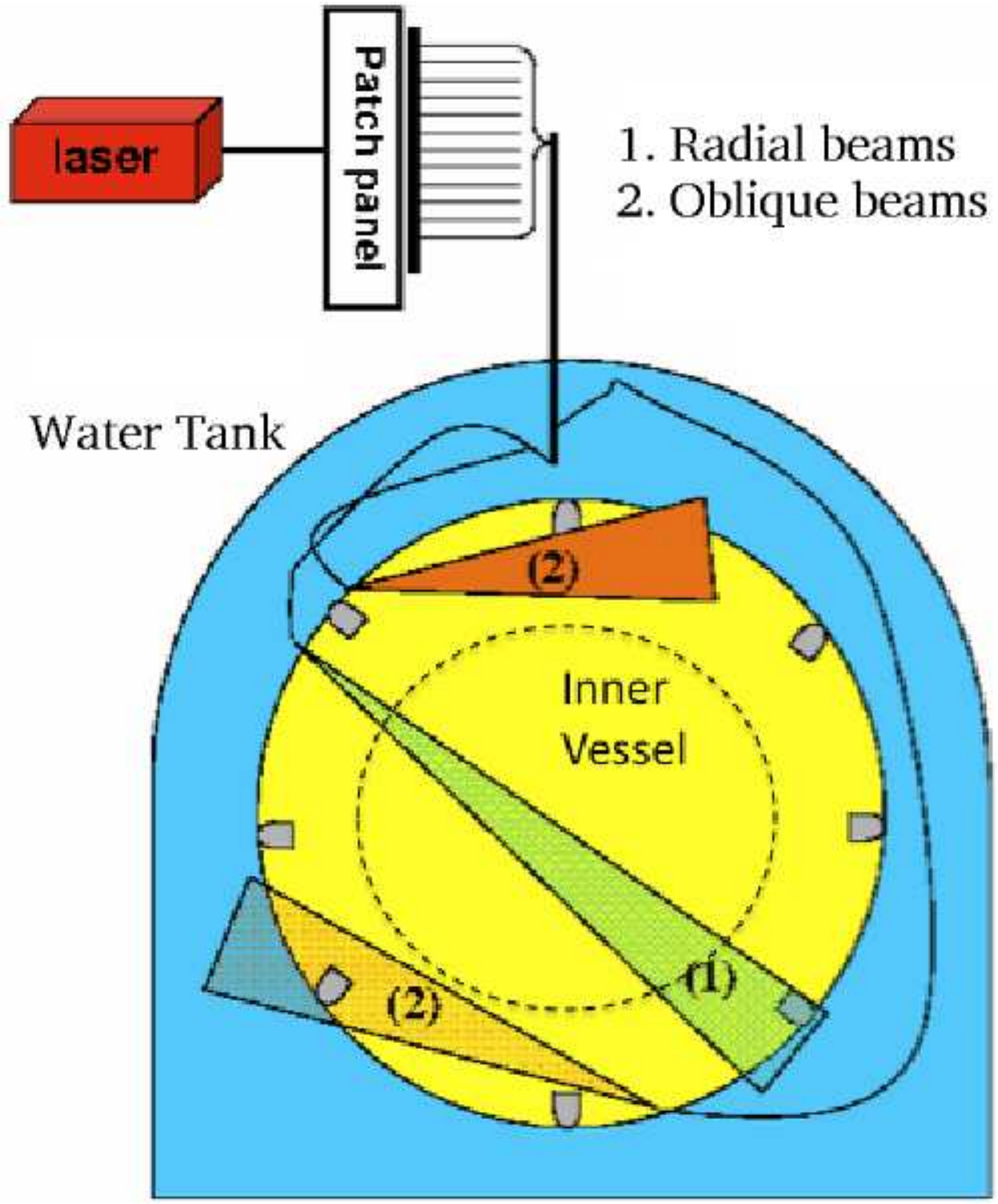}
\caption{ Scheme of the system to monitor detector transparency: the laser light is transmitted by 
means of optical fibers  through the Water Tank into the SSS. 
The picture sketches the light beam in case of  $radial~feed-through$ (1) 
and $oblique~feedthroughs$ (2).}  
\label{fig:transparency}
\end{figure}
\end{center}

\begin{center}	
 \begin{figure}[hb]
\includegraphics[width=0.5\textwidth,bbllx=0,bblly=0,bburx=460,bbury=340,clip=]{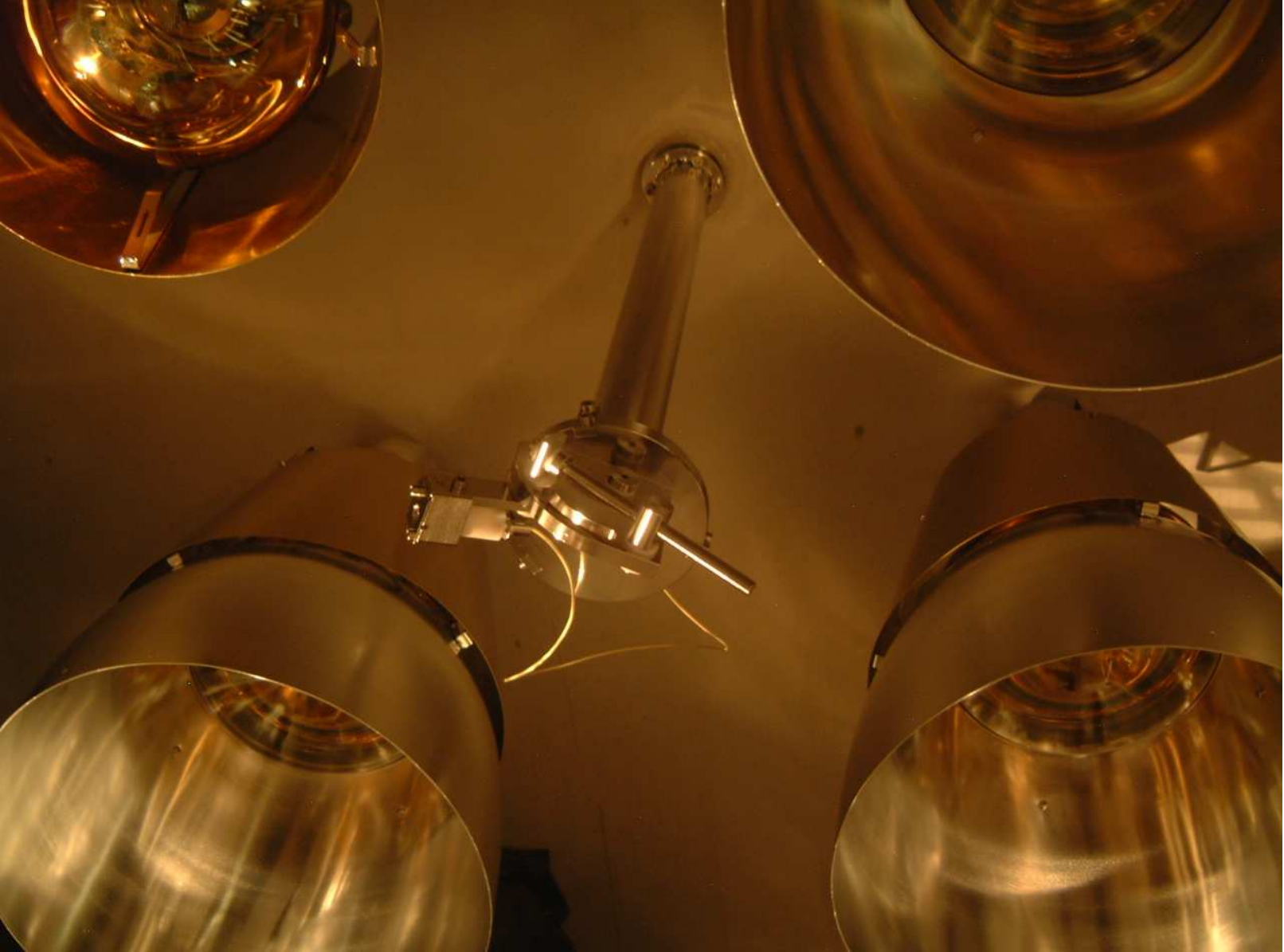}
\caption{ The pointing device for the $oblique$ type of feed-through: the photo shows a 
module already installed on the Borexino 
Stainless Steel Sphere and surrounded by PMTs. The device is designed to point laser 
light at an angle towards a specific target phototube. The angle is chosen so that the light 
crosses only the buffer liquid and not the active scintillator.}  
\label{fig:pointing}
\end{figure}
\end{center}
 
For these reasons, we have implemented a system which allows a non-invasive and easy 
monitoring of the optical properties of the detector liquids by sending laser light of 
different frequencies through different portions of the detector.
A schematic view of the  system is depicted  in Fig.~\ref{fig:transparency}. 
It consists mainly of three parts: 
\begin{enumerate}
\item Lasers. In order to probe the optical properties of the detector at different wavelengths, two
lasers are available: one is the diode laser already described in section 
\ref{sec:laser_timing} ($\lambda =394$~nm). 
The second one is a solid state laser which emits light at $\lambda$ = 355 nm;

\item Optical fibers. The laser light is carried inside the detector by means of 31 single fibers which 
enter the Stainless Steel Sphere in 31 different positions. The system is designed to 
work with only one fiber at a time,
in order to collect data with the light coming from a single point on the sphere:
the selection of the fiber is performed via an external patch-panel;

\item Feed-throughs on the SSS. They are of two types: 
the $radial~type$ and the $oblique~type$ (see Fig.~\ref{fig:transparency}). 
There are 11 $radial$ feed-throughs and they
are designed to shoot light radially through the detector. In order to collimate the light beam, 
they are equipped with a small lens and a pin-hole aperture.
The $oblique$ feed-throughs are designed to shoot light at an angle such that it crosses the 
buffer liquid only and not the active scintillator. In order to do so, the feed-throughs 
are equipped with a pointing device (Fig.~\ref{fig:pointing})
which aims precisely at a selected target photomultiplier (PMT).
The target PMTs have been chosen at different distances from the feed-throughs
in order to probe attenuation of light after different path lengths: depending on the 
feed-through, the distance which the light crosses before reaching the target 
PMT ranges from a minimum of 2.5~m to a maximum of 7.88~m.

\end{enumerate}

The monitoring strategy dictates that we perform, periodically ($\sim$ once per month) some
dedicated runs shooting laser light both through the buffer liquid using the 
$oblique$ feed-throughs and through the buffer + scintillator using the 
$radial$ feed-throughs described above.
In this way, it is possible to study the transparency of the two media while undergoing
both absorption and elastic scattering processes: we recall that absorption reduces
the total amount of collected light (and therefore affects energy reconstruction), 
while scattering changes the time distribution of detected photons thus affecting the position
reconstruction and $\alpha/\beta$ discrimination capability.
The laser light at $\lambda$ = 394~nm  can be used both for $radial$ and 
$oblique$ beams since in this wavelength region scattering on pseudocumene is 
the dominant process (with an interaction length of approximately $\sim$ 5~m). 
The laser light at $\lambda$ = 355~nm can be only used to
study the buffer transparency and not the scintillator one  
since the absorption and  re-emission process on PPO at this wavelength would 
completely block light crossing the scintillator.
At this wavelength the expected scattering length of PC is $\sim$ 2.5~m.
Thanks to the fact that both the radial and the oblique beams are collimated (depending on the
type of feed-through, the opening angles range between 
2$^\circ$ and 6$^\circ$), it is possible to measure the 
scattered and transmitted light separately thus performing a periodic monitoring of 
the detector's optical response.
Collecting data with $oblique$ beams with target photomultipliers at different distances
also allows measurement of the absolute PC attenuation length.
In order to have redundancy of information and to keep the systematics 
under control, the $oblique$ system has more than one beam crossing the same distance.
For more details about this monitoring and calibration system see \cite{bib:JManeiraThesis}.

\section{Internal Source Calibration System}
\label{sec:insertion}

\subsection{Motivation}

In order to complete the scientific program, and particularly to measure the 
solar neutrino fluxes, the Borexino detector must be carefully calibrated.
  
As shown by several studies done with Monte Carlo simulations, and confirmed
by our first measurement of \ber\ neutrinos \cite{bib:be7paper}, 
the main sources of uncertainty in the measurement of the solar neutrino fluxes are: 

\begin{enumerate}

\item The knowledge of the fiducial mass, {\it i.e.} the inner most part of the
scintillator that is used as the neutrino target and whose volume is defined by
means of a cut on the reconstructed position of the events. An accuracy 
of $\pm$ 2 cm is required for the knowledge of the radius of the 
fiducial volume.

\item The knowledge of the energy scale for $\alpha$, $\beta$, and $\gamma$ 
particles.

\item The knowledge of the detector energy response as a function of the 
event position within the fiducial volume, of the particle type and of the 
energy itself (non linear effects).

\end{enumerate}

Additionally, the optimization of the $\alpha$/$\beta$ separation, and a precise 
knowledge of its efficiency, as a function of particle energy 
is crucial for several physics analyses.

It is worth noting that some of these problems can be addressed by studying 
some internal radioactive contaminants in the scintillator. However, the
bulk activity is generally not sufficient to study the detector response 
well enough.

For these reasons, we have developed a system for the insertion of radioactive
and light sources in the Borexino bulk. 
This system must fulfill stringent requirements in terms of radiopurity, cleanliness, 
mechanical strength and reliability, and must guarantee complete air tightness 
while inserting, operating, and removing the sources from the detector.

With the use of this system, a suitable set of $\alpha$, $\beta$ and $\gamma$ sources 
can be put into various accurately known positions within, and at the border of, the
fiducial volume, so that the complete detector response can be determined.

\subsection{Cleanliness requirements}

The chief concern of an extensive calibration campaign must be the minimization of the
risk of detector contamination. The system described in section \ref{sec:locate} was 
designed to insert, position, locate, and remove a radioactive or fiber optic calibration 
source in a manner that does not pose undue risk to the vessels or contamination of the 
scintillator.  All components that come into contact with Inner Vessel scintillator have 
been  electropolished, pickled and/or passivated with proper acids, washed with 
suitable detergents, rinsed with ultra-pure water until particulate counting reached 
class level\footnote{MIL-STD 1246C} 20. This class level 20 corresponds to a contamination of
about 2 $\cdot$ $10^{-16}$ g/g when the contaminant is present in the dust at the ppm level,
the typical value for LNGS dust both in Thorium and in Uranium.  

The most critical components are housed inside of a glove box under a Low Argon and 
Krypton Nitrogen (LAKN) atmosphere \cite{bib:lakn} in order to prevent accumulation of Radon and 
other isotopes on their surfaces when not in use.  In addition to the strict cleaning 
procedures, redundancies in critical instrumentation, and a very detailed operational 
procedure have been implemented in order to reduce the risks of contamination to a 
suitable level.  

\subsection{Description of the system}
\label{sec:locate}

The system can be effectively broken down into two subsystems -- insertion and location.  
The insertion system provides a means to place a source at any location inside the 
Inner Vessel, while the latter is used to provide feedback on the precise location 
($\pm$2 cm) of the source.  A brief description of each follows.

\subsubsection{Insertion system}

The insertion system was designed to insert three devices into the detector: 
radioactive calibration sources, fiber optic calibration sources, and a scintillator 
sampling tube that can be used for detector monitoring purposes.  
The radioactive sources are small quartz spheres (radius 2.5~cm) containing 
the desired isotopes as well as a small volume of scintillator,  whereas the fiber 
optic sources are cylindrical quartz vials which contain the fiber optic terminations 
and also the LED used for location.

The sources are attached to the end of a series of neutrally buoyant stainless steel 
rods, one meter in length, with an optional hinged section which allows the assembly 
to be rotated up to 90$^\circ$.  The rotation of the hinge is accomplished by pulling on a 
Teflon tether tube attached to the source coupler.  The entire assembly can also be rotated 
azimuthally 180$^\circ$ in either direction to map out cylinders in the Inner Vessel.

\begin{figure}[ht]
\centering
\includegraphics[width=0.48\textwidth]{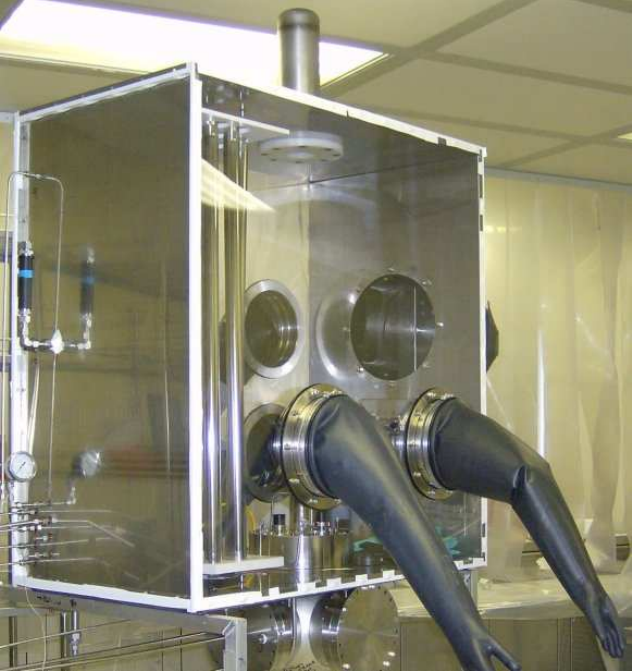}
\caption{The glove box used for source insertion.  Some of the insertion rods are 
visible at the left of the glove box.}
\label{fig:gbpic}
\end{figure}

The insertion operations are performed through a custom built glove box (see Fig. 
\ref{fig:gbpic}) which resides in a class 10 clean room atop the Borexino Water Tank. 
Located between the glove box and the Inner Vessel fill tube is a load lock where the 
sources are attached to a spring loaded rod coupler.  The insertion rods and Teflon
tether tube pass through sliding seals since there is a significant pressure difference
between the glove box and the load lock.  The glovebox, load lock, and the sliding rod 
seal, are continuously purged with LAKN during operations to avoid introducing any 
radioactive contaminants into Borexino during a calibration operation.  

All of the instrumentation for the insertion system is computer controlled and 
monitored, with several interlocks and alarms to alert operators before any damage 
is caused to Borexino.  Once the source has been inserted and positioned into the
predetermined site, the location system (described below) goes to work.

\subsubsection{Location system}
The system described above does not provide enough feedback to determine the position 
of the source to the level of precision required.  Thus it was decided to equip every 
source with a light source, and use a system of seven digital cameras to find the position of 
the source.  The cameras used are consumer grade Kodak DC290 digital cameras, each of 
which is equipped with a Nikon FC-E8 fish eye lens.  During 2002, this system was 
tested with a string of LEDs suspended in the center of Borexino and the ability to locate 
the sources to within 2 cm was verified (see \cite{bib:BackThesis} for details).

\begin{figure}[ht]
\centering
\includegraphics[width=0.48\textwidth]{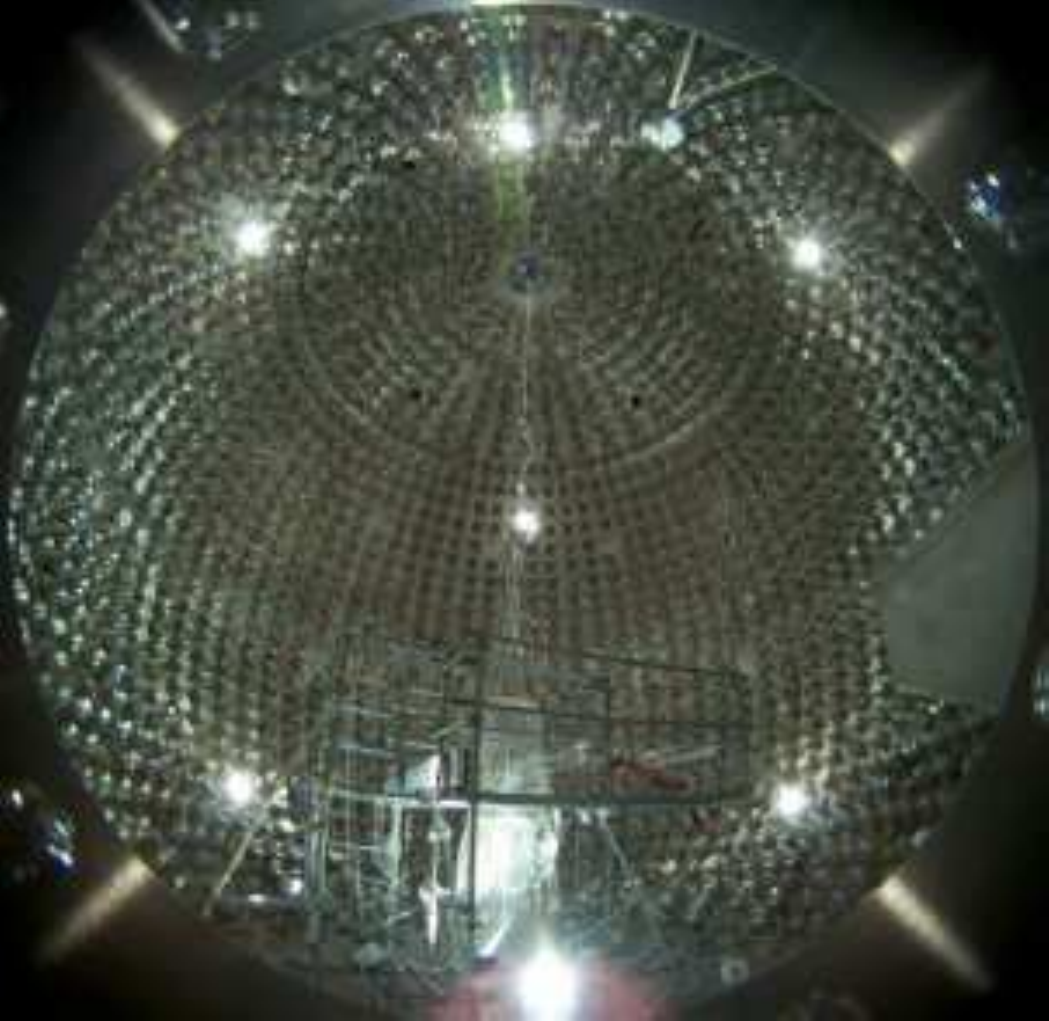}
\caption{The sphere during the nylon vessels installation. The picture also shows 
 the position of six out of the seven cameras (the bright spots). 
The 7$^{th}$ is of course the one that took the picture.} 
\label{fig:cameras}
\end{figure}

The camera system has also proved to be a valuable tool for vessel monitoring, 
particularly during inflation and filling operations; all pictures of the internal part of the
SSS that are shown in this publication and elsewhere are taken with this system. A 
software package was written to present the collaboration with many analysis tools for 
deconvolving the images taken during the water and scintillator filling processes. Fig.
\ref{fig:cameras} shows the position of 6 out of 7 cameras, the 7$^{th}$ being the one 
that took the picture.

\begin{figure}[ht]
\centering
\includegraphics[width=0.48\textwidth]{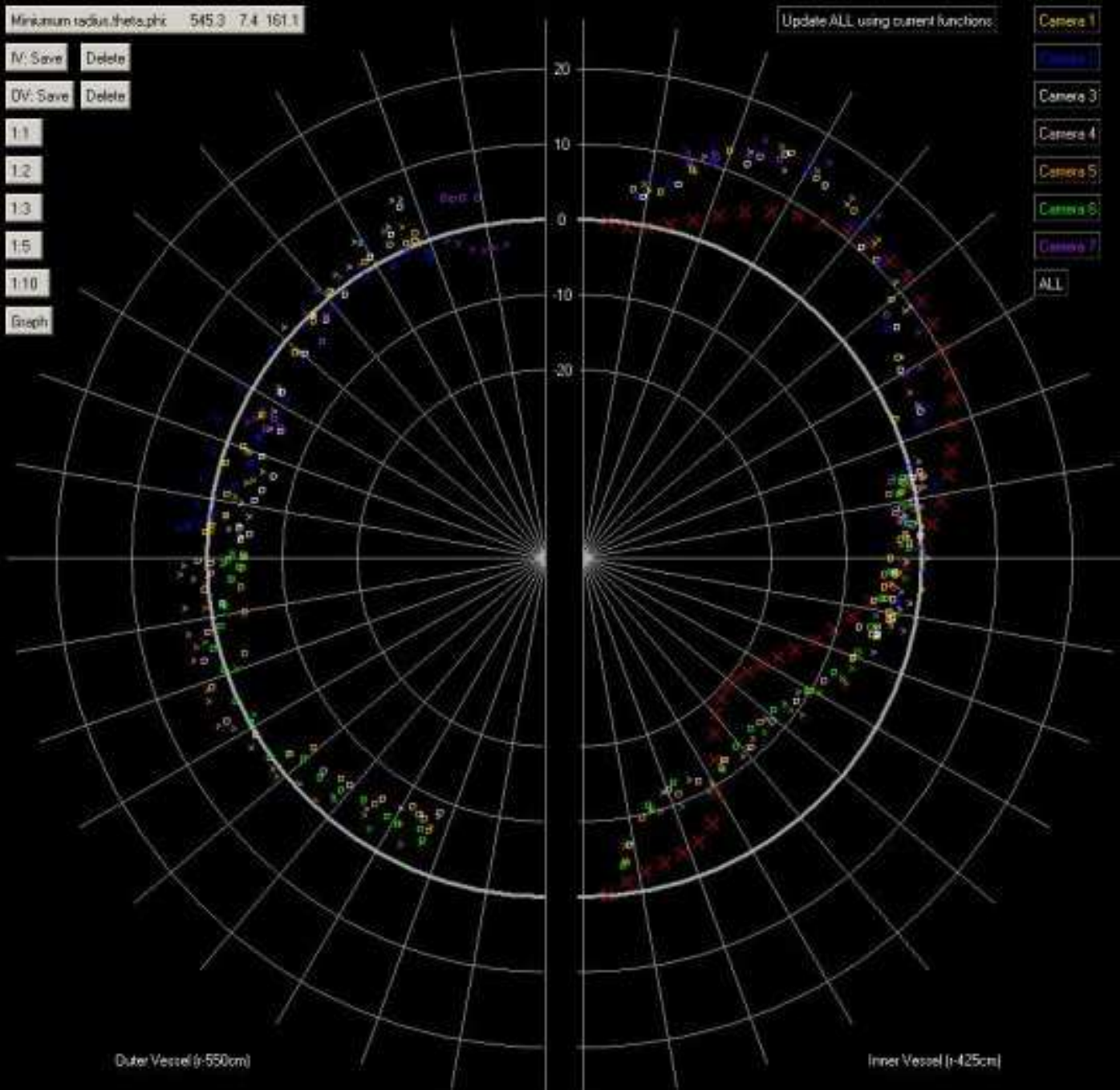}
\caption{This image shows the deviation in cm of the radius of the nylon vessels for each of
the seven cameras. Each color refers to a different camera (see Fig. \ref{fig:cameras}). 
The white line is the Inner Vessel nominal position projected in the plane orthogonal to each camera.
The points show the reconstructed position in the same plane. The maximum deviation 
from the ideal spherical shape is less than 20 cm in radius (each white circle in the figure
corresponds to 10 cm deviation from ideal shape). } 
\label{fig:vessel_pos}
\end{figure}

The camera pictures can be analyzed with an associated image processing software. This software
reconstructs the position of the nylon vessels with an accuracy of a few
cm. Fig. \ref{fig:vessel_pos} shows the reconstructed position of the Inner Vessel
as viewed by the different cameras and its deviation from the nominal shape.

\section{Detector performance}
\label{sec:Results}

A preliminary study of the performance of the Borexino detector has been carried out during 
several test runs (the so called  {\it Air runs}) before the detector was filled. These runs,
which occurred between 2002 and 2005, offered a unique opportunity to debug and tune--up the 
whole detector. During these runs it was possible to check the photomultiplier tube status 
and the full read-out chain performance. It was also possible to make a complete test of the online
and DAQ system. The trigger system was finalized and the possible trigger configurations 
(see section \ref{sec:Trigger}) were defined. 
Several dedicated runs with the laser were made, in order to develop and validate the timing and 
charge calibration procedure.

\subsection{PMT failure rate}
\label{sec:pmts_results}
All 2212 PMTs except for a few tens located on the floor of the SSS were installed in 
Borexino in the years 2001-2002. 
The remaining PMTs were installed in 2004 right after the installation on the nylon
vessels and immediately before closing the SSS.
Because of the well known problems at the Gran Sasso Laboratory, in the
years 2002-2006, the PMTs remained in air, being only 
occasionally operated in a few air runs. 
During this time, about 50 PMTs were lost, most of them showing flashing problem,
i.e. the emission of light from the bulb due to vacuum loss. 

The filling of the SSS with ultra-pure water began on August 2006 and finished in
November 2006. During this period about 80 PMTs were lost, most likely because
of defects or cracks in the sealing. We observed a clear infancy effect, i.e. the failure rate 
was much higher at the beginning, and most PMTs died shortly after having been 
submerged by water.

During the beginning of January 2007, until May 2007, the water in the 
SSS has been replaced by PC and, at the same time, the Water Tank was also filled
with ultra-pure water. Even in this case, we observed a high failure rate for a few weeks,
mostly because of imperfect or damaged connectors or cables in the 
Water Tank (we recall here that the PMT connectors pass through the SSS and are 
immersed in the water of the Water Tank).

The total number of dead PMTs at the beginning of data taking was 175. Since then,
the failure rate has gone down significantly, reaching the current plateau of about
3 lost PMTs per month.  At the time of writing (May 2008) the number of dead PMTs 
is 206 out of 2212.

\subsection{Charge and time calibration}
\label{sec:Calib}
\begin{figure}[!ht]
\begin{center}
\includegraphics[width=0.53\textwidth,clip=true]{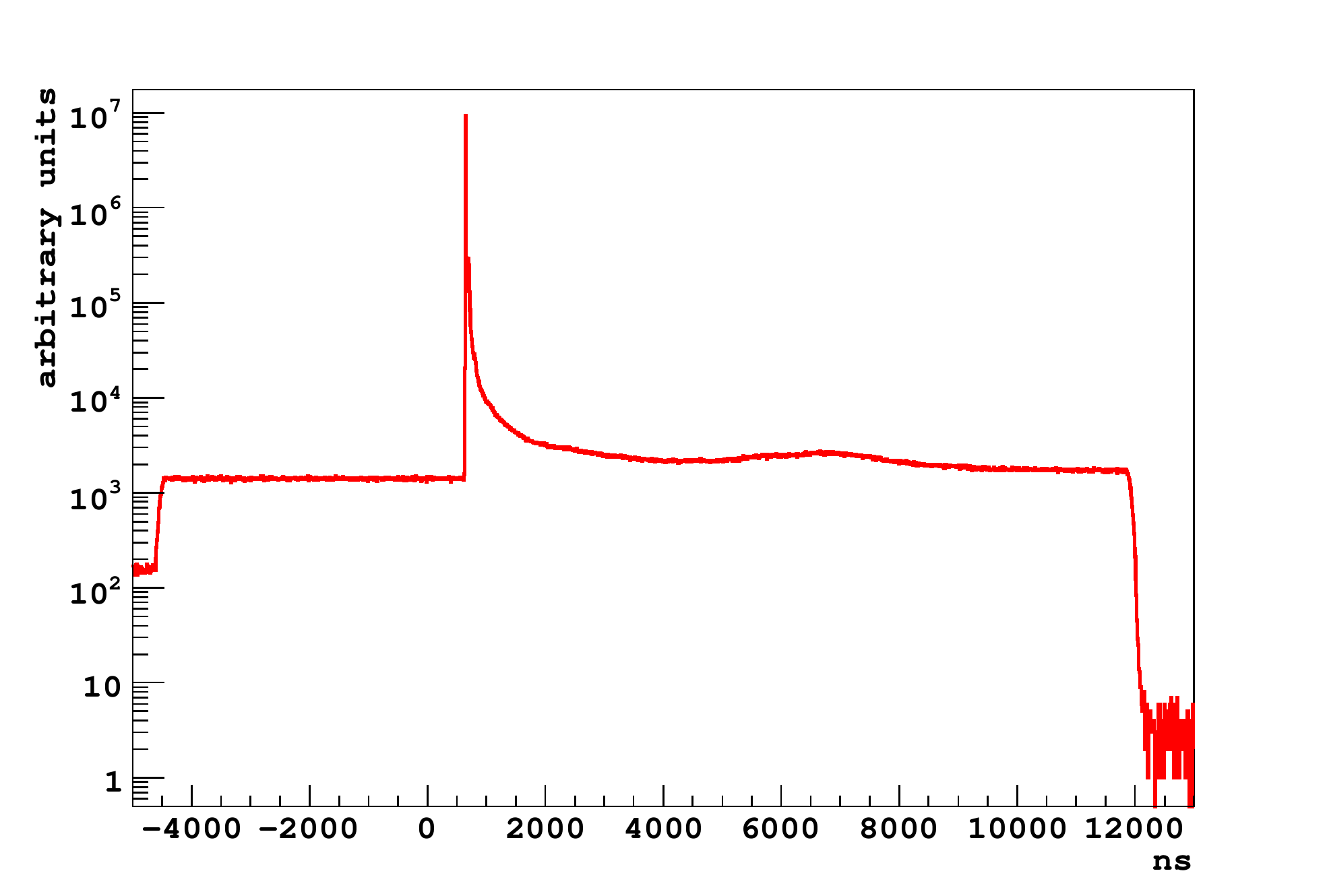}
\caption{Time response of all photomultipliers to the laser pulse in a typical 
laser run before any time alignment: the laser peak is clearly visible on a plateau of 
uncorrelated random noise hits. The data acquisition gate of about 16.5 $\mu$s is also visible. 
Although the intrinsic laser pulse is very narrow in time, the peak has a width of several tens 
of ns (with tails up to hundreds of ns) because of the light reflection and propagation on
the PMT glass and on the SSS surface. Note the vertical log scale. }
\label{fig:laser_raw}
\end{center}
\end{figure}

\begin{figure}[!ht]
\begin{center}
\includegraphics[width=0.53\textwidth,clip=true]{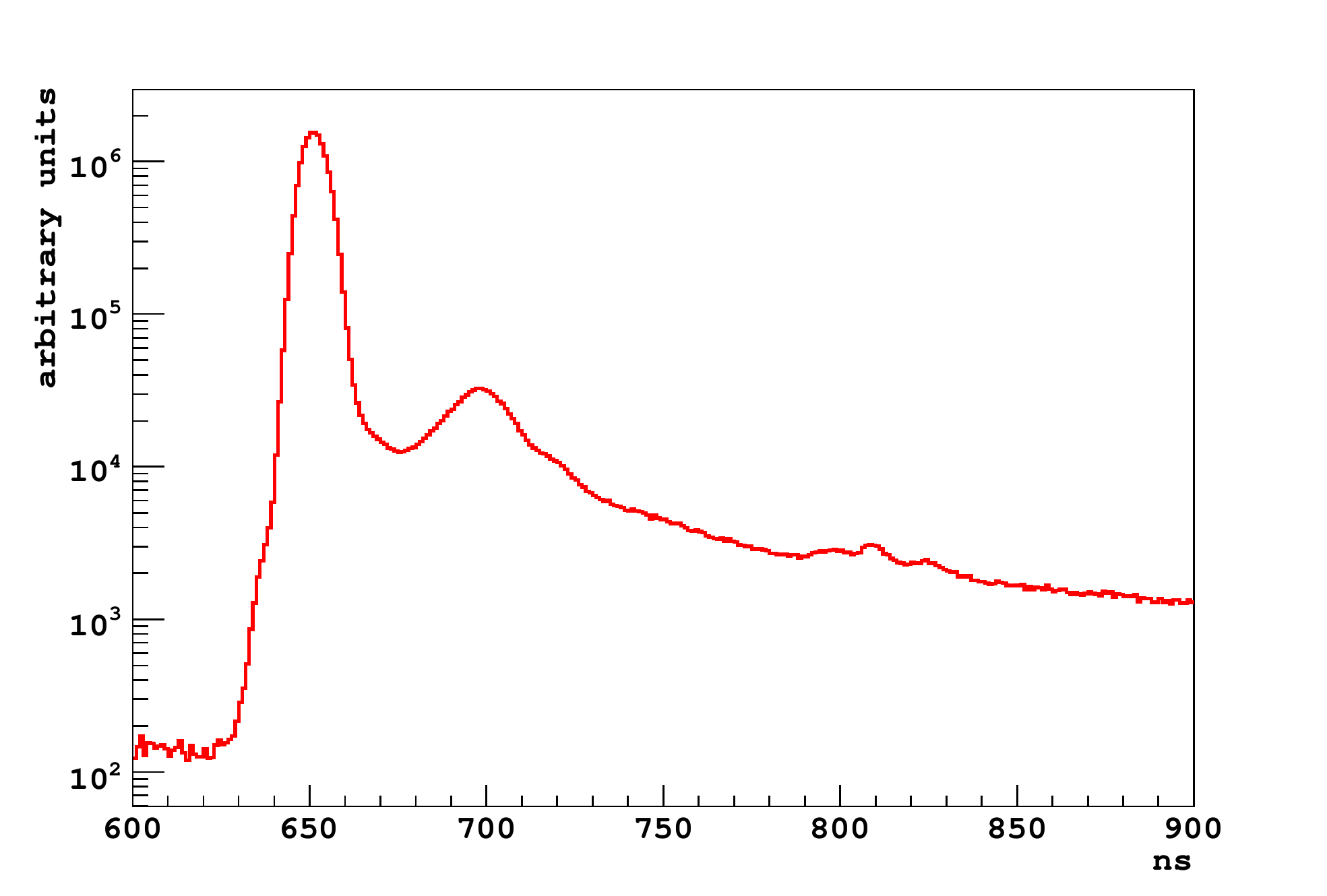}
\caption{This figure is the zoom of Fig. \ref{fig:laser_raw} between 600 and 900 ns within
the gate. The main laser peak is clearly shown, together with 2 additional peaks that
are due to reflection of light on the glass and its propagation through the SSS. The diameter
of the SSS is 13 m and the refractive index of PC is 1.5. The position of the second peak
is in good agreement with the expected maximum value of 64 ns. }
\label{fig:laser_raw_zoom}
\end{center}
\end{figure}

As discussed in section \ref{sec:CalibHardware}, the precise determination of 
photon arrival time at the photomultipliers (PMTs)
is crucial for Borexino to perform an accurate  position reconstruction of the events
and  for pulse shape discrimination purposes. Also, the measurement of total collected charge is  
used to determine the energy of the events.
For these reasons, an accurate and periodic time and charge calibration of the photomultipliers 
is of utmost importance. 
To accomplish this, a multiplexed optical fiber system has been designed which was described in
detail in section  \ref{sec:CalibHardware}.
The calibration strategy adopted for Borexino \cite{bib:MMonzaniThesis} requires that dedicated runs 
are taken periodically (at least once per week) with the laser on. 
During these runs a NIM timing unit simultaneously delivers
a driving signal to the laser and a trigger signal to the main Borexino electronics. 
A copy of this signal is also sent to dedicated electronics channels to precisely 
measure the time delay between the laser pulse and the arrival time of photons at the PMTs. 
The laser is usually driven at a repetition rate of 100 Hz.
The light intensity is selected in order to work at the single photoelectron 
level on each phototube. 
In these conditions the statistics for a good calibration run can be collected 
in less than 1 hour.
The typical plot of the time difference between the laser trigger and the 
response of the phototubes to the laser light is shown in Fig.~\ref{fig:laser_raw} 
(summing several thousands events and including all PMTs): 
the plot shows the $16.5~\mu$s  acquisition gate where the laser signal is 
clearly visible on top of a much smaller plateau of hits due to uncorrelated 
dark noise of the PMTs\footnote{In 2007 the acquisition gate was 7.2 $\mu$s and has been
changed since January 2008 in order to reduce the dead time.}. In Fig. \ref{fig:laser_raw_zoom}
(a zoom of Fig. \ref{fig:laser_raw}) the time spread of the laser peak is visible. The second
peak about 60 ns after the first one is due the light that is detected on the other side of the
SSS after being reflected by the PMT glass.

\begin{figure}[!ht]
\begin{center}
\includegraphics[width=0.53\textwidth,clip=true]{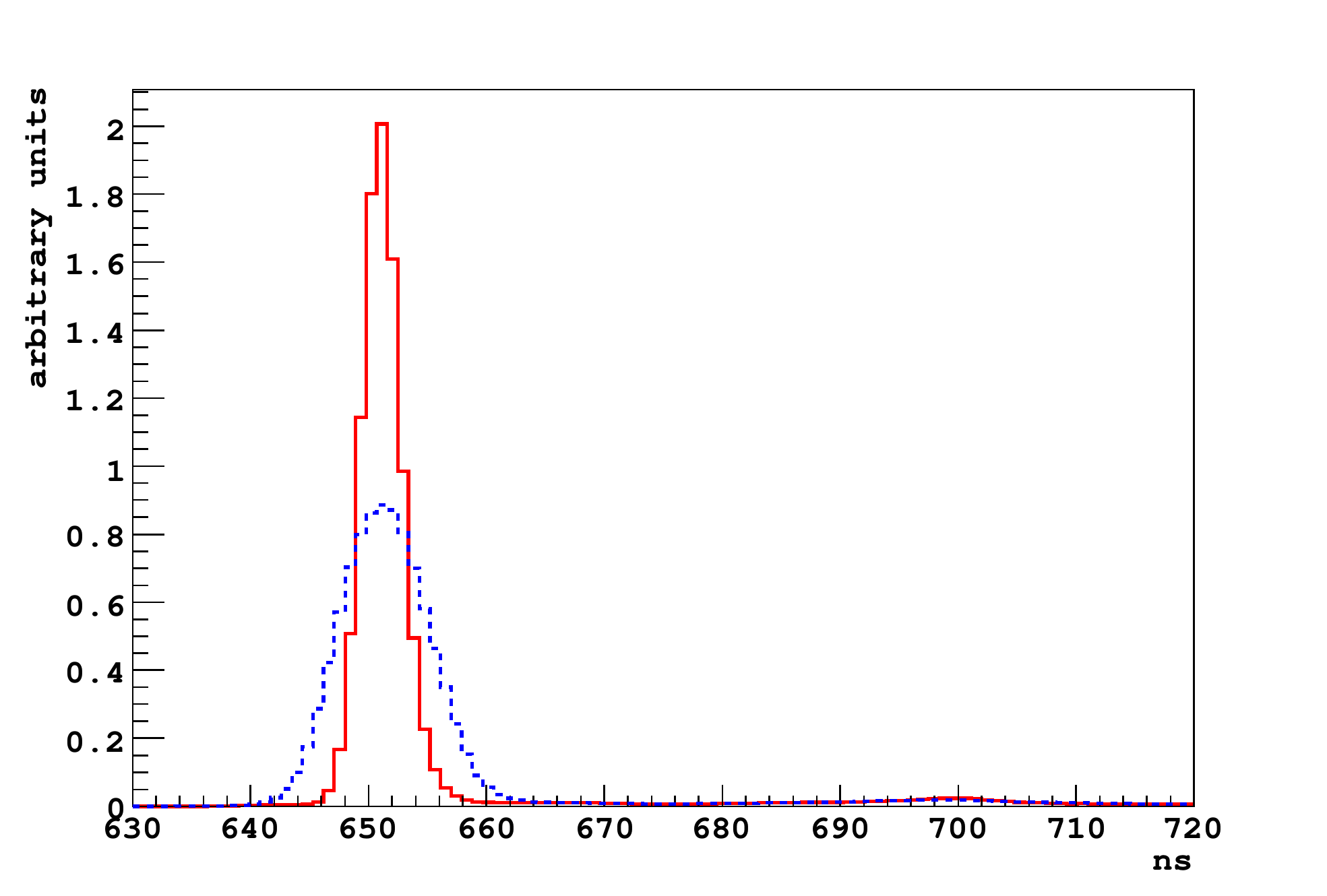}
\caption{Time response of the phototubes to the laser pulse before time alignment (blue dotted curve) 
and after time alignment (solid red curve). The width of the peak is reduced to the expected 
value of 1.6~ns. The wide shallow peak around 700 ns is due to the reflection of light on 
the PMT glass. This light is then detected on the other side of the SSS. }
\label{laser_corrected}
\end{center}
\end{figure}

The width of the laser peak is an indication of the time spread of the PMT 
response before any time alignment, since the laser light is sent simultaneously 
to all phototubes with an accuracy much better than 1 ns. 
The sigma of the distribution is  approximately 4~ns before alignment. 
In order to reduce this spread, the time response is studied individually for each
PMT and the position of the laser peak is calculated via a gaussian fit; 
the computed calibration constants are then written in the Borexino database 
and are used during the data processing to correct the response of each phototube 
on a channel-by-channel basis. Fig. \ref{laser_corrected} shows the response to 
the laser light for all phototubes before and   
after time alignment: the width of the gaussian is reduced by the calibration procedure 
to 1.6~ns, an excellent result  which is consistent with the unavoidable 
residual effects of PMT time jitter and intrinsic time resolution of the electronics.
\newline

\begin{figure}[!ht]
\begin{center}
\includegraphics[width=0.53\textwidth,clip=true]{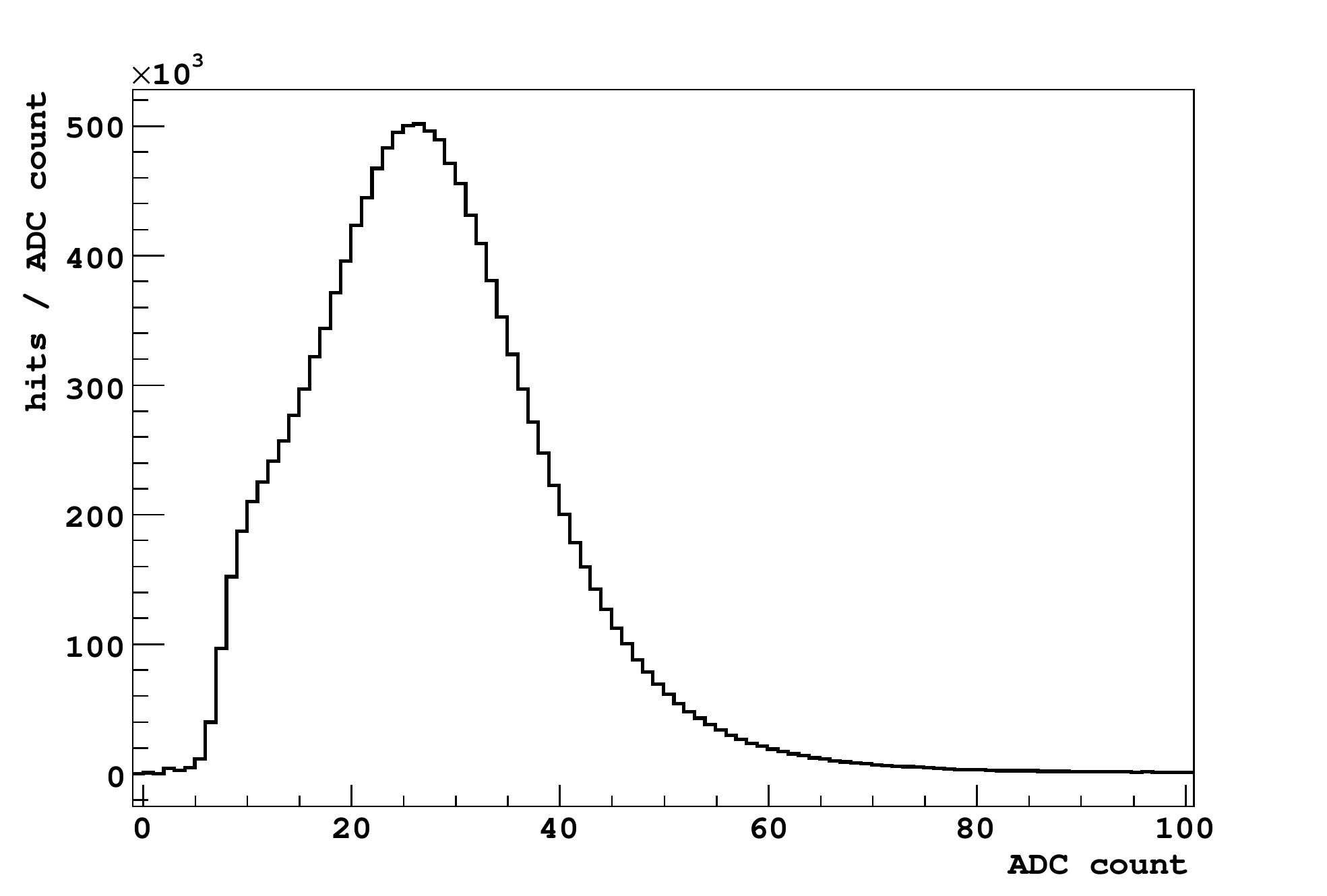}
\caption{Charge response of the phototubes (in ADC units) for a typical laser run. The
distribution is the sum of all live PMTs (about 2050). One photoelectron is approximately 
25 ADC counts.}
\label{fig:charge}
\end{center}
\end{figure}

\begin{figure}[!ht]
\begin{center}
\includegraphics[width=0.53\textwidth,clip=true]{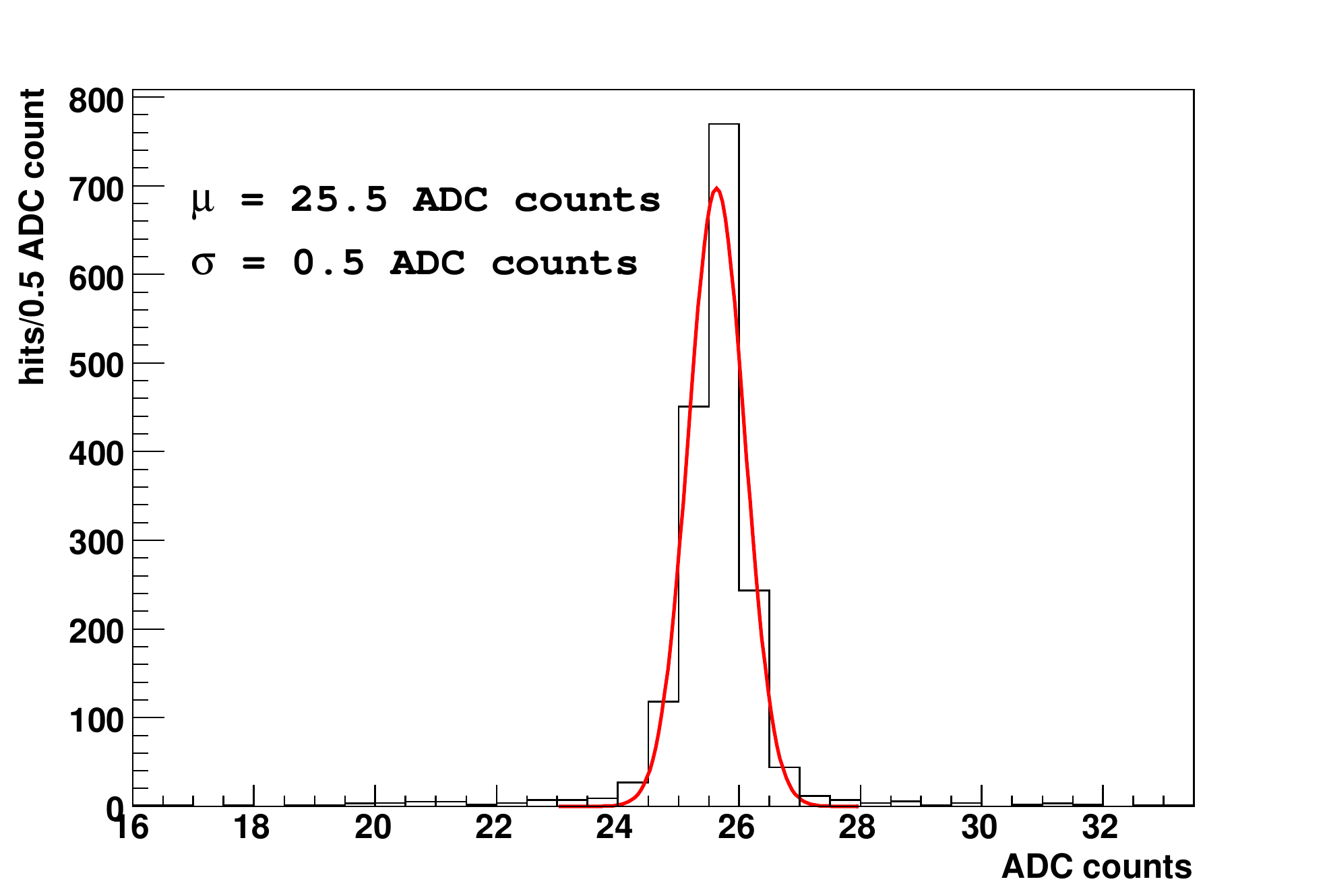}
\caption{Distribution of the position of the single photoelectron peak for about 2000 PMTs.
The position is computed by fitting the single photoelectron peak using laser events. This
uniformity was achieved by choosing the PMT High Voltage appropriately. No software
correction is applied to the charge gain. }
\label{fig:charge2}
\end{center}
\end{figure}

The same laser pulses are used to monitor the charge response of each phototube. 
Fig.~\ref{fig:charge} shows the charge spectrum (in photoelectron units) 
for all phototubes after calibration. The gain of all PMTs have been equalized by
choosing the optimal value of their High Voltage. Fig. \ref{fig:charge2} demonstrates that 
we managed to adjust the gain to be uniform within a sigma of less than 1 ADC count.

\section{Data Analysis}
\label{sec:per}
In January 2007, Borexino began filling with scintillator. A few weeks later, the detector 
was turned on and the first data were collected. Data taking continued during most of the 
filling and allowed detector monitoring and final hardware and software tuning, and
shifter training. 
Borexino filling was completed on May 15$^{th}$, 2007. The first available data have been very 
useful to understand the overall performance of the detector, in terms of energy and 
position reconstruction, the capability to tag delayed coincidence events, the
capability to disentangle $\alpha$ and $\beta$ particles, and the muon tagging 
efficiency. In the following, we report a few analysis results whose only
purpose is to show that the hardware described in this paper is working as expected. The following
is neither comprehensive nor should be regarded as a list of final results. A future paper
will report the details of the analysis techniques and the final detector performance.

\begin{figure}[!ht]
\begin{center}
\includegraphics[width=0.53\textwidth,clip=true]{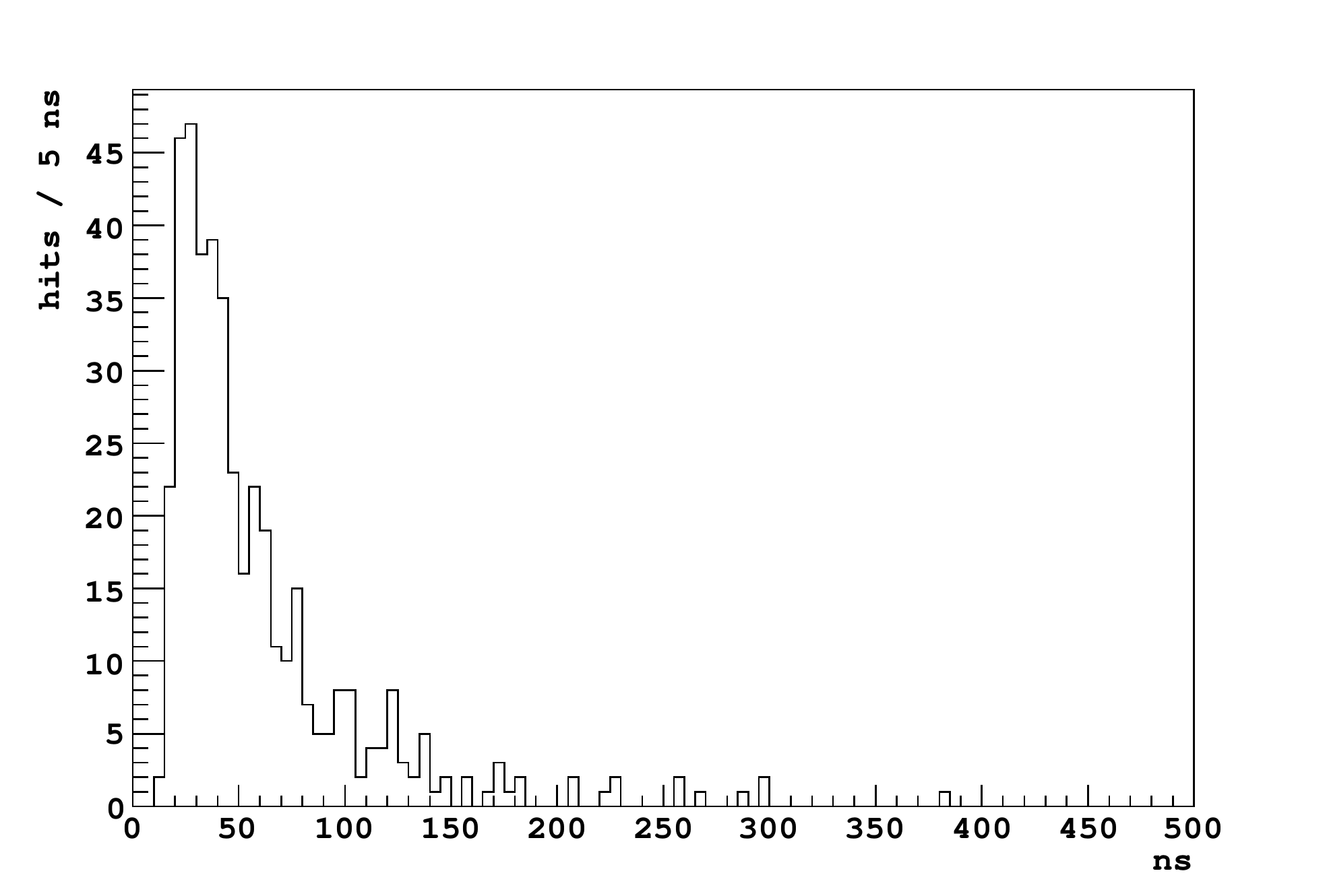}
\caption{Time distribution of the PMT hits in a single cluster event.}
\label{fig:cluster}
\end{center}
\end{figure}

\subsection{Scintillation events}
An event in Borexino is a collection of PMT hits occurring within a time window of a 
few tens of ns with a long tail that can reach a few hundreds ns. When the 
trigger fires, however, all hits occurring within the DGS of 16 $\mu$s are recorded (see
section \ref{sec:Laben} for details). In this long gate both the signal hits and the
random noise hits of the 2212 PMTs are recorded. Although the number of random
noise hits is relatively small (about 1 per $\mu$s to be compared with a signal of about 100
hits in less than 200 ns at 200 keV),  a software procedure is necessary to identify both 
the beginning of the scintillation pulse and its end. We define a {\it cluster} 
a collection of hits belonging to the same scintillation event, and we call {\it clustering}
the offline software procedure that is applied to identify the {\it cluster}.   A typical 
cluster is shown in Fig. \ref{fig:cluster}. 

The {\it clustering} procedure determines the time position of the rising edge of each {\it cluster}
with a resolution of less than 1 ns, allowing the measurement of the time difference between
close events with that precision. 

\begin{figure}[!ht]
\begin{center}
\includegraphics[width=0.53\textwidth,clip=true]{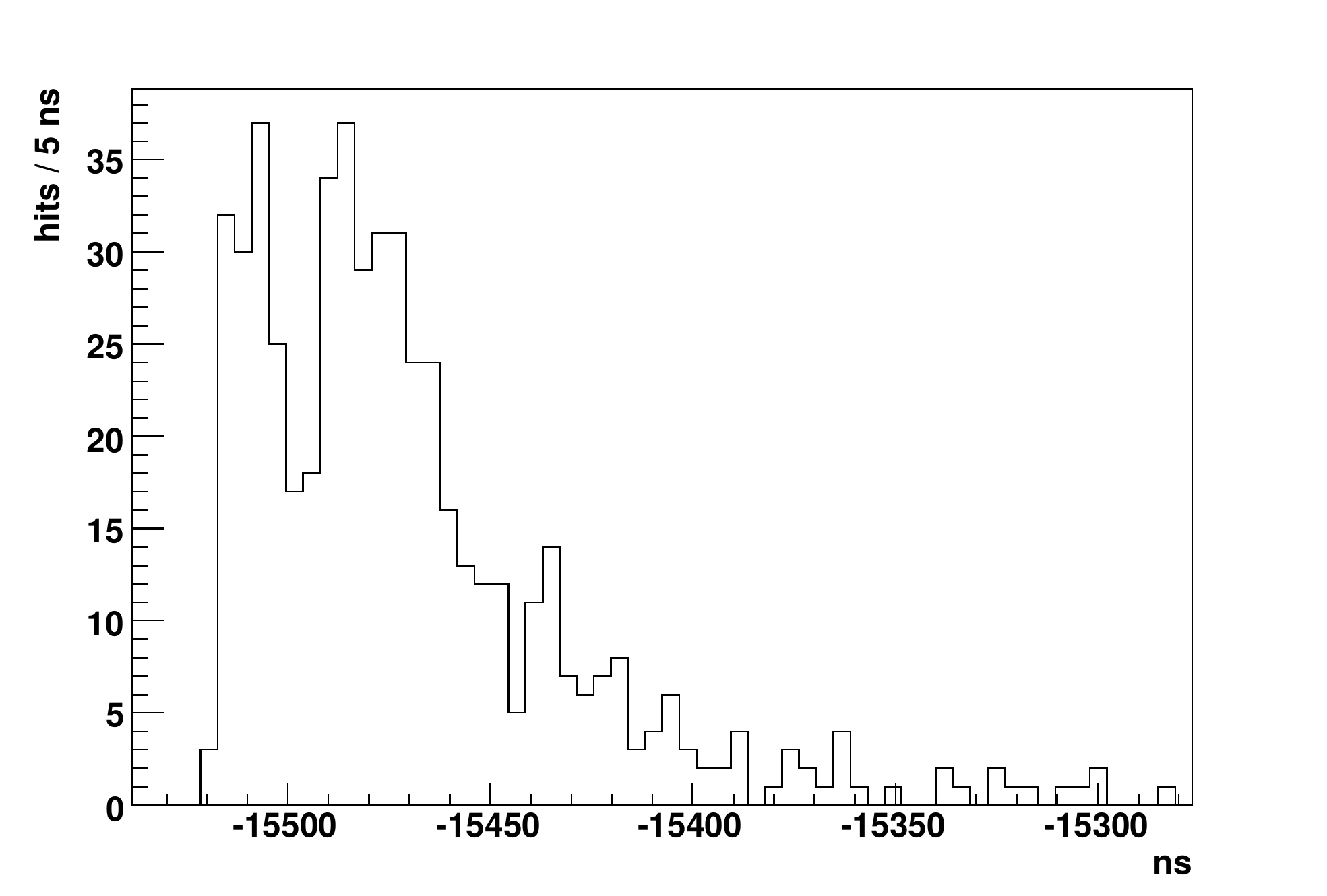}
\caption{Two clusters partially overlapped and identified by the splitting software. The time
distance of the two clusters is in this case about 25 ns. Clusters
as close as 15 ns can be efficiently identified.}
\label{fig:cluster2}
\end{center}
\end{figure}

Each {\it cluster} is also analyzed to check whether it is compatible with being a unique 
scintillation event or it is the overlap of more than one event. This is important both to 
reject pile-up events and to identify very fast concidences like those due to 
\bipo\ decay or $^{85}$Kr. Fig. \ref{fig:cluster2} show an example of a two cluster event,
a good \bipo\ candidate. 

The time distribution of the hits within the cluster allow the reconstruction 
of the position of the event, the determination of the total energy deposited 
in the scintillator and the discrimination of $\alpha$ and $\beta$ particles.
The following sections describe these points in more details. 

\subsection{Energy and position reconstruction }
 
\begin{figure}[!ht]
\begin{center}
\includegraphics[width=0.5\textwidth,clip=true]{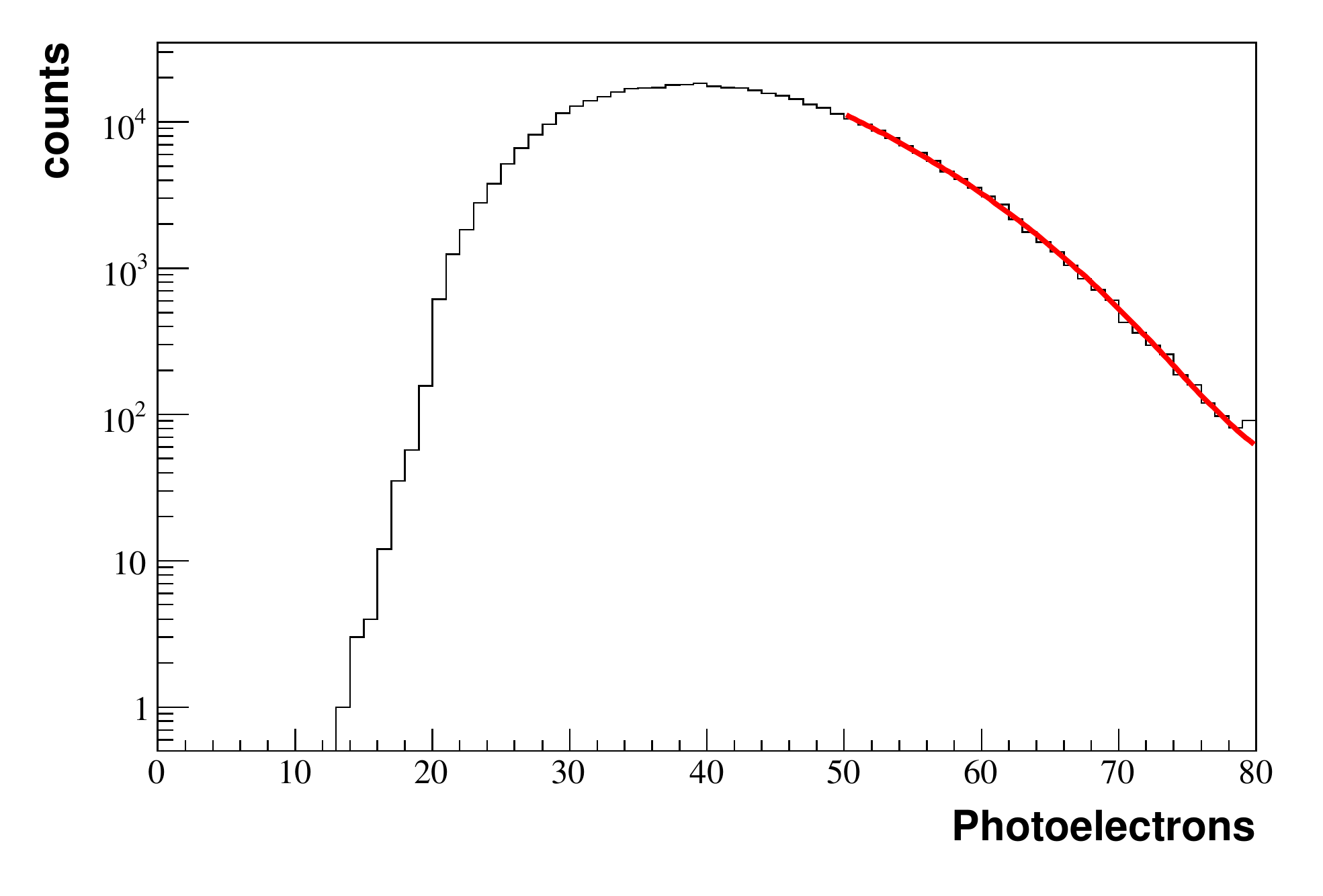}
\caption{ Low energy portion of the Borexino spectrum: a fit is performed to the $^{14}$C beta 
spectrum including the effect of beta ionization quenching, the beta decay form factors
and the background shape. The spectrum does not start from 0 photoelectrons 
because of the trigger threshold. }
\label{fig:c14}
\end{center}
\end{figure}

One of the most crucial points for Borexino is the energy resolution, which ultimately depends on
the light yield, {\it i.e.} number of photoelectrons per MeV collected by the PMTs.
A preliminary measurement of the detector light yield was obtained  by fitting the 
shape of the $^{14}$C spectrum: 
we recall  that the $^{14}$C isotope is present in an organic liquid like pseudocumene 
and is therefore an unavoidable background which affects the low energy part of the 
Borexino spectrum ($\beta$ decay with an endpoint of 156 keV).
Fig.~\ref{fig:c14} shows the beta spectrum of $^{14}$C and the result of the fit: the fitting function
includes the effect of energy resolution, beta ionization quenching 
(through the Birks' factor K$_B$, see for example \cite{bib:Birks}), 
the beta decay form factors and background parameterization (exponential + one gaussian).
The resulting conversion factor, normalized to 2000 phototubes, 
is found to be approximately 500 photoelectrons/MeV. 
The systematic error associated with the fit is estimated to
be of the order of 5\%.

\begin{figure}[!ht]
\begin{center}
\includegraphics[width=0.53\textwidth,clip=true]{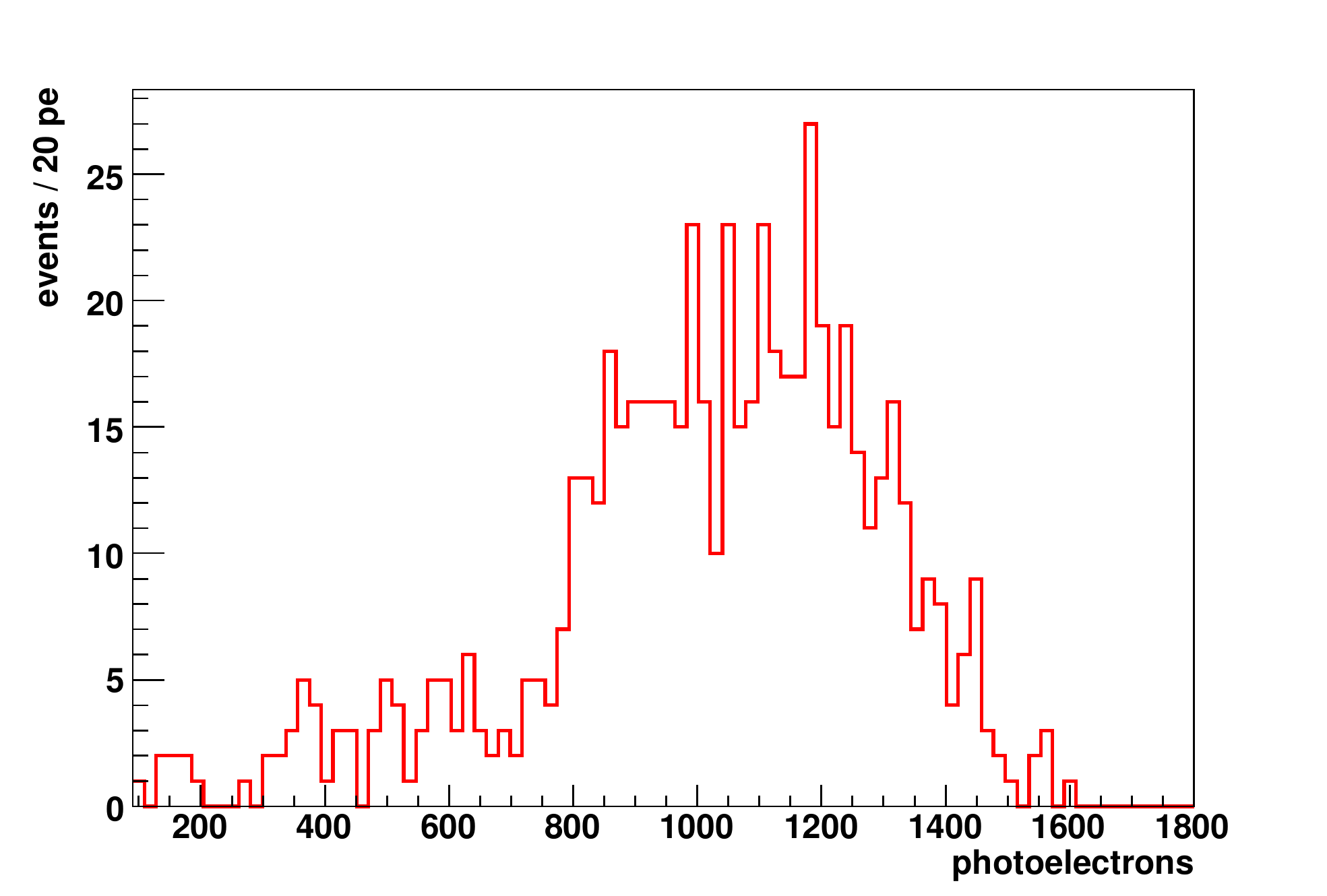} \\
\includegraphics[width=0.53\textwidth,clip=true]{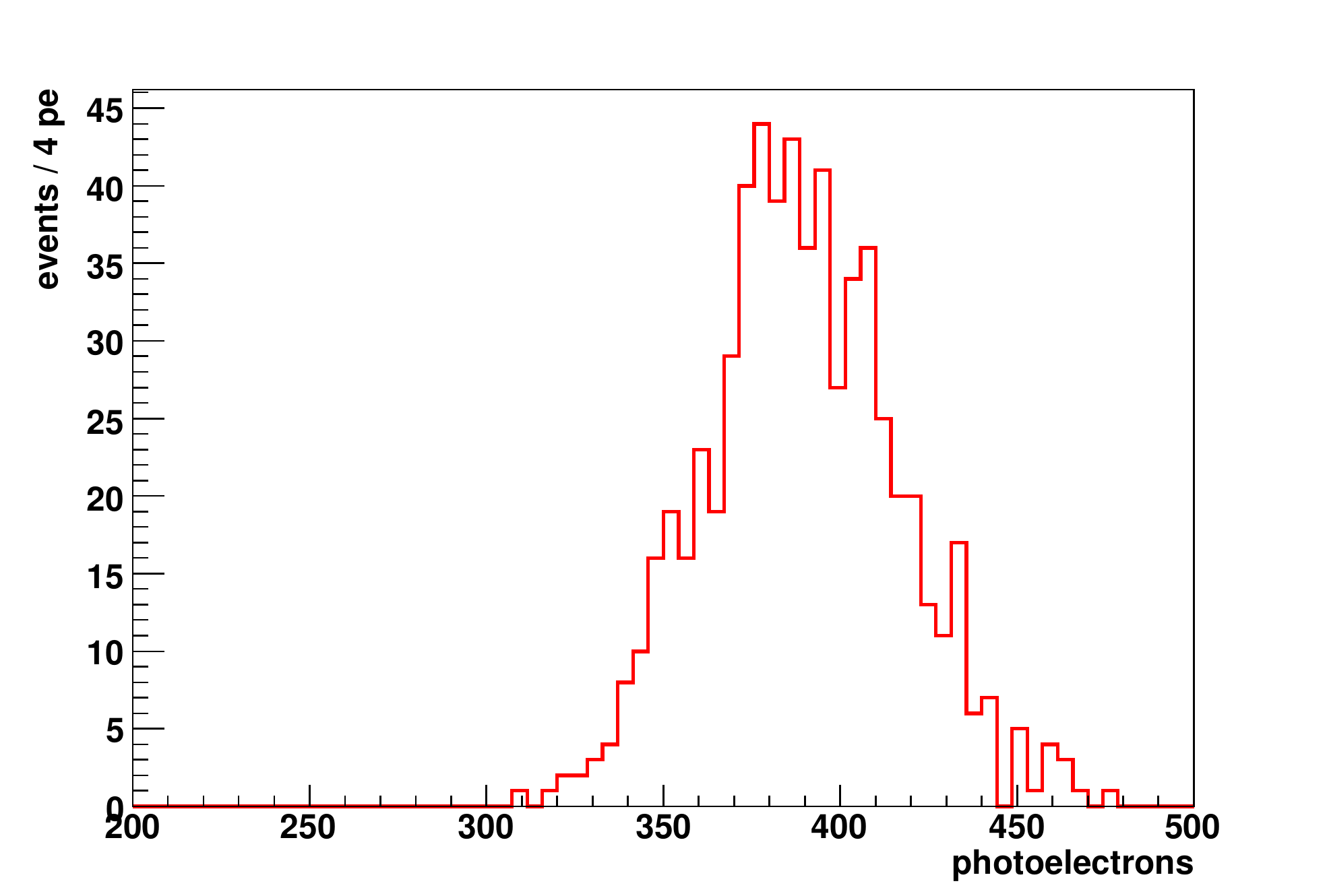}
\caption{  Energy spectrum for the first (top) and second (bottom) event which pass the coincidence
selection cuts. The first plot is compatible with the expected $\beta$ + $\gamma$ spectrum of $^{214}$Bi. The second plot clearly shows  the $^{214}$Po alpha peak. The events are all 
selected in the fiducial volume (100 t nominal) and correspond to about 1 year of data. 
The relatively poor statistics is a consequence of the extreme purity of the scintillator.}
\label{BiPo_spectra}
\end{center}
\end{figure}

\begin{figure}[!ht]
\begin{center}
\includegraphics[width=0.53\textwidth,clip=true]{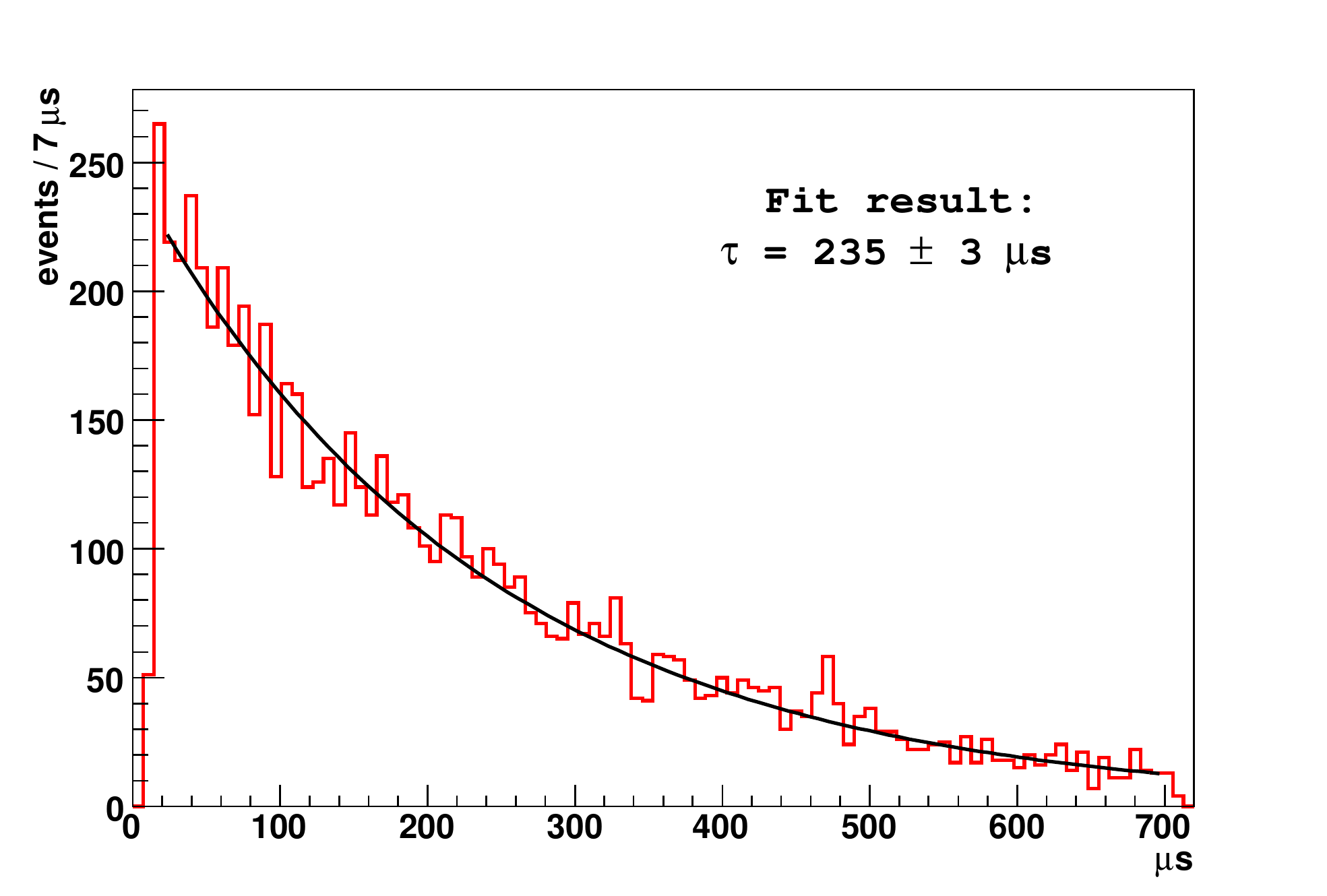}
\caption{Time difference between the first and second event of the \Bipo ~coincidence sample 
(see text for selection cut details). A fit with an exponential + a constant is performed. The resulting
lifetime is $\tau$ = 235 $\pm$ 3  $\mu$s.}
\label{BiPo_lifetime}
\end{center}
\end{figure}

The presence of some radon caused by a small air leak (a few liters) which developed in the system
during filling afforded the opportunity to further understand the performance of the detector: 
we recall that $^{214}$Bi and $^{214}$Po are two isotopes produced in sequence in the radon chain. 
$^{214}$Bi decays by beta or beta+gammas (Q value = 3.23 MeV) emission, 
while $^{214}$Po decays via alpha emission (E$_{\alpha}$ = 7.7 MeV).
The short lifetime of $^{214}$Po ($\tau$ = 236.6 $\mu$s) gives the opportunity to tag the 
two event sequence very efficiently 
providing a sample virtually free of background. 
Fig.~\ref{BiPo_spectra}  shows the energy of $^{214}$Bi and $^{214}$Po candidates after
requiring that the two events have occurred within a time window smaller than 5 times the 
$^{214}$Po lifetime. The first plot is compatible with the $\beta + \gamma$ spectrum of
$^{214}$Bi, while the second plot clearly shows the $^{214}$Po alpha peak. 
The reconstructed energy resolution is 7 \% (1 $\sigma$).
It is worthwhile to notice that the
position of this peak corresponds to an equivalent energy on the beta scale which is 
approximately 1/10th of the nominal alpha energy; this reduction of light in the case of alpha particles
is due to the well-known light quenching 
effect \cite{bib:Birks}.
Selecting only events under the $^{214}$Po peak and plotting the time difference 
between the first and second event, one obtains 
Fig.~\ref{BiPo_lifetime}: a fit to the time difference with an exponential
plus a constant (to account for the background of random coincidences mostly coming 
from $^{14}$C events) gives $\tau = (235\pm 3) \mu$s which is in  very good agreement 
with the expected $^{214}$Po lifetime.
The sample of \Bipo ~coincidences also provides a useful tool to study the performance of spatial
reconstruction algorithms. Since the $^{214}Po$ lifetime is very short compared to the typical 
diffusion time, the two events occur exactly in the same position. Therefore the reconstructed 
distance between the two events depends only on the spatial resolution of the reconstruction 
algorithm\footnote{ Some care should be taken because $^{214}Bi$ decays 
81~\% of the time emitting a $\gamma$, and therefore the light barycenter could move from the 
decay position.}. 

\begin{figure}[!ht]
\begin{center}
\includegraphics[width=0.53\textwidth]{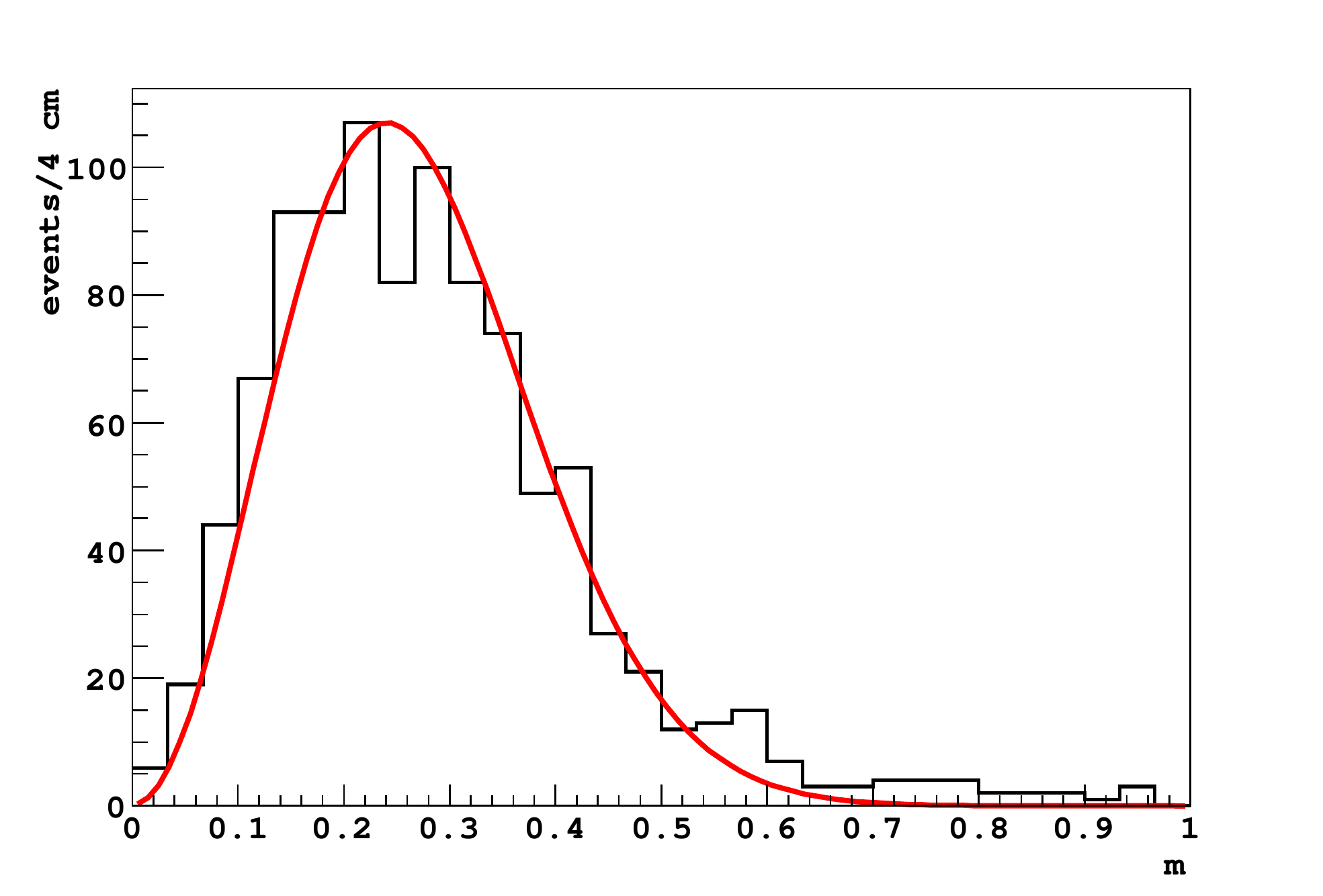}
\caption{Distribution of the distance between the two events of a \Bipo\ coincidence. From this
plot, a resolution of 16 cm is obtained by fitting the distribution with the correct expected
curve (not a gaussian).}
\label{fig:BiPo_spectra}
\end{center}
\end{figure}

Fig.~\ref{fig:BiPo_spectra} shows the  histogram of reconstructed distance for the clean sample of 
coincidence events. By fitting the distribution with the correct non-gaussian distribution, 
the position resolution at these energies is found to be 16~cm, in good agreement with expectations
and Monte Carlo simulations.

\begin{figure}[!ht]
\begin{center}
\includegraphics[angle=+90,width=0.53\textwidth]{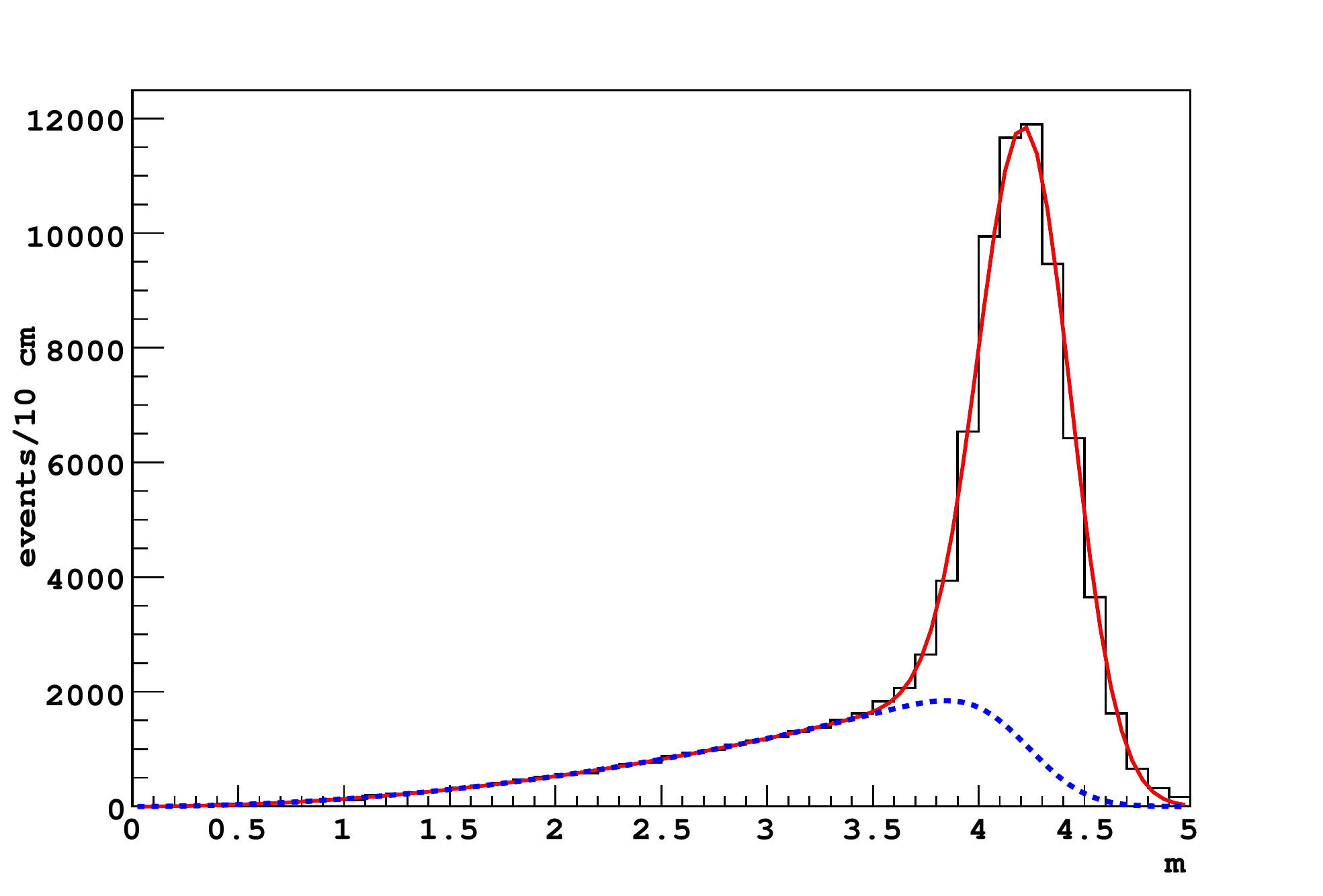}
\caption{Reconstructed radius for events in the energy region between 100 and 200 
photoelectrons. The distribution can be fit with two contributions: the internal one 
(blue dotted line) and the external one (red continuous line)}
\label{fig:radial_fit}
\end{center}
\end{figure}

The position reconstruction is the crucial tool to define the fiducial volume by means of 
a software cut.  
As an example, Fig. \ref{fig:radial_fit} shows the radial distribution of events detected 
in the energy region between (100-200)photoelectrons.  
The plot can be fit as the sum of two contributions: the internal volumetric distribution (mainly
due, in this case, to $^{210}$Po alphas) which
goes as r$^2$ and the external background distribution (mainly due to PMT's radioactivity) 
which is well approximated by a gaussian centered at the vessel border. 

\subsection{$\alpha$-$\beta$ discrimination}
\begin{figure}[!ht]
\begin{center}
\includegraphics[width=0.53\textwidth]{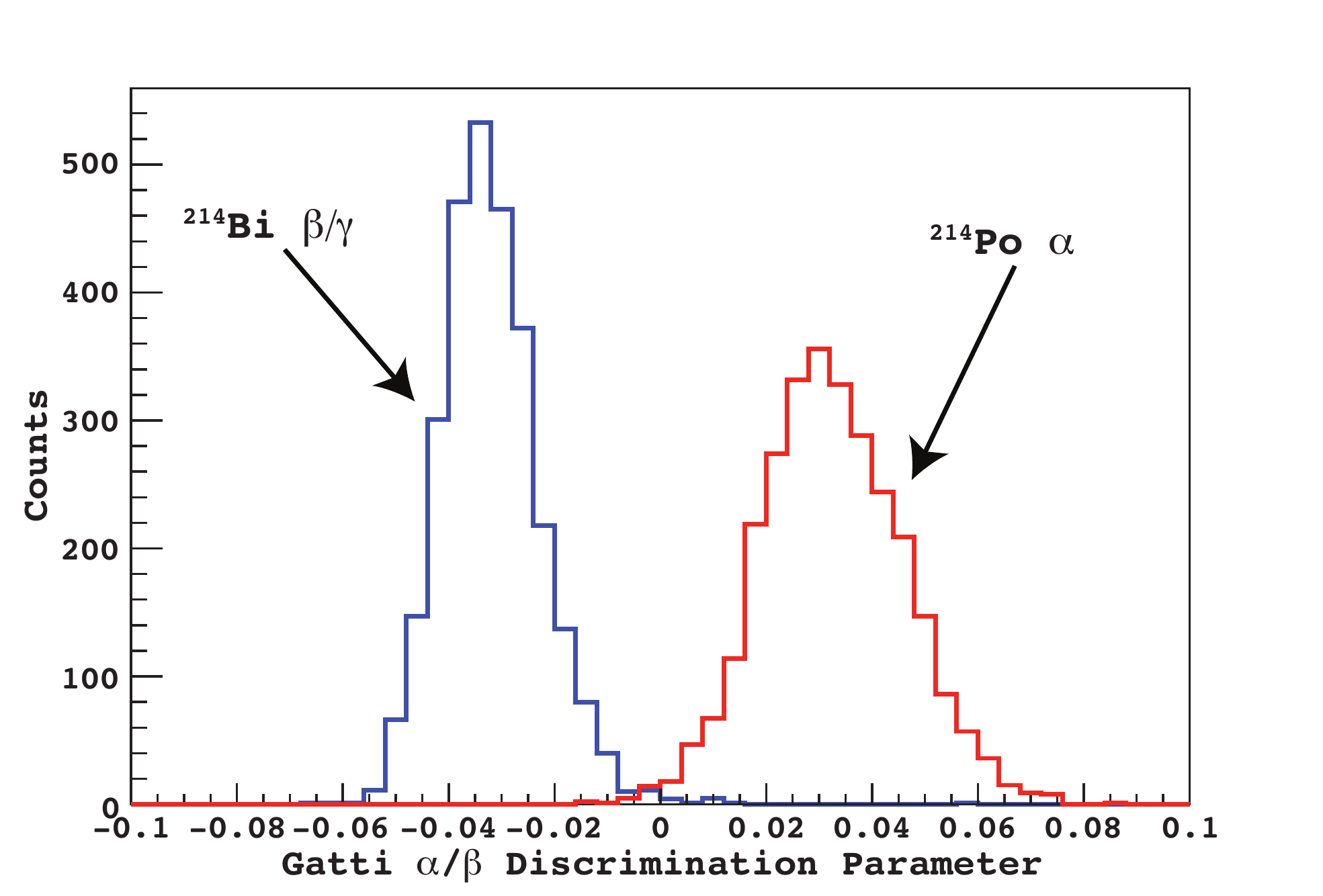}
\caption{Gatti variable for \Bipo\ events. The two components are clearly separated. }
\label{fig:gatti}
\end{center}
\end{figure}

The \Bipo\ delayed coincidences provide a nice and clean sample of 
$\alpha$ and $\beta$ events that can be used to develop and tune the $\alpha$/$\beta$ 
discrimination algorithms. We recall that these algorithms are based on the different 
time response of the scintillator 
to $\alpha$ and $\beta$ particles (see Fig.~\ref{fig:alphabeta}). 
In this paper we just 
show the results of the effectiveness of $\alpha$/$\beta$ separation based on the Gatti optimal filter 
\cite{bib:gatti}. Fig. \ref{fig:gatti} shows the Gatti variable for the $\alpha$ and $\beta$ samples 
obtained from the \Bipo\ analysis: the two curves are very well separated.
It should be pointed out that the $\alpha$/$\beta$ discrimination
power is expected (and verified) to be energy dependent and to get worse at lower energy. 
A comprehensive study on this issue will be reported in a future paper.

\subsection{Muons and neutrons}

\begin{figure}[!ht]
\begin{center}
\includegraphics[width=0.53\textwidth]{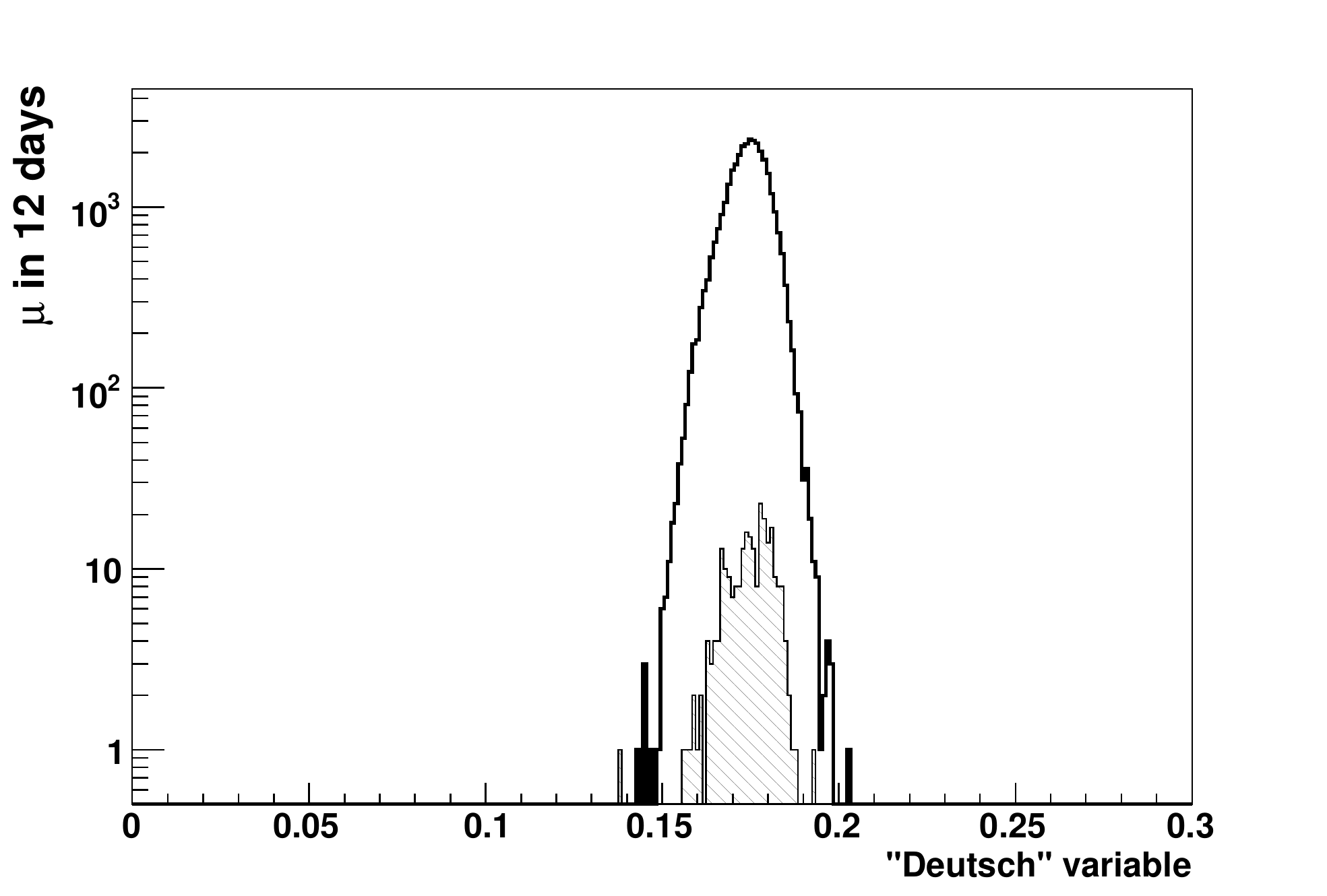}
\caption{Outer detector muon tag efficiency. The full curve is the distribution of muon events selected
by means of the inner detector requiring at least 6000 photoelectrons collected by the PMTs. The shaded
area is the distribution of muons that do not have an outer detector tag. From the figure, an efficiency
of $>$ 99\% is measured. The x axis variable ("Deutsch" variable) is the ratio of the collected charge
seen by inner PMTs without light cone divided by the total inner detector charge. }  
\label{fig:muoneff}
\end{center}
\end{figure}

Finally, we have performed a preliminary study of the muon tagging efficiency. Both inner and outer 
detector can identify muons. The outer detector does so by means of the \che\ light produced in the external Water Tank by the crossing muons. A study of the detector behavior 
shows that, with a threshold of 6 photoelectrons, the
muons are very efficiently detected with negligible random noise. The inner detector can
separate muon events from scintillation events by means of several shape variables. In this paper
we do not discuss this matter further, except to show the outer detector efficiency measured by
means of a clean sample of muons that cross the inner detector. This sample is selected requiring
a huge energy deposit in the scintillator ($>$6000 photoelectrons). Fig. \ref{fig:muoneff} shows
that the efficiency of the outer detector for this class of events is $>$ 99\%. 
In the physics analysis, the muon rejection is done using both outer and inner detector capabilities. 
The detailed study of the overall muon tagging efficiency will be reported in a future paper. 

\begin{figure}[!ht]
\begin{center}
\includegraphics[width=0.52\textwidth]{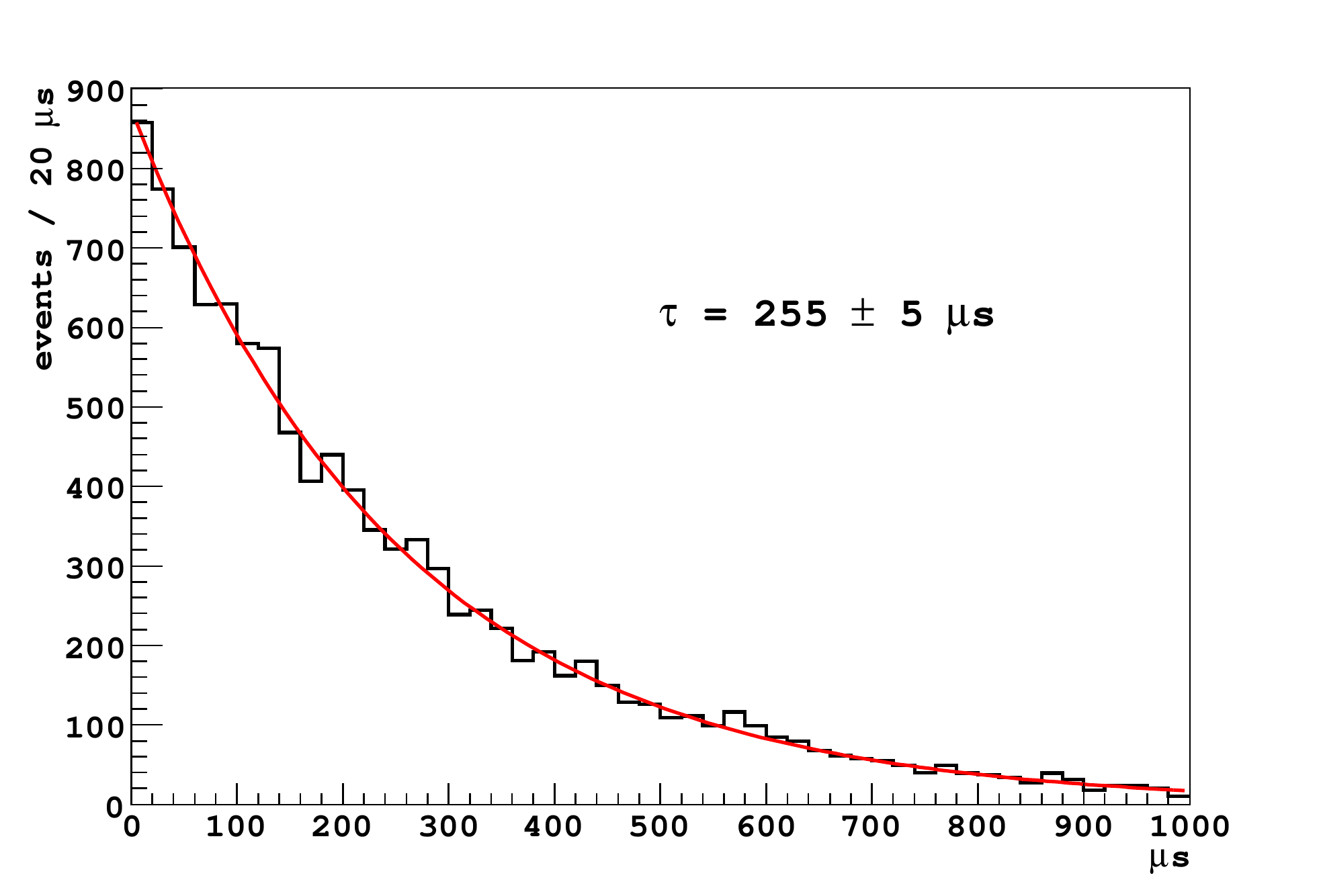}
\caption{Distribution of the spallation neutron capture time in the scintillator. Muons crossing the
scintillator or the buffer occasionally produce spallation neutrons that are then captured by
protons and detected by means of the emitted gamma from deuteron. }  
\label{fig:neutron}
\end{center}
\end{figure}

We can also efficiently detect spallation neutrons produced by the muons crossing the 
scintillator volume. The neutrons are quickly thermalized in the scintillator and are captured
by protons with the emission of the characteristic 2.26 MeV gamma. 
Fig. \ref{fig:neutron} shows that the capture time distribution is in good agreement 
with the expected value. This result was obtained after the modification to the triggering
scheme described in section \ref{sec:Trigger}. 

\section{Conclusions}
\label{sec:Conclu}
The construction and the commissioning of the Borexino detector is 
completed. Data taking has begun on May 15th, 2007 and is going to continue for several years. 
This paper shows that the detector meets, or in some cases  exceeds, 
the expected performance. The radioactive background is lower than the design values
for several contaminants, particularly for the $^{238}$U and $^{232}$Th daughters. The
PMTs of both inner and outer detectors, the electronics, and the trigger system work 
as expected.  

\section*{Acknowledgements}
We sincerely thank the funding agencies: INFN (Italy), NSF (USA), BMBF, DFG and MPG (Germany), 
Rosnauka (Russia), MNiSW (Poland), and we acknowledge the generous support of the Laboratori Nazionali del Gran Sasso.
This work was also supported by the ILIAS integrating activity
(Contract No.RII3-CT-2004-506222) as part of the EU FP6 program.

This paper is dedicated to the memory of  Cristina Arpesella, Martin Deutsch, 
Burkhard Freudiger, Andrei Martemianov and Sandro Vitale, and to John Bahcall, 
a friend and strong supporter of Borexino. 
  
%
%

\end{document}